\documentclass[twocolumn]{aastex63}
\usepackage[cmex10]{amsmath}
\usepackage{amssymb}	

\newcommand\kms{km\,s$^{-1}$}
\newcommand{\asec}{$^{\prime\prime}$}
\newcommand{\amin}{$^{\prime}$}
\newcommand{\chg}{ } 

\received{18 Oct. 2023}
\revised{20 Nov. 2023}
\accepted{20 Nov. 2023}
\submitjournal{ApJ}
\shorttitle{MeerKAT Supernova Remnants}
\shortauthors{Cotton et al.}

\begin{document}

\title{MeerKAT 1.3 GHz Observations of Supernova Remnants}

\correspondingauthor{William Cotton}
\email{bcotton@nrao.edu}

\author[0000-0001-7363-6489]{W.~D.~Cotton}
\affiliation{National Radio Astronomy Observatory \\
520 Edgemont Road \\
Charlottesville, VA 22903, USA}
\affiliation{South African Radio Astronomy Observatory \\
2 Fir Street \\
Cape Town, 7925, South Africa} 

\author[0000-0001-5953-0100]{R.~Kothes}
\affiliation{Dominion Radio Astrophysical Observatory,  \\
Herzberg Astronomy and Astrophysics Research Centre,  \\
National Research Council Canada, PO Box 248, Penticton, \\
BC V2A 6J9, Canada}  

\author[0000-0002-1873-3718]{F.~Camilo}
\affiliation{South African Radio Astronomy Observatory \\
2 Fir Street \\
Cape Town, 7925, South Africa}

\author[0000-0002-0844-6563]{P.~Chandra}
\affiliation{National Radio Astronomy Observatory \\
520 Edgemont Road \\
Charlottesville, VA 22903, USA}  

\author{S.~Buchner}
\affiliation{South African Radio Astronomy Observatory \\
2 Fir Street \\
Cape Town, 7925, South Africa}

\author[0000-0002-8973-4072]{M.~Nyamai}
\affiliation{South African Radio Astronomy Observatory \\
2 Fir Street \\
Cape Town, 7925, South Africa}



\begin{abstract}
We present full Stokes MeerKAT L band (856--1712\,MHz) observations of \chg{36} high latitude supernova remnants.  Sensitive, high dynamic range images show a wealth of structure.  G15.1$-$1.6 appears to be an HII region rather than an SNR. G30.7$-$2.0 consists of three background extragalactic sources which appear to form an arc when imaged with much lower resolution. At least half of the remnants in the sample contain ``blowouts'', or ``ears'' showing these to be a common feature. Analysis of the polarimetric data reveals details of the magnetic field structure in the emitting regions of the remnants as well as magnetized thermal plasma in front of polarized emission.  The chance alignment of G327.6+14.6 with a background AGN with very extended polarized jets allows testing for the presence of Faraday effects in the interior of the remnant.  Scant evidence of Faraday rotating material is found in the interior of this remnant.
\end{abstract}

\keywords{Supernova Remnants}


\section{Introduction} 
Supernova remnants (SNRs) form when  envelopes of massive progenitor stars are ejected with supersonic speeds ($\sim$ 5,000-20,000 \kms) in supernova  explosions. A massive star can end its life in two ways - 1:) thermonuclear supernovae where a C-O white dwarf resulting from a 4--8 \,$M_\odot$ progenitor star accretes matter from its companion in a binary system and eventually leading to thermonuclear runaway reaction; 2:) core collapse supernovae resulting from collapse of $\ge 8 \, M_\odot$ massive stars once the core has exhausted its nuclear fuel, leaving a compact remnant, most often a neutron star. The supersonic explosion ejecta interacts with the surrounding 
circumstellar medium (CSM) or interstellar medium (ISM) resulting in a shock front at the outer edge. 
In general SNRs have three major evolutionary stages;  first, ejecta dominated (ED) stage where ejected mass and explosion energy define their observed properties; when the mass swept up by the shock exceeds the \chg{ejecta} mass, the ED phase transitions into Sedov–Taylor (ST) phase. The  explosion energy is still an important parameter in this phase, and the radiative losses are negligible.  For older SNRs, post ST phase, radiative losses become  important resulting in  formation of a  cool dense shell just behind the shock front. This is called the pressure dominated phase as a SNR is driven by the interior pressure on the cool dense shell. As the shock front slows down with time, the shock front speed becomes slow enough to become indistinguishable from the ISM.

Supernova remnants play a crucial role in the evolution of galaxies by enriching the interstellar medium with heavy elements and contributing to  formation of new stars. They provide important clues to the evolution leading to the dying moments of the stars, and energy injection to the Galaxy produced via supernovae explosion\chg{s}.

The radio emission observed from SNRs primarily arises via synchrotron radiation, emitted  when high-energy electrons accelerated during the supernova explosion, spiral along magnetic field lines. 
In some cases, thermal radio emission is produced when an expanding SNR shockwave encounters denser surrounding medium. 
The radio emission from SNRs can vary depending on various factors, including the age of the remnant, the energy of the explosion, the properties of the surrounding interstellar medium, and the interaction of the shockwave with nearby clouds or structures, \chg{and} hence carries a plethora of information.

\citet{Green2019} has presented a sample of 294 Galactic SNRs. This list has been recently revised by \citet{rl23}, who presented a revised table of 390 Galactic SNRs, however, some of them are SNR candidates.
This is much smaller than the predicted population of a few thousand Galactic SNRs \citep{rl22}.
 However, amongst the detected ones, only a small fraction have been well characterized to understand their explosion type, age and evolutionary state, as well as explosion energy \citep{lw17}. 

 Several mature SNRs in radio bands reveal bilateral or barrel-shaped structures which can be characterized by a symmetry axis and show a connection with the Galactic magnetic field \citep{vanderLaan1962}. Small scale structures which are commonly seen is some radio SNRs are protrusions, which are also called ``ears''. There are several scenarios proposed for the formation of ears. \chg{The t}wo most popular scenarios are, via the  launch of two jets during or post supernova explosion causing protrusion in the forward shock \citep[e.g.,][]{Gaensler1998,Bear2017}, and via the interaction of the SNR with a bipolar circumstellar environment \citep{Chiotellis2021}.

SNRs are considered one of the primary sources of Galactic cosmic rays 
as seen in several cases via high energy emission  \citep{giuliani11, ackermann13, mckee13}.
GHz to sub-GHz radio observations are especially useful in SNR studies. They allow us to study the spectral aging of the SNR radio emission, providing insights into the dynamics, energetics, and evolution of the SNR.

The importance of magnetic fields has been emphasized in SNR evolution \citep{pavlovic18}.
GHz radio observations can provide information about the strength, structure, and orientation of the magnetic field. 
Polarization measurements of SNRs  in radio bands can reveal the geometry and orientation of the magnetic fields within SNRs.  The polarization of radio emission can also provide information about the diffusion and transport of cosmic rays within SNRs.  

Overall,  radio observations of SNRs play a crucial role in investigating the dynamics, magnetic fields, particle acceleration, and evolutionary processes associated with them.

Here we  present a comprehensive study of 36 Galactic supernova remnants observed with the MeerKAT array. Most of them have not been studied well previously. We describe the observations in  \S \ref{Observations}. The data analysis and imaging are presented in  \S \ref{Analysis} and \S \ref{Imaging}.  Data products are described in \S \ref{Products}. The total intensity images and a description of individual SNRs are given in  \S \ref{Results}.  Remnants  with higher spectral resolution polarimetric analysis are presented in  \S \ref{HiResPol}.  The results are discussed in  \S \ref{Discussion} with a summary in  \S \ref{Summary}.

\section{Observations\label{Observations}}
We present a selection of Galactic SNRs observed using
the MeerKAT array.
This telescope is described in more detail in \cite{Jonas2016} and \cite{DEEP2}.
The target SNRs were selected from the catalog of \cite{Green2019}, restricted to those laying outside the area already covered by the SARAO MeerKAT Galactic Plane Survey (SMGPS) within $\pm 1.5^\circ$ of the Galactic plane (Goedhart et al., in preparation), and furthermore having a diameter of $<1^\circ$ in order to fit within the MeerKAT primary beam at $\sim$ 1.4~GHz. SNR~G57.2+0.8 was observed in error but is included in this compilation.

The observations were either of a single remnant in a session or groups of 3 or 4 observed cyclically during longer sessions.
The bulk of the observations were between Feb--Jun 2022 Feb with some re-observations in 2023.
The date, pointing centers, time on each SNR (``Exposure"), the total time of the session, including calibration,  and whether higher spectral resolution polarization analysis was done are given in Table \ref{ObsTab}.
Most observations used project code SSV-20220221-SA-01.
We also include here two other SNRs that had been previously observed with MeerKAT but were unpublished: G21.8$-$3.0 (project code SSV-20200310-FC-01) and G296.5+10.0 (code SSV-20181016-FC-01).
Each session included either PKS~B1934$-$638 or PKS~0408$-$65 as the flux density/group delay/bandpass calibrator and a nearby astrometric calibrator.
When possible, either 3C138 or 3C286 were included as a polarization calibrator. 
Observations were done at L band (856 to 1712 MHz) recording all four combinations of the two linearly polarized feeds and using 8 second integrations in 4096 spectral channels.
\begin{table*}
\caption{Observations}
\label{ObsTab}
\centering
\begin{tabular}{|l|c|c|c|r|r|c|}
\hline\hline
SNR & Date       & RA   & Dec & Exposure$^1$ & Total$^2$ & Poln.$^3$\\
    & yyyy-mm-dd & J2000 & J2000     & hr.     & hr. & \\
\hline
G4.2$-$3.5  &  2022-02-26 & 18 08 55.00&$-$27 03 00.0  & 2.0 & 9.5 &\\
G4.5+6.8  &  2022-02-26 & 17 30 42.00&$-$21 29 00.0  & 2.0 & 9.5 &\\
G4.8+6.2  &  2022-03-26 & 17 33 25.00&$-$21 34 00.0  & 2.0 & 9.5 &X\\
G5.2$-$2.6  &  2022-02-26 & 18 07 30.00&$-$25 45 00.0  & 2.0 & 9.5 &\\
G5.9+3.1  &  2022-02-26 & 17 47 20.00&$-$22 16 00.0  & 2.0 & 9.5 &\\
G6.4+4.0  &  2022-03-26 & 17 45 10.00&$-$21 22 00.0  & 2.0 & 9.5 &\\
G7.7$-$3.7  &  2022-05-22 & 18 17 25.00&$-$24 04 00.0  & 2.1 & 9.5 &X\\
G8.7-5.0  &  2022-05-22 & 18 24 10.00&$-$23 48 00.0  & 2.0 & 9.5 &\\
G15.1$-$1.6 &  2022-03-13 & 18 24 00.00&$-$16 34 00.0  & 2.0 & 10.0 &\\
G16.2$-$2.7 &  2022-03-13 & 18 29 40.00&$-$16 08 00.0  & 2.0 & 10.0 &\\
G17.4$-$2.3 &  2022-03-13 & 18 30 55.00&$-$14 52 00.0  & 2.0 & 10.0 &\\
G17.8$-$2.6 &  2022-03-13 & 18 32 50.00&$-$14 39 00.0  & 2.0 & 10.0 &\\
G21.8$-$3.0 &  2020-03-15 & 18 41 46.50&$-$11 16 17.3  & 7.4 & 10.0 &\\
G30.7$-$2.0 &  2022-06-24 & 18 54 25.00&$-$02 54 00.0  & 3.0 & 3.5 &\\
G30.7$-$2.2 &  2023-05-18 &     "      &     "       & 3.0 & 3.5 &\\
G36.6+2.6 &  2022-06-25 & 18 48 49.00& 04 26 00.0  & 3.0 & 3.5 &\\
G53.6$-$2.2 &  2022-06-27 & 19 38 50.00& 17 14 00.0  & 2.1 & 7.5 &X\\
G55.7+3.4 &  2022-06-27 & 19 21 20.00& 21 44 00.0  & 2.1 & 7.5 &\\
G57.2+0.8 &  2022-06-27 & 19 34 59.00& 21 57 00.0  & 2.0 & 7.5 &\\
G261.9+5.5&  2022-03-30 & 09 04 20.00&$-$38 42 00.0  & 2.0 & 3.5 &\\
G272.2$-$3.2&  2023-04-06 & 09 06 50.00&$-$52 07 00.0  & 2.8 & 3.25 &\\
G284.3$-$1.8&  2022-03-29 & 10 18 15.00&$-$59 00 00.0  & 2.0 & 2.5 &\\
G292.0+1.8&  2022-05-16 & 11 24 36.00&$-$59 16 00.0  & 3.0 & 3.75 &\\
G296.5+10.0& 2018-10-27 & 12 10 00.91&$-$52 26 28.4  & 10.3  & 15.25 &\\
G299.2$-$2.9&  2022-04-05 & 12 15 13.00&$-$65 30 00.0  & 3.1 & 3.5 &\\
G312.5$-$3.0&  2022-05-14 & 14 21 00.00&$-$64 12 00.0  & 2.0 & 10.0 &\\
G315.4$-$2.3&  2022-05-14 & 14 43 00.00&$-$62 30 00.0  & 2.0 & 10.0 &X\\
G326.3$-$1.8&  2022-05-14 & 15 53 00.00&$-$56 10 00.0  & 2.0 & 10.0 &X\\
G327.6+14.6& 2022-05-25 & 15 02 50.00&$-$41 56 00.0  & 6.0 &  7.0 &X\\
G332.5$-$5.6&  2022-05-14 & 16 43 20.00&$-$54 30 00.0  & 2.0 & 10.0 &\\
G343.1$-$2.3&  2022-03-29 & 17 08 00.00&$-$44 16 00.0  & 2.0 & 7.5 &\\
G350.0$-$2.0&  2022-03-26 & 17 27 50.00&$-$38 32 00.0  & 2.0 & 9.5 &\\
G351.0$-$5.4&  2022-03-26 & 17 46 00.00&$-$39 25 00.0  & 2.0 & 9.5 &\\
G353.9$-$2.0&  2022-05-22 & 17 38 55.00&$-$35 11 00.0  & 2.1 & 9.5 &\\
G355.9$-$2.5&  2022-05-22 & 17 45 53.00&$-$33 43 00.0  & 2.1 & 9.5 &X\\
G356.2+4.5&  2022-03-29 & 17 19 00.00&$-$29 40 00.0  & 2.0 & 7.5 &X\\
G358.0+3.8&  2022-03-29 & 17 26 00.00&$-$28 36 00.0  & 2.0 & 7.5 &X\\
\hline
\end{tabular}
\hfill\break
Notes:\hfill\break
$^1$ Time observing this SNR.\hfill\break
$^2$ Total time, including calibration, in the observing session.\hfill\break
$^3$ ``X" indicates that this SNR had higher spectral resolution polarimetry analysis.\hfill\break
\end{table*}

\section{Data Analysis\label{Analysis}}
The observations were processed following the basic method described in \cite{DEEP2},  \cite{XGalaxy} and \cite{Knowles2022} using the {\small Obit} software package \citep{OBIT}\footnote{http://www.cv.nrao.edu/$\sim$bcotton/Obit.html}.  
\subsection{ Calibration\label{Calibration}}
Each observing session was preceeded by a ``noise diode'' calibration which was used to measure and remove the bulk of the instrumental ``X-Y'' phase difference between the two parallel hand visibility systems. 
The end channels of the bandpass were trimmed and the remainder divided into 8 spectral windows for calibration purposes.
Frequency ranges known to contain strong, persistent interference were blanked. 
In subsequent processing calibration and editing steps were interleaved.
The final visibility data were averaged to 952 channels (119 in each of 8 spectral windows) and subjected to baseline dependent time averaging not exceeding 32 seconds and not reducing the amplitudes more than 1\% at the edge of the field of view.
\subsection{Polarization Calibration\label{PolCal}}
All datasets were initially calibrated using the noise diode calibration to remove most of the instrumental X-Y phase differences.
The remainder of the polarization calibration is relatively stable.
When a known polarized calibrator (3C138 or 3C286) was included in a session, the instrumental ``leakage'' terms and residual X-Y phases were determined from observations in that session.
PKS~B1934$-$638 and PKS~0408$-$65 were assumed unpolarized and the polarization states of the astrometric/gain calibrators were determined together with the final instrumental calibration parameters from a joint solution as a function of frequency.
If no polarized calibrator was included, the calibration parameters applied were the medians of those determined from sessions including polarization calibrators.

\section{Imaging \label{Imaging}}
Stokes I, Q, U and V were imaged using the Obit task MFImage which is described more fully in \cite{Cotton2018}.
The spectrum is divided into multiple, constant fractional bandwidth, frequency sub-bands which are imaged independently and CLEANed jointly to accommodate the antenna gain and sky variations with frequency.
The curvature of the sky was corrected using faceting; this allows fully covering a given field of view while including outlying facets around stronger sources. 

Imaging used 5\% fractional bandwidth sub-bands (variable widths in Hz), fully imaged to a radius of 1.0$^\circ$ with outlying facets to 1.5$^\circ$ centered on sources estimated from the NVSS \citep{NVSS} or 843~MHz SUMSS catalog
\citep{mau03} to appear brighter than 1 mJy.
The center frequencies of the subbands are given in Table \ref{tab:subband}.
All pointings were subjected to two phase self calibrations.
A Briggs Robust factor of $-1.5$ (AIPS/Obit usage) was used to result in a typical resolution of 7.5\asec.
\begin{table}[h]
    \centering
    \begin{tabular}{c|c|c}
    Subband & Frequency & Comment \\
             & MHz      & \\
             \hline
       1 &  908.0 & \\
       2 &  952.3 & \\
       3 &  996.6 & \\
       4 & 1043.4 & \\
       5 & 1092.8 & \\
       6 & 1144.6 & \\
       7 & 1198.9 & Blanked\\
       8 & 1255.8 & Blanked\\
       9 & 1317.2 & \\
       10 & 1381.2 & \\
       11 & 1448.1 & \\
       12 & 1519.9 & \\
       13 & 1593.9 & \\
       14 & 1656.2 & \\
       \hline
    \end{tabular}
    \caption{MeerKAT subband central frequencies.}
    \label{tab:subband}
\end{table}

CLEANing used a gain of 0.03 to improve imaging of extended features and proceeded to a maximum of 2,500,000 components or a residual of 35 $\mu$Jy beam$^{-1}$ in Stokes I, 80,000 components or a residual of 20 $\mu$Jy beam$^{-1}$ in Stokes Q and U and 3000 components or a residual of 20 $\mu$Jy beam$^{-1}$ in Stokes V. 

The procedure described above makes only direction independent gain corrections and is insufficient to produce noise limited images in fields with bright and extended emission.
For a number of cases, ``Peeling'' \citep{Noordam2004,Smirnov2015} was applied to allow direction dependent gain correction to reduce artifacts from the nearby bright sources in the field of view.

\subsection{Short Baseline UV Coverage}
The remnants in the sample presented here are generally large enough that the smoother parts of the emission will be completely resolved out even on the shortest baselines. 
What matters for this is the baseline length in wavelengths which for the 2:1 \chg{bandpass} used will result in a factor of two difference from the top of the band to the bottom for MeerKAT's shortest (29 m) baselines.
If uncorrected, this effect can render in-band spectral indices wildly in error (e.g. spectral indices in error by 6 or 8).
To reduce this effect, the initial imaging used an inner Gaussian taper \citep{XGalaxy} to approximately equalize the short baseline uv coverage across the bandpass.
This technique has the unfortunate side effect of reducing the sensitivity to the largest scale structures which are detectable only in the lower portions of the bandpass.
\subsection{Polarization}
The Stokes Q and U images were derived using the same 5\% fractional bandwidth bins giving 14 channels across the band, two of which are totally blanked by interference.
This spectral resolution gives reasonable sensitivity to Faraday rotation up to $\pm 300$ rad m$^{-2}$.
All remnants were subjected to a Faraday depth search as described in \citet{XGalaxy} which produces estimates of the peak depth of the Faraday spectrum, the polarized intensity at the peak and the polarization angle.
The search range was $\pm 150$ rad m$^{-2}$.

For a subset of the remnants (those with significant polarized emission associated with the remnant), the data were reimaged in Stokes Q and U using 1\% fractional bandwidth (68 channels) and using a joint Q and U deconvolution. 
This deconvolution uses the band average of the channel polarized intensity in each pixel to drive the CLEAN.  
The search range in the Faraday analysis was $\pm 400$ rad m$^{-2}$ and in some cases wider. These are indicated in Table \ref{ObsTab} and described in Section \ref{HiResPol}.

\subsection{Reference Frequency\label{refFreq}}
The fundamental product of the imaging is a cube of subband images and these need to be combined into a broadband image.
In order to get the same reference frequency at the center of all images, a weighted subband combination is the obvious option.
An optimal weighting uses the $1/\sigma^2$ in each subband as the weights.
However, the steep spectrum of the Milky Way's disk emission adds substantially to the system noise in the lower parts of the bandpass causing the $1/\sigma^2$ weighting to basically ignore the bottom of the band.
Therefore a  $1/\sigma$ weighting is adopted.
The weights chosen are the average of those determined from each remnant field; this gives an effective reference frequency of 1335.3 MHz.

Away from the pointing center, the gain of the antennas drops faster at higher frequency that lower ones.
This causes the effective frequency to drop with increasing distance from the pointing center. 
If we define a weighting matrix for each subband:
$$ w_j(x,y)\ =\ {{Weight_j}\over{PB_j(x,y)}}\ ,$$
where $Weight_j$ is the subband  $j$ weight and $PB_j(x,y)$ is the antenna primary gain in pixel $(x,y)$ of subband $j$; then the broadband effective frequency in pixel $(x,y)$ is
$$\nu _{eff}(x,y)\ =\ {{\sum_{j=0}^n{w_j(x,y)\ \nu_j}\over{\sum_{j=0}^n{w_j(x,y)}}}}$$
where $n$ is the number of subbands and $\nu_j$ is the subband frequency.  A first order correction of the broadband image can be obtained assuming a spectral index $\alpha$ by:
$$I_{corr}(x,y)\ =\ I_{PB}(x,y)\ exp\Big({-\alpha\ ln{{{\nu_{eff}(x,y)\over{\nu_0}}}}}\Big),$$
where $I_{PB}(x,y)$ is the primary beam corrected image and $\nu_0$ is the broadband reference frequency.  This should result in approximately the same reference frequency in each pixel.  For purposes of this correction, a value of $\alpha=-0.6$ was used.
The primary beam gain, $PB_j$, is a symmetric cosine beam \citep[][eq 3.95]{CondonRansom}, given by
$$PB_j = cos (\pi\ \rho_r)/(1-4{\rho_r}^2))$$
where
$$\rho_r\ =\ 1.18896\ \theta / (1.4375\ \nu_j^{-1}),$$
$\theta$ is the offset from the pointing center in degrees  and $\nu_j$ is the subband frequency in GHz.

\subsection{Data Limitations\label{Limits}}
\begin{itemize}
\item {\bf Large scale structure:}  Interferometers are high pass spatial frequency filters; structures larger than a certain size will give an increasingly attenuated response.
The shortest MeerKAT baselines are 29 m corresponding to a fringe spacing of about 25\amin\ at band center.
Emission with scale sizes this large or greater will suffer significant loss.
Several of the objects discussed in this paper have structure on size scales  that are not well represented in the images presented.
The lack of sufficiently short baselines may result in negative bowls around brighter extended sources with missing flux density.
\item {\bf Astrometry:} Astrometric errors of up to an arcsecond are possible due to the limited accuracy of the geometric model used for the observations.
See \cite{Knowles2022} for details.
\item {\bf Dynamic Range:}  The excellent uv coverage of MeerKAT generally gives good dynamic range but bright sources and direction dependent gain effects can still limit the dynamic range.
All data were phase self calibrated which largely removes direction independent effects.
Artifacts resulting from direction dependent gain effects can be reduced using the ``Peeling" process \citep{Noordam2004,Smirnov2015}.
Two of the SNRs presented here (G16.2$-$2.7, G55.7+3.4) were affected by artifacts from these direction dependent gain effects and were subjected to the peeling process.
This will greatly reduce the artifacts in the images distributed but not from the raw visibility data.
\item {\bf Off-axis instrumental polarization:}  Off-axis instrumental polarization becomes increasing important with increasing distance from the pointing center and with increasing frequency; see \cite{deVilliers2022} for details.
\end{itemize}

\section{Data Products \label{Products}}
A number of products from this project are available; visibility data can be obtained from the SARAO archive and image products from DOI \url{https://doi.org/10.48479/nz0n-p845}.
\begin{itemize}
\item {\bf Visibility data:}  The raw visibility data is available
  under project codes SSV-20220221-SA-01, SSV-20200310-FC-01 (G21.8$-$3.0), and SSV-20181016-FC-01 (G296.5+10.0) from the SARAO archive at \url{https://archive.sarao.ac.za/}.
\item {\bf Stokes I images:}  Primary beam corrected images with adjustments to a constant reference frequency as described in Section \ref{refFreq} are provided in two forms.  The first form is a 16 plane cube described in \cite{MFImage} with plane 1 being the broadband image at the reference frequency of 1335.3 MHz; this image is the weighted average of the subband images corrected for primary beam attenuation and adjusted to a constant reference frequency using a spectral index of $\alpha=-0.6$.  The second plane is the fitted spectral index for pixels with adequate signal-to-noise for a least squares fit and total intensity in excess of 100 $\mu$Jy beam$^{-1}$; other pixels are blanked.  The subsequent 14 planes are the subband images without primary beam corrections, two of which are totally blanked by RFI.  The frequencies are given in Table \ref{tab:subband}.  These files contain ``I\_Cube" in the name.

The second form has the same broadband and spectral index planes as described in the above but the third and fourth planes are least squares error estimates for the total intensity and spectral index.  The fifth plane is the $\chi^2$ of the fit.  These files have ``I\_Fit" in the name.
\item {\bf Rotation measure images:} Rotation measure cubes derived from a 14 channel deconvolution and searching $\pm$150 rad m$^{-2}$ are given for all SNRs.  These were derived from a direct search of Faraday depth (RM) space using an increment of 0.5 rad m$^{-2}$; the RM giving the maximum unwrapped polarized intensity was the RM in each pixel.  The cubes consist of 1) peak Faraday depth (rad m$^{-2}$), 2) Electric Vector Position Angle (EVPA) evaluated at wavelength=0 (radians), 3) unwrapped polarized intensity at Faraday depth RM (Jy) and 4) the $\chi^2$ of the solution.  These cubes have names including ``ShallowRM" and have been cropped to the regions of the SNRs.  The RM and EVPA planes have been blanked where the peak polarized intensity is less than 50 $\mu$Jy beam$^{-1}$.

For a select subset of SNRs, a higher resolution CLEAN in Q and U, with 1\% fractional bandwidth=68 channels, was performed with joint Q and U deconvolution and a search in Faraday depth of $\pm$400 rad m$^{-2}$. The form of these files is the same as for the ``ShallowRM" files but are labeled ``DeepRM".
\end{itemize}

\section{Results\label{Results}}

We studied 36 Galactic SNRs, listed in Green's catalog of Galactic supernova remnants \citep{Green2019}, all but one at latitude $|b| > 1.5\degr$, accessible in the MeerKAT sky. 21 of those have not been well studied in the radio and most of them do not even have a proper radio image published. More than half of the sources were discovered in surveys or source catalogs. Ten SNRs have been first published in a catalog of new SNR candidates \citep[EffbgCat,][]{Reich1988}, which have been identified through a comparison of data from two Effelsberg surveys of the Galactic plane at 11~cm \citep{Reich1990-11} and at 21~cm \citep{Reich1990-21}. For most of them, there have not been any follow-up observations until now. Five were first published in a SNR catalog by \citet[ParkesCat][]{Duncan1997}, which is based on a radio continuum survey of the southern Galactic plane with the Parkes telescope at 2.4~GHz. None of them have been properly followed up in the radio. Three more have been confirmed in Parkes follow-up observations of shell-like sources at 1410~MHz and 2650~MHz by \citet[HillCat][]{Hill1967}. The others were single discoveries.

\subsection{Stokes I\label{IPol}}

Basic characteristics of the observed supernova remnants, such as name, center coordinates, angular extent, and integrated flux densities are listed in Table~\ref{tab:snrchar}. Images of all SNRs are shown in Figures~\ref{fig:Ig4.2-3.5} through \ref{fig:Ig358.0}. In most cases we used a linear transfer function for the colour scheme, but in same cases we used log(0.5) to give the image more dynamic range and enhance faint emission relative to the bright parts. 

Flux densities $S$ have been integrated in concentric rings starting at the source center (see Table~~\ref{tab:snrchar}). The background is determined in the ring just outside the source, which is then subtracted. The uncertainty $\Delta S$ for this ring integration can be calculated via \citep{Klein1981}:
\begin{equation}
    \Delta S = \sigma \cdot \sqrt{\frac{N^2_{source}}{N_{back}} + N_{source}},
\end{equation}
here $\sigma$ is the rms noise in the map of the SNR, $N_{source}$ is the number of pixels integrated on the SNR and $N_{back}$ is the number of pixels in the ring used for the background determination. As the rms noise is highly variable over the images, the ring used for the background determination contains additional systematic variations due to the fact that the source might be sitting in a bowl as a result of missing short spacings (see Section~\ref{Limits}), and we use the cleaning residual of $35~\mu$Jy~beam$^{-1}$ (see Section~\ref{Imaging}) for $\sigma$.
For most sources this background estimate is the largest source of uncertainty and is therefore reflected in the error of the integrated flux density. A calibration uncertainty of 5\% is added quadratically to this uncertainty. Spectral indices $\alpha$ are defined via:
\begin{equation}
    S \propto \nu^\alpha,
\end{equation}
where $\nu$ is the frequency.
The extent listed in Table~\ref{tab:snrchar} was measured in the MeerKAT images of the SNRs. It is the maximum size in the Galactic coordinate system in Longitude and Latitude.

\begin{table*}
    \centering
    \caption{\label{tab:snrchar} Characteristics of the SNRs determined from the radio total power images. Flux densities which, compared to the literature, have been more than 90\% recovered are indicated by an asterisk.}
    \begin{tabular}{|l|cc|cc|cc|c|l|}
    \hline \hline
    SNR & \multicolumn{4}{c|}{Center Coordinates} & \multicolumn{2}{c|}{extent} & $S_{1335}$ & Notes\\
     & GLong[\degr] & GLat[\degr] & RA J2000 & DEC J2000 & GLong[$'$] & GLat[$'$] & Jy & \\ \hline
    G4.2$-$3.5 & 4.12 & $-$3.57 & 17 41 40.6 & $-$23 33 34 & 23 & 26 & 1.1$\pm$0.2 & \\
    G4.5+6.8  & 4.52 & +6.82 & 17 30 41.0 & $-$21 29 30 & 4.0 & 4.7 & 15.9$\pm$0.8$^*$ & Kepler's SNR\\
    G4.8+6.2 & 4.78 & +6.24 & 17 33 23.6 & $-$21 35 12 & 16 & 16 & 1.7$\pm$0.1 & \\
    G5.2$-$2.6 & 5.19 & $-$2.61 & 18 07 31.6 & $-$25 45 39 & 16 & 14 & 1.3$\pm$0.2 & \\
    G5.9+3.1 & 5.86 & +3.15 & 17 47 11.0 & $-$22 17 37 & 19 & 20 & $1.2\pm 0.1$ & \\
    G6.4+4.0 & 6.39 & +4.00 & 17 45 11.8 & $-$21 24 18 & 32 & 31 & $0.61\pm 0.20$ & \\
    G7.7$-$3.7 & 7.73 & $-3.80$ & 18 17 28.9 & $-$24 06 05 & 25 & 28 & $5.2\pm 0.4$ & SN~AD386? \\
    G8.7$-$5.0 & 8.74 & $-4.99$ & 18 24 08.6 & $-$23 45 55 & 26 & 26 & $4.1\pm 0.4^*$ & \\
    G15.1$-$1.6 & 15.10 & $-$1.63 & 18 24 03.0 & $-$16 34 57 & 33 & 20 & - & mostly thermal\\
    G16.2$-$2.7 & 16.12 & $-2.63$ & 18 29 43.0 & $-$16 08 41 & 18 & 18 & $1.64\pm 0.19$ & \\
    G17.4$-$2.3 & 17.41 & $-2.18$ & 18 30 31.8 & $-$14 47 37 & 65 & 63 & - & \\
    G17.8$-$2.6 & 17.80 & $-$2.64 & 18 32 57.7 & $-$14 39 36 & 26 & 26 & $2.9\pm 0.3$ & \\
    G21.8$-$3.0 & 21.80 & $-$2.92 & 18 41 29.1 & $-$11 14 00 & 44 & 44 & $1.75\pm 0.30$ & \\
    G30.7$-$2.0 & - & - & - & - & - & - & - & no SNR\\
    G36.6+2.6 & 36.64 & +2.60 & 18 48 55.8 & +04 29 03 & 13 & 15 & $0.60\pm 0.08^*$ & \\
    G53.6$-$2.2 & 53.65 & $-$2.23 & 19 38 46.3 & +17 16 11 & 30 & 32 & $3.8\pm 0.3$ & 3C400.2 NRAO611\\
    G55.7$+$3.4 & 55.66 & +3.40 & 19 21 53.7 & +21 44 01 & 24 & 22 & $1.25\pm 0.12^*$ & \\
    G57.2+0.8 & 57.24 & +0.15 & 19 34 54.6 & +21 53 28 & 11 & 12 & $1.25\pm 0.1^*$ & 4C21.53 \\
    G261.9+5.5 & 261.93 & +5.51 & 09 04 23.8 & $-$38 39 57 & 36 & 38 & $2.7\pm 0.4$ &  \\
    G272.2$-$3.2 & 272.20 & $-3.18$ & 09 06 45.8 & $-$52 06 13 & 14 & 15 & $0.38\pm 0.05^*$ &  \\
    G284.3$-$1.8 & 284.4 & $-$1.8 & 10 18 46.2 & $-$59 03 52 & 45 & 45 & $8.9\pm 0.8^*$ & MSH~10$-$53 \\
    G292.0+1.8 & 292.03 & 1.75 & 11 24 35.2 & $-$59 16 16 & 9.5 & 10 & $5.5\pm 0.6^*$ & Plateau MSH~11$-$54\\
    G292.0+1.8 & 292.031 & 1.755 & 11 24 36.4 & $-$59 16 00 & 4.0 & 3.9 & $5.8\pm 0.4^*$ & Core MSH~11$-$54\\
    G296.5+10.0& 296.45 & +9.95 & 12 09 30.5 & $-$52 23 39 & 68  &  97     & $3.3 \pm 0.6$  & PKS~1209$-$51/52 \\
    G299.2$-$2.9 & 299.14 & $-$2.87 & 12 14 49.1 & $-$65 28 19 & 19 & 17 & $0.54\pm 0.05$ & \\
    G312.5$-$3.0 & 312.48 & $-$2.99 & 14 20 53.4 & $-$64 11 27 & 22 & 21 & $1.1\pm 0.1$ & \\
    G315.4$-$2.3 & 315.39 & $-$2.34 & 14 42 42.2 & $-$62 29 22 & 45 & 47 & $15.4\pm 1.0$ & RCW86 MSH~14$-$63\\
    G326.3$-$1.8 & 326.32 & $-$1.74 & 15 53 01.8 & $-$56 07 59 & 40 & 36 & $40\pm 4$ & MSH~15$-$56\\
    \chg{G326.3$-$1.8} & \chg{326.20} & \chg{$-$1.75} & \chg{15 52 24.5} & \chg{$-$56 13 00} & \chg{10} & \chg{11} & \chg{$20\pm 7^*$}& \chg{PWN}\\
    G327.6+14.6 & 327.58 & +14.55 & 15 02 54.7 & $-$41 56 38 & 34 & 31 & $8.2\pm 0.7$ & SN1006 PKS~1459$-$41\\
    G332.5$-$5.6 & 332.47 & $-$5.62 & 16 43 04.7 & $-$54 36 43 & 43 & 38 & $1.6\pm 0.1^*$ & \\
    G343.1$-$2.3 & 343.10 & $-$2.38 & 17 08 20.7 & $-$44 18 07 & $\ge 35$ & $\ge 35$ & - & Shell \\
    G343.1$-$2.3 & 343.07 & $-$2.65 & 17 09 27.5 & $-$44 29 12 & 8.5 & 8.5 & $0.043\pm 0.005^*$ & PWN \\
    G350.0$-$2.0 & 349.99 & $-$2.02 & 17 27 55.0 & $-$38 27 46 & 55 & 47 & $5.8\pm 0.5$ & \\
    G351.0$-$5.4 & 351.01 & $-$5.52 & 17 45 56.2 & $-$39 28 25 & 48 & 50 & - & \\
    G353.9$-$2.0 & 353.94 & $-$2.10 & 17 38 58.5 & $-$35 11 32 & 13 & 13 & $\chg{0.43}\pm 0.04^*$ & \\
    G355.9$-$2.5 & 355.95 & $-$2.55 & 17 45 57.3 & $-$33 43 14 & 13 & 14 & $5.1\pm 0.4^*$ & \\
    G356.2$+$4.5 & 356.21 & 4.46 & 17 18 59.7 & $-$29 40 31 & 20 & 20 & $3.1\pm 0.3^*$ & \\
    G358.0$+$3.8 & 357.98 & 3.80 & 17 26 03.0 & $-$28 35 20 & 35 & 36 & $0.80\pm 0.15^*$ & \\
    \hline
    \end{tabular}
\end{table*}

\subsubsection{G4.2$-$3.5}
\begin{figure}[h]
    \centerline{\includegraphics[width=0.50\textwidth]{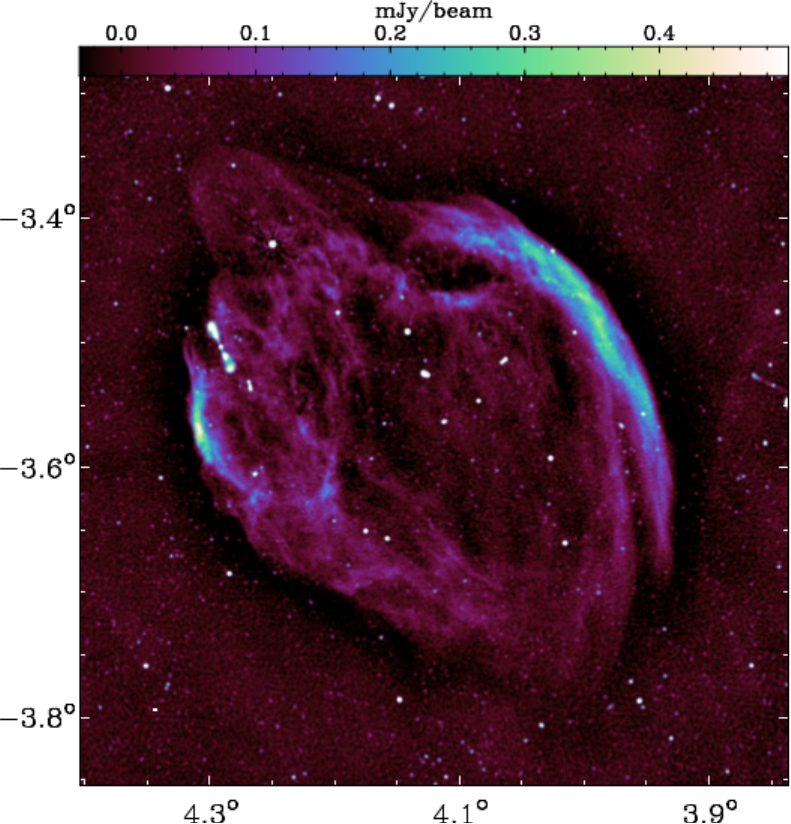}}
    \caption{Total power MeerKAT image of the SNR G4.2$-$3.5 at 1335~MHz in Galactic coordinates. The resolution of the image is 10\asec\ as indicated by the white circle in the lower left corner. \label{fig:Ig4.2-3.5}}
\end{figure}
The supernova remnant G4.2$-$3.5 was first mentioned in the catalog of new SNR candidates by \citet{Reich1988}. 
Based on those observations the radio spectral index is about $\alpha = -0.6$. There have not been any follow-up radio observations until now. 
The best to-date image is the 11~cm image from the Effelsberg survey at a resolution of 4.4\amin\ \citep{Reich1990-11}. 
Our new MeerKAT image at 1335~MHz with a resolution of 10\asec\ is shown in Figure~\ref{fig:Ig4.2-3.5}.

Overall, the SNR shows the typical bilateral or barrel-shaped structure indicative of mature SNRs expanding in an approximately uniform ambient medium and magnetic field. However, the eastern shell seen in the 11~cm Effelsberg image is splitting up into many smaller and fainter filamentary structures inside the main shell and there seems to be a break-out to the north-east, similar to the ``ears" that have been found in many shell-type SNRs.

The integrated flux density is $S_{1335} = 1.1 \pm 0.2$~Jy, which is only about 40\% of the value expected from the literature \citep{Reich1988}. At a diameter of about 25\amin\ this is indicative of missing large-scale emission, filtered out by the interferometer. 

\subsubsection{G4.5$+$6.8 (Kepler's SNR)}
\begin{figure}[h]
    \centerline{\includegraphics[width=0.50\textwidth]{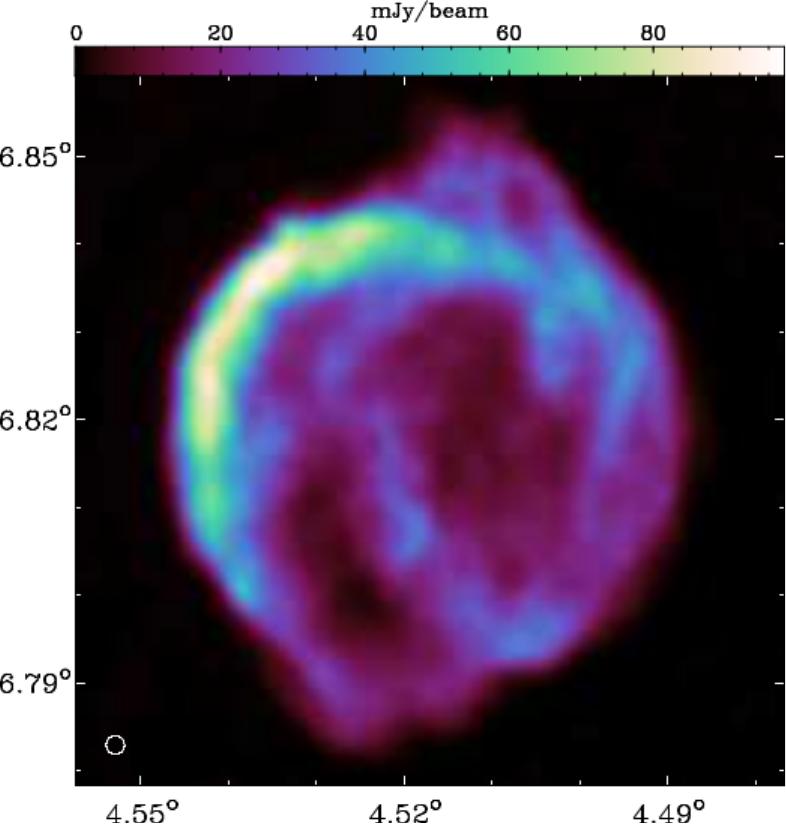}}
    \caption{Total power MeerKAT image of Kepler's SNR (G4.5$+$6.8) at 1335~MHz in Galactic coordinates. The resolution of the image is 8\asec\ as indicated by the white circle in the lower left corner. \label{fig:Ikepler}}
\end{figure}
G4.5$+$6.8, also called SN1604 or Kepler's SNR, is the remnant of the most recent supernova in our Galaxy that was unquestionably observable with the naked eye. The supernova was discovered by Johannes Kepler in 1604 and visible for more than a year \citep{Kepler}. This SNR has been studied at all wavelengths many times. The most recent radio spectrum was published by \citet{casteletti2021}, with a radio spectral index of $\alpha = -0.66$. The expected flux density at 1335~MHz is 15.8~Jy, which agrees very well with our result of $S_{1335} = 15.9 \pm 0.8$~Jy. The radio image in Figure ~\ref{fig:Ikepler} displays the typical partial radio shell with the well-known ears at the top and bottom in the image. The average diameter in our radio image is 4.3\amin.

\subsubsection{G4.8$+$6.2}
\begin{figure}[h]
    \centerline{\includegraphics[width=0.50\textwidth]{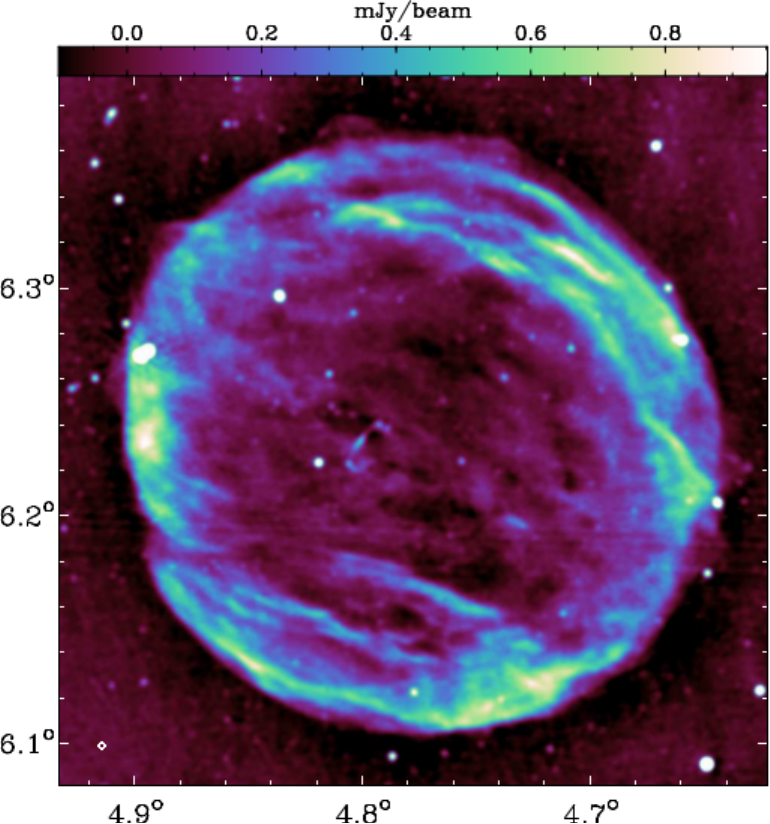}}
    \caption{Total power MeerKAT image of the SNR G4.8$+$6.2 at 1335~MHz in Galactic coordinates. The resolution of the image is 10\asec\ as indicated by the white circle in the lower left corner. \label{fig:Ig4.8}}
\end{figure}
G4.8$+$6.2 was first listed as a SNR candidate in a Parkes 2.4~GHz southern Galactic plane survey by \citet{Duncan1995}. Follow-up observations by \citet{Bhatnagar2000} with the Giant Metrewave Radio Telescope (GMRT) confirmed this source as a SNR. Based on those observations the radio spectral index is about $\alpha = -0.6$. The best to-date image was taken with the GMRT at 327~MHz \citep{Bhatnagar2000}. Our new MeerKAT image at 1335~MHz with a resolution of 10\asec\ is shown in Figure~\ref{fig:Ig4.8}.

G4.8+6.2 is an almost circular SNR with multiple shells showing the typical barrel-shaped structure indicative of SNRs expanding in an approximately uniform ambient medium with a relatively uniform magnetic field. In contrast to most barrel-shaped SNRs it shows emission all around its perimeter. The high sensitivity of our images also reveals about half a dozen small breakouts or ears in the north-east and south-west quadrants.

The integrated flux density is $S_{1335} = 1.7 \pm 0.1$~Jy, which is only about 70~\% of the value expected from the literature \citep{Bhatnagar2000}. At an average diameter of about 16\amin\ this is indicative of missing large-scale emission, filtered out by the interferometer. The fact that we see radio emission from all around the SNR may indicate that the ambient magnetic field has a large angle with the plane of the sky ($\ge 60\degr$) \citep{Kothes2009}.

\subsubsection{G5.2$-$2.6}
\begin{figure}[h]
    \centerline{\includegraphics[width=0.50\textwidth]{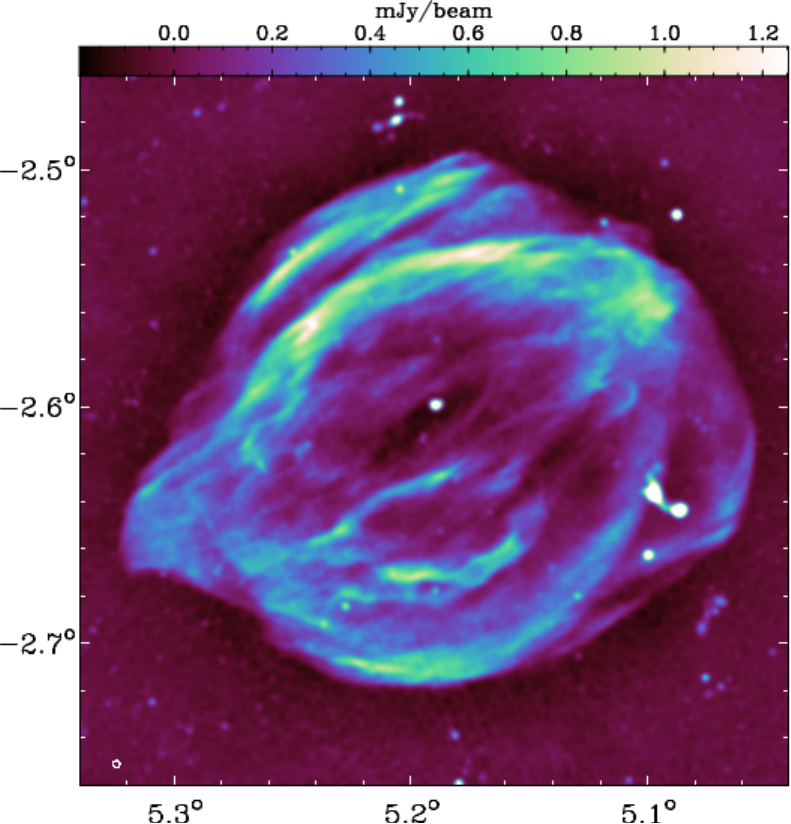}}
    \caption{Total power MeerKAT image of the SNR G5.2$-$2.6 at 1335~MHz in Galactic coordinates. The resolution of the image is 10\asec\ as indicated by the white circle in the lower left corner. \label{fig:Ig5.2}}
\end{figure}
The supernova remnant G5.2$-$2.6 was first mentioned in the catalog of new SNR candidates by \citet{Reich1988}. Based on those observations the radio spectral index is about $\alpha = -0.6$. There have not been any follow-up radio observations until now. The best to-date image is the 11~cm image from the Effelsberg survey at a resolution of 4.4\amin\ \citep{Reich1990-11}. Our new MeerKAT image at 1335~MHz with a resolution of 10\asec\ is shown in Figure~\ref{fig:Ig5.2}.

G5.2$-$2.6 looks very much like Kepler's SNR rotated clock-wise by about 40\degr. It is an almost circular barrel-shaped SNR with, in contrast to Kepler's SNR, multiple shells all around and only one prominent ear in the south-west.

The integrated flux density is $S_{1335} = 1.3 \pm 0.2$~Jy, which is only about 60\% of the value expected from the literature \citep{Reich1988}. At an average diameter of about 15\amin\ this is indicative of missing large-scale emission, filtered out by the interferometer. 

\subsubsection{G5.9$+$3.1}
\begin{figure}[h]
    \centerline{\includegraphics[width=0.50\textwidth]{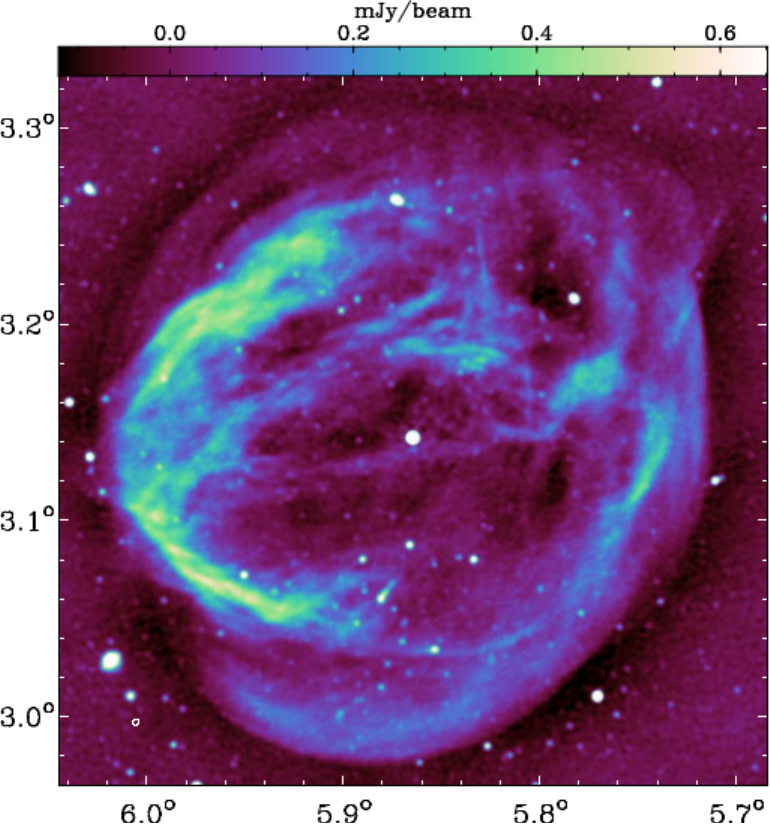}}
    \caption{Total power MeerKAT image of the SNR G5.9$+$3.1 at 1335~MHz in Galactic coordinates. The resolution of the image is 10\asec\ as indicated by the white circle in the lower left corner. \label{fig:Ig5.9}}
\end{figure}
The supernova remnant G5.9$+$3.1 was listed in the catalog of new SNR candidates by \citet{Reich1988}. There were follow-up radio observations with the Murchison Widefield Array \citep[MWA, ][]{MWA2019}, who published a spectral index of $\alpha = -0.42$. The best to-date image is from this paper at 200~MHz and a resolution of 2.4\amin. Our new MeerKAT image at 1335~MHz with a resolution of 10\asec\ is shown in Figure~\ref{fig:Ig5.9}.

G5.9$+$3.1 shows asymmetric shells and does not have a clear bilateral symmetry as it is found in barrel-shaped SNRs. It shows a brighter shell to the east and fainter filaments in the other areas of the source. A blowout feature or an ear are not present.

The integrated flux density is $S_{1335} = 1.2 \pm 0.1$~Jy, which is only about 50\% of the value expected from the literature \citep{Reich1988}. At an average diameter of about 20\amin\ this is indicative of missing large-scale emission, filtered out by the interferometer. 

\subsubsection{G6.4$+$4.0}
\begin{figure}[h]
    \centerline{\includegraphics[width=0.50\textwidth]{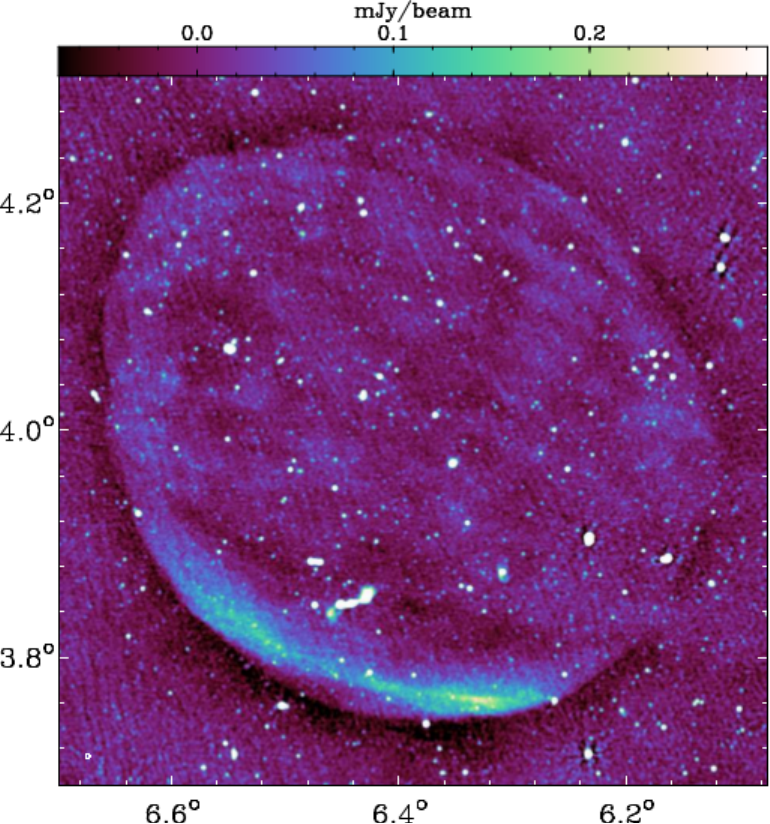}}
    \caption{Total power MeerKAT image of the SNR G6.4$+$4.0 at 1335~MHz in Galactic coordinates. The resolution of the image is 10\asec\ as indicated by the white circle in the lower left corner. \label{fig:Ig6.4}}
\end{figure}
The supernova remnant G6.4$+$4.0 was first mentioned in the catalog of new SNR candidates by \citet{Reich1988}. Based on those observations the radio spectral index is about $\alpha = -0.4$. There have not been any follow-up radio observations until now. The best to-date image is the 11~cm image from the Effelsberg survey at a resolution of 4.4\amin\ \citep{Reich1990-11}. Our new MeerKAT image at 1335~MHz with a resolution of 10\asec\ is shown in Figure~\ref{fig:Ig6.4}.

G6.4$+$4.0 is probably the lowest surface brightness SNR in our sample. It shows a clear bilateral structure with one prominent shell in the south-east. A blowout feature or an ear are not present.

The integrated flux density is $S_{1335} = 0.61 \pm 0.20$~Jy, which is only about 50\% of the value expected from the literature \citep{Reich1988}. At an average diameter of about 32\amin\ this is indicative of missing large-scale emission, filtered out by the interferometer. 

\subsubsection{G7.7$-$3.7}
\begin{figure}[h]
    \centerline{\includegraphics[width=0.50\textwidth]{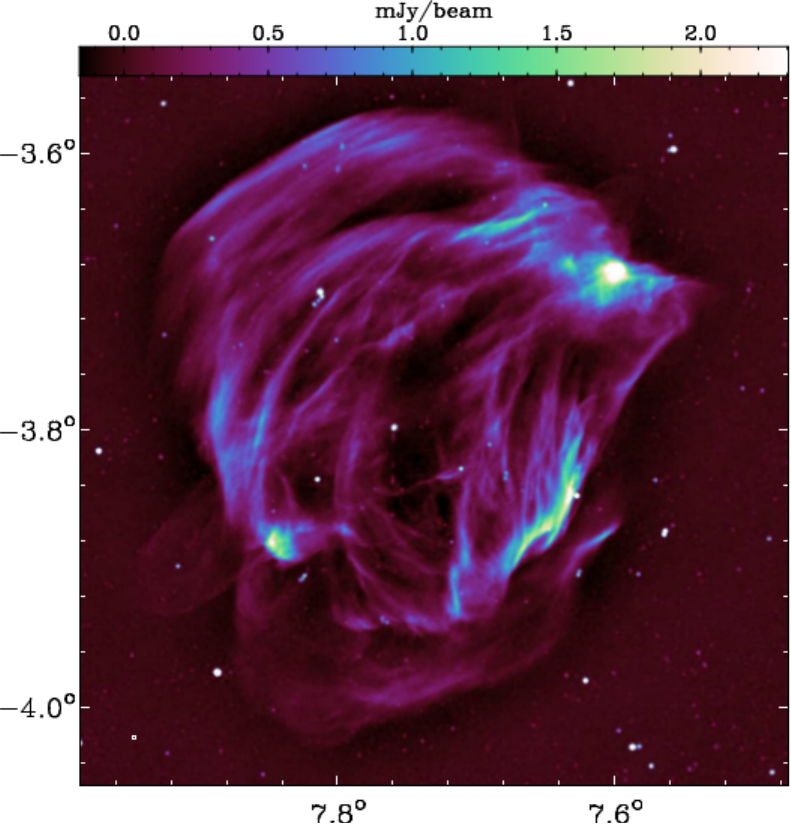}}
    \caption{Total power MeerKAT image of the SNR G7.7$-$3.7 at 1335~MHz in Galactic coordinates. The resolution of the image is 10\asec\ as indicated by the white circle in the lower left corner. \label{fig:Ig7.7}}
\end{figure}

G7.7$-$3.7 was first identified as a SNR by \citet{Milne1974}. This SNR has been associated with a supernova observed in AD386 \citep{Zhou2018}. The latest radio study of this object by \citet{Dubner1996} determined a radio spectral index of $\alpha \approx -0.32$. They also published the best-to-date image, revealing a complex multi-shell structure at a resolution of 71\asec\ $\times$ 35\asec. 

Our 10\asec\ MeerKAT image at 1335~MHz in Figure~\ref{fig:Ig7.7} resolves most of these shells into many very thin filaments. In the south-eastern area there is a very faint blowout or ear with a point source in its center.

The integrated flux density is $S_{1335} = 5.2 \pm 0.40$~Jy, which is only about 60\% of the value expected from the literature \citep{Dubner1996}. At an average diameter of about 27\amin\ this is indicative of missing large-scale emission, filtered out by the interferometer. 

\subsubsection{G8.7$-$5.0}
\begin{figure}[h]
    \centerline{\includegraphics[width=0.50\textwidth]{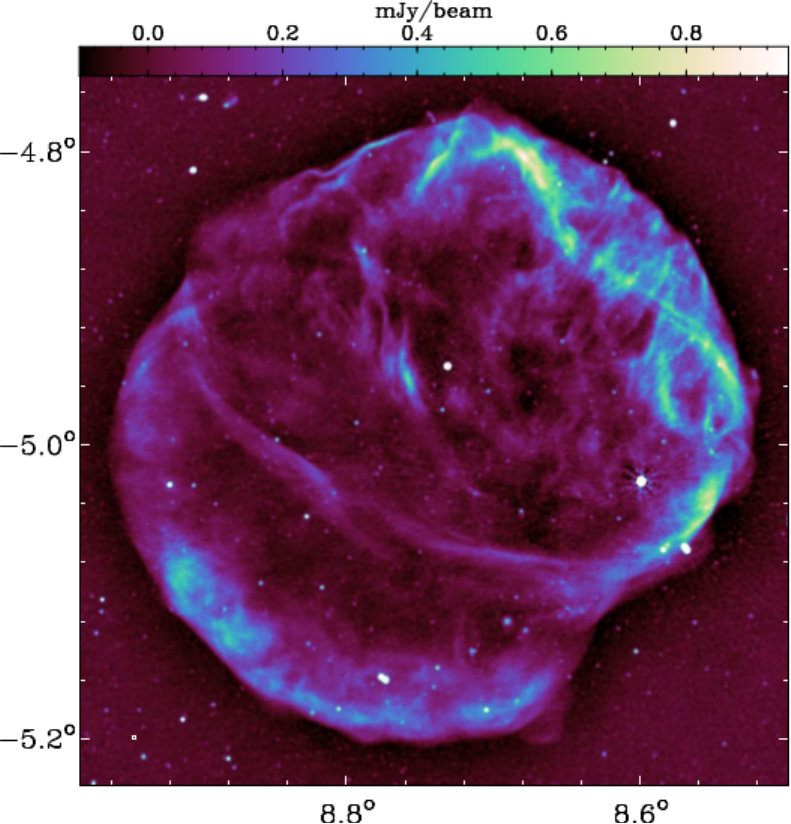}}
    \caption{Total power MeerKAT image of the SNR G8.7$-$5.0 at 1335~MHz in Galactic coordinates. The resolution of the image is 10\asec\ as indicated by the white circle in the lower left corner. \label{fig:Ig8.7}}
\end{figure}
The supernova remnant G8.7$-$5.0 was first mentioned in the catalog of new SNR candidates by \citet{Reich1988}. Based on those observations the radio spectral index is about $\alpha = -0.3$. There have not been any follow-up radio observations until now. The best to-date image is the 11~cm image from the Effelsberg survey at a resolution of 4.4\amin\ \citep{Reich1990-11}. Our new MeerKAT image at 1335~MHz with a resolution of 10\asec\ is shown in Figure~\ref{fig:Ig8.7}.

G8.7$-$5.0 shows a clear bilateral structure with one quite smooth and regular shell in the south-east. The bright and smooth shell visible in the 11-cm image in the north-west consists of a lot of filaments and a loop feature to the west. Another shell feature just below the center is also visible that seems to mark a shell moving to the south-east and towards us or away from us. Above this shell there are smooth cloud-like emission features. Those features could indicate that the shells moving towards us and away from us are breaking up into smaller features, indicating a well-structured environment there.

The integrated flux density is $S_{1335} = 4.1 \pm 0.4$~Jy, which agrees with the value expected from the literature \citep{Reich1988}. At an average diameter of about 26\amin\ this is, compared to other SNRs of the same angular dimension, unexpected. We would have expected missing extended emission coming from the shells moving towards us and away from us. This seems to indicate that those shells moving towards us and away from us are interacting with a well-structured medium which moves emission from large-scales into smaller scales.

\subsubsection{G15.1$-$1.6\label{IPol_G15.1}}
\begin{figure}[h]
    \centerline{\includegraphics[width=0.50\textwidth]{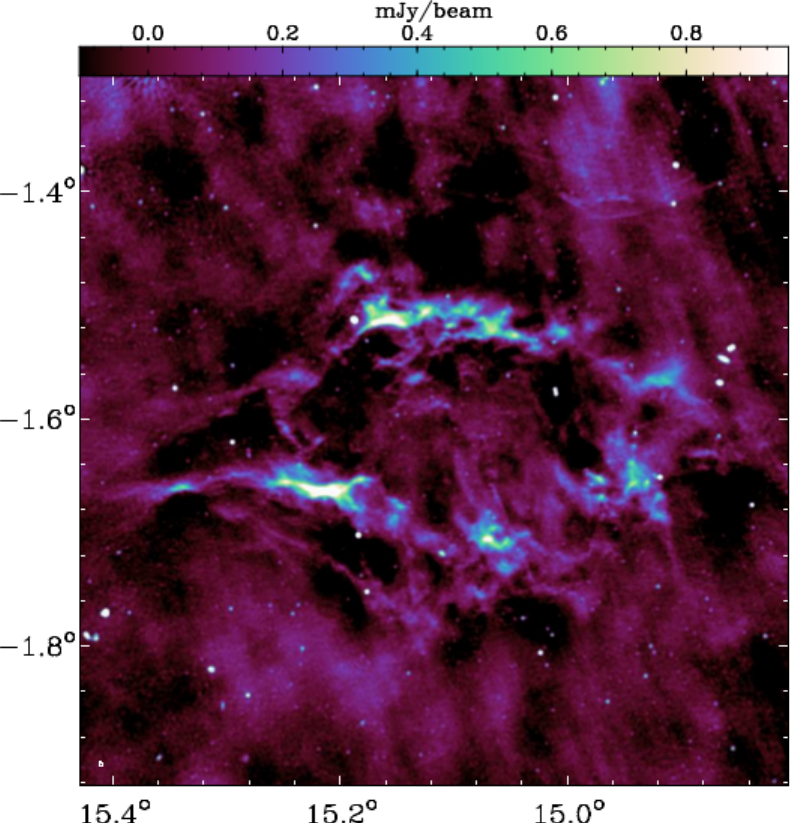}}
    \caption{Total power MeerKAT image of the SNR G15.1$-$1.6 at 1335~MHz in Galactic coordinates. The resolution of the image is 10\asec\ as indicated by the white circle in the lower left corner. \label{fig:Ig15.1}}
\end{figure}
G15.1$-$1.6 was first listed as a SNR candidate in the catalog by \citet{Reich1988}. The latest radio study of this object by \citet{Sun2011} determined a radio spectral index of $\alpha \approx -0.01$, more indicative of thermal emission from an HII region. A follow-up study in the optical by \citet{Boumis2008} found a lot of diffuse optical emission and only in one small area indication for shock-heated gas from optical line ratios. They also found strong coincidence of the optical emission with IRAS images at $60~\mu$m, another indication for thermal emission. There is also no X-ray emission coming from this object.

The best to-date image is the 11~cm image from the Effelsberg survey at a resolution of 4.4\amin\ \citep{Reich1990-11}. This image reveals a smooth double shell structure. Our new MeerKAT image at 1335~MHz with a resolution of 10\asec\ is shown in Figure~\ref{fig:Ig15.1}. In this image the two smooth shells are resolved into a complex of filaments. We were not able to determine a flux density due to the low surface brightness and the highly fluctuating background. 

Overall, the available evidence points to a HII region rather than a SNR. The only place where optical line ratios were found by \citet{Boumis2008} to indicate shock-heated gas is at about $15.15\degr$, $-1.52\degr$ in Galactic coordinates, which is where a small bright shell-like feature is visible in our MeerKAT image (Fig.~\ref{fig:Ig15.1}). We propose that this object is an HII region that contains a small non-thermal filament that could be coming from the shock-wave of a SNR that exploded in this environment or a stellar wind zone.

\subsubsection{G16.2$-$2.7}
\begin{figure}[h]
    \centerline{\includegraphics[width=0.50\textwidth]{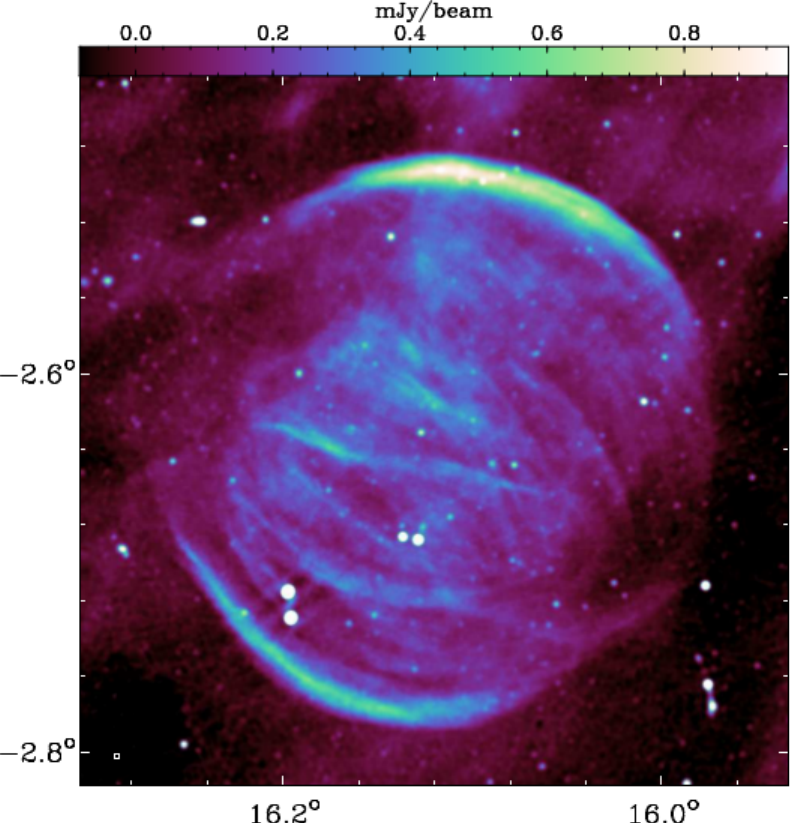}}
    \caption{Total power MeerKAT image of the SNR G16.2$-$2.7 at 1335~MHz in Galactic coordinates. The resolution of the image is 10\asec\ as indicated by the white circle in the lower left corner. \label{fig:Ig16.2}}
\end{figure}
The SNR G16.2$-$2.7 was discovered by \citet{Trushkin1999} following up the detection of a double shell source in the NRAO VLA Sky Survey \citep[NVSS, ][]{NVSS} with observations with the RATAN-600 radio telescope. The latest radio observations of G16.2$-$2.7 by \citet{Sun2011} give a spectral index of $\alpha = -0.42$, ignoring the original values of \citet{Trushkin1999}, while \citet{Trushkin1999} found a spectral index of $\alpha = -0.51$. It is difficult to decide who is correct, because \citet{Sun2011} use observations that barely resolve this SNR at all and the observations by \citet{Trushkin1999} result in a pencil beam which resolves the SNR well in one direction and not at all in the other.

Our new MeerKAT image at 1335~MHz with a resolution of 10\asec\ is shown in Figure~\ref{fig:Ig16.2}. The integrated flux density is $S_{1335} = 1.64 \pm 0.19$~Jy, which agrees within errors with the value expected from \citet{Trushkin1999}, but reaches only about 75\% of the flux expected from \citet{Sun2011}. The SNR is a picture-book example of a bilateral barrel-shaped remnant with diffuse emission coming from the center. It looks almost identical to the recently discovered SNR G181.1+9.5 \citep{Kothes2017} or DA\,530 \citep[G93.3+6.9, e.g.][]{Booth2022}, which is always seen as the typically-shaped supernova remnant. It is circular projected to the plane of the sky with a diameter of 18\amin. Comparing G16.2$-$2.7 visually with other SNRs in our sample, that do show missing short spacings, we expect that all spatial frequencies are covered for this SNR. Therefore we favour the spectrum published by \citet{Trushkin1999}. We did not find any ears or blowouts.

\subsubsection{G17.4$-$2.3}
\begin{figure}[h]
    \centerline{\includegraphics[width=0.50\textwidth]{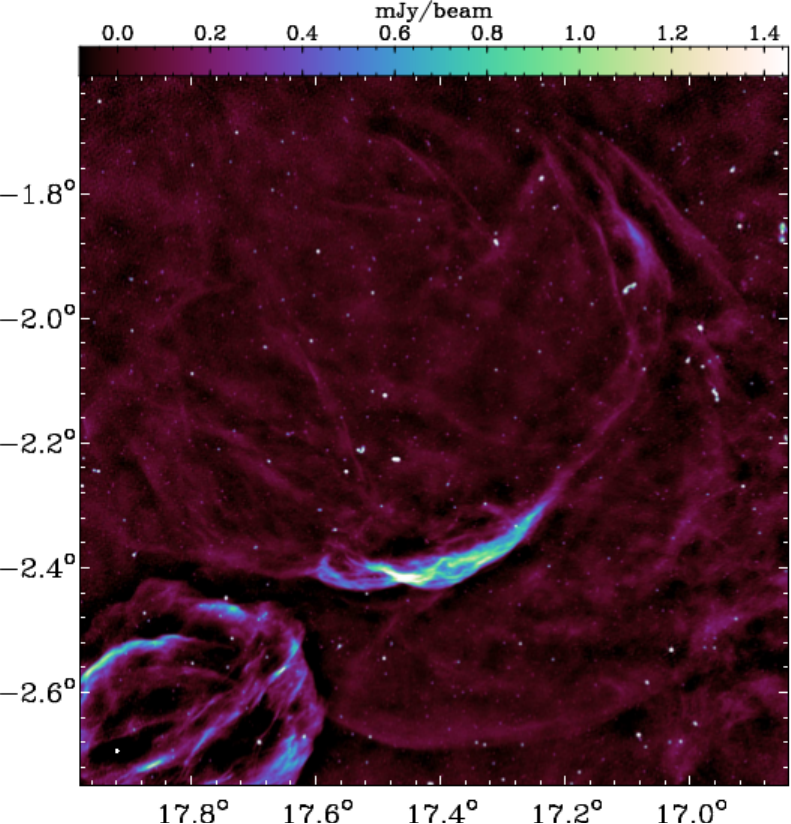}}
    \caption{Total power MeerKAT image of the SNR G17.4$-$2.3 at 1335~MHz in Galactic coordinates. The resolution of the image is 10\asec\ as indicated by the white circle in the lower left corner. \label{fig:Ig17.4}}
\end{figure}
G17.4$-$2.3 was first listed as a SNR candidate in the catalog by \citet{Reich1988}. The latest radio study of this object by \citet{Sun2011} determined a radio spectral index of $\alpha \approx -0.46$, which is typical for non-thermal emission from a mature supernova remnant. The best to-date image is the 11~cm image from the Effelsberg survey at a resolution of 4.4\amin\ \citep{Reich1990-11}. In this image only one shell is visible and the source seems to be about the same size as the next SNR in our list G17.8$-$2.6, which is very close by.

Our new MeerKAT image at 1335~MHz with a resolution of 10\asec\ is shown in Figure~\ref{fig:Ig17.4}. The SNR is clearly much larger than anticipated from the earlier observations. The shell visible at 11~cm in the Effelsberg survey \citep{Reich1990-11} seems to be much larger extending in both directions, disappearing in the noise. There are also multiple additional filaments outside that shell following a similar curvature. It looks like we did not cover the whole remnant with our observations. The part that is detected in our observations has a diameter of a little over a degree. The SNR G17.8$-$2.6 is seen in the lower left corner of our image (Figure~\ref{fig:Ig17.4}). It is not clear whether those two SNRs are interacting or are just overlapping along the line of sight.

We were not able to determine a flux density as the SNR is of very low surface brightness, its environment seems to be complex, and we do not even know its full extent. 

\subsubsection{G17.8$-$2.6}
\begin{figure}[h]
    \centerline{\includegraphics[width=0.50\textwidth]{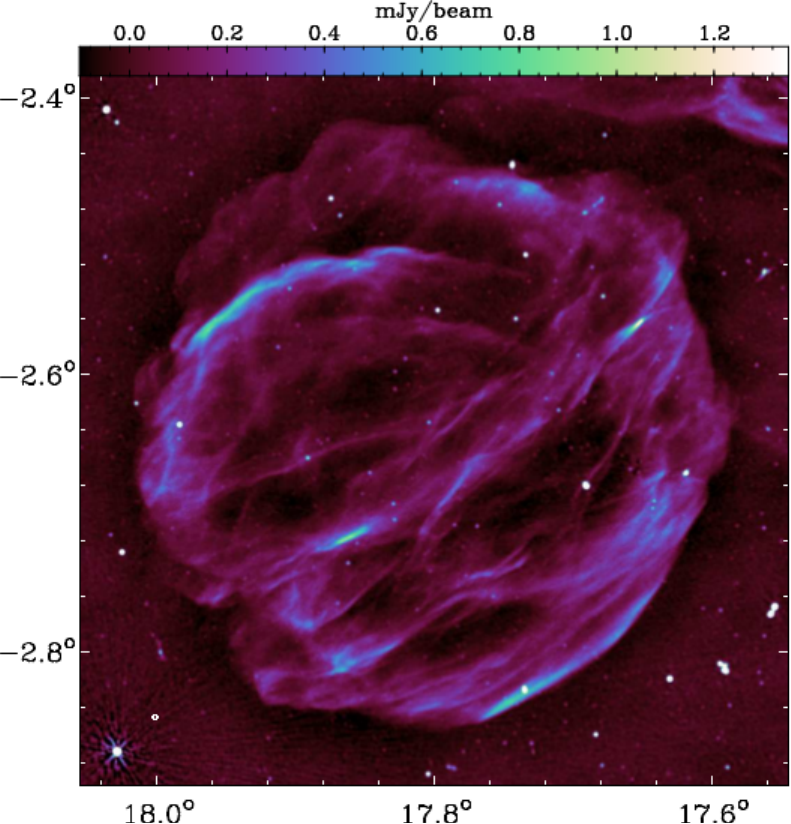}}
    \caption{Total power MeerKAT image of the SNR G17.8$-$2.6 at 1335~MHz in Galactic coordinates. The resolution of the image is 8\asec\ as indicated by the white circle in the lower left corner. \label{fig:Ig17.8}}
\end{figure}
The supernova remnant G17.8$-$2.6 was first mentioned in the catalog of new SNR candidates by \citet{Reich1988}. The latest radio study of this object by \citet{Sun2011} reveals a radio spectral index of $\alpha \approx -0.50$, which is typical for non-thermal radio emission from a mature supernova remnant. The best to-date image is the 11~cm image from the Effelsberg survey at a resolution of 4.4\amin\ \citep{Reich1990-11}. In this it looks like a smooth radio shell with multiple segments. 

Our new MeerKAT image at 1335~MHz with a resolution of 8\asec\ is shown in Figure~\ref{fig:Ig17.8}. This SNR shows a bilateral structure with multiple filaments at the edges and also in the central area. The central filaments are likely part of the shells that are moving away and towards us. 

The integrated flux density from our data is $S_{1335} = 2.9 \pm 0.3$~Jy, which is about 70\% of the value expected \citep{Sun2011}, which indicates missing large-scale emission. Its edge is circular at a diameter of 26\amin. Ears or blowouts are not visible, although that could be the result of the complex structure with multiple filaments.

\subsubsection{G21.8$-$3.0}
\begin{figure}[h]
    \centerline{\includegraphics[width=0.50\textwidth]{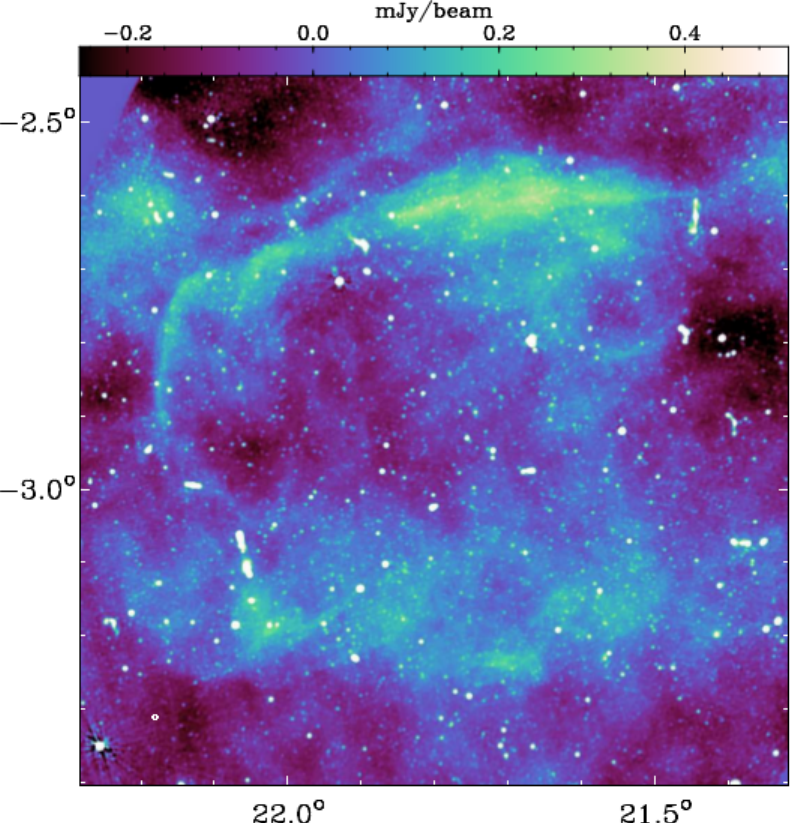}}
    \caption{Total power MeerKAT image of the SNR G21.8$-$3.0 at 1335~MHz in Galactic coordinates. The resolution of the image is 15\asec\ as indicated by the white circle in the lower left corner. \label{fig:Ig21.8}}
\end{figure}
The SNR G21.8$-$3.0 was discovered in a dedicated study by \citet{Gao2020} through a comparison of polarimetric observations at 11~cm with the Effelsberg 100-m radio telescope and at 6~cm taken from the Urumqi 25-m telescope \citep{Gao2010}. They found a spectral index of $\alpha \approx -0.7$. The best to-date image is the 11~cm Effelsberg image from their study at a resolution of 4.4\amin.

Our new MeerKAT image at 1335~MHz with a resolution of 15\asec\ is shown in Figure~\ref{fig:Ig21.8}. It shows a bilateral structure almost parallel to the Galactic plane, confirming the images published by \citet{Gao2010}. This is a very low surface brightness SNR even for MeerKAT. The integrated flux density from our data is $S_{1335} = 1.75 \pm 0.30$~Jy, which is a little more than 40\% of the value expected \citep{Gao2020}, which indicates missing large-scale emission. This is not surprising as the SNR has a diameter of about 44\amin. \citet{Gao2020} determined a diameter of 1\degr, but in their images it looks more like about 45\amin, confirming our estimate.

\subsubsection{G30.7$-$2.0}
\begin{figure}[h]
    \centerline{\includegraphics[width=0.50\textwidth]{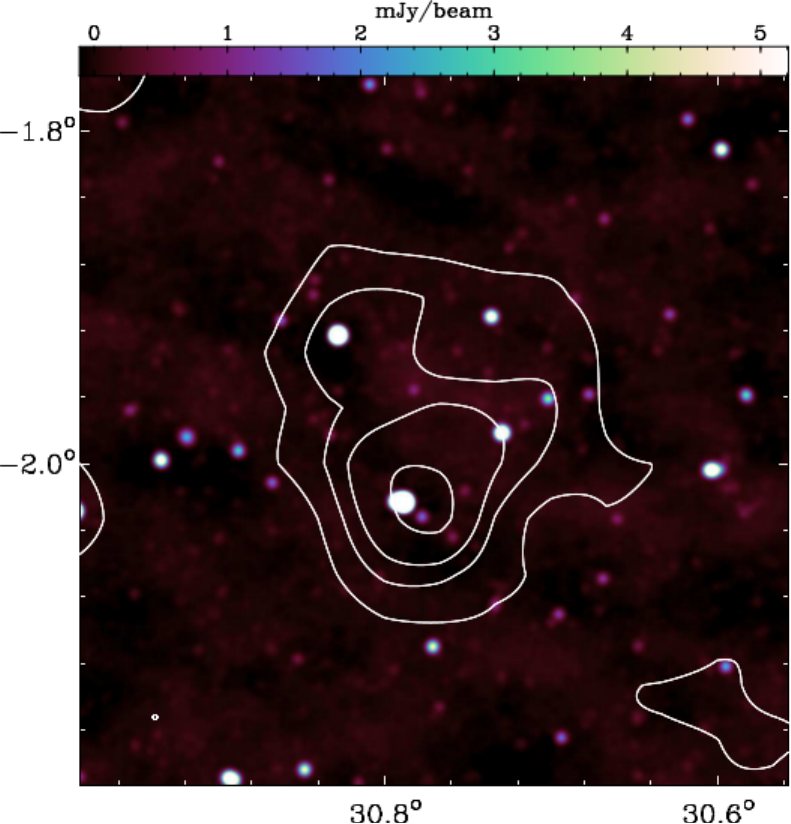}}
    \caption{Total power MeerKAT image of the SNR G30.7$-$2.0 at 1335~MHz in Galactic coordinates. The resolution of the image is 20\asec\ as indicated by the white circle in the lower left corner. The contours are from the Effelsberg image at 11 cm \citep{Reich1990-11}. \label{fig:Ig30.7}}
\end{figure}
G30.7$-$2.0 was first mentioned in a catalog of new SNR candidates by \citet{Reich1988}. A 1335~MHz total power image of this area is shown in Figure~\ref{fig:Ig30.7} together with white contours taken from the 11~cm Effelsberg Survey \citep{Reich1990-11}. It is clear that there is no SNR visible but three bright extra-galactic compact sources that were smoothed together with the 4.4\amin\ Effelsberg beam to look like an extended shell-like object. The three bright point sources add up to more than 250~mJy and the 21~cm flux of this object in the Effelsberg Survey of 9.4\amin\ resolution is about 400~mJy. The estimated spectral index of $\alpha \approx -0.7$ also looks close to expected extra-galactic values. We conclude that G30.7$-$2.0 is not a supernova remnant.

\subsubsection{G36.6$+$2.6}
\begin{figure}[h]
    \centerline{\includegraphics[width=0.50\textwidth]{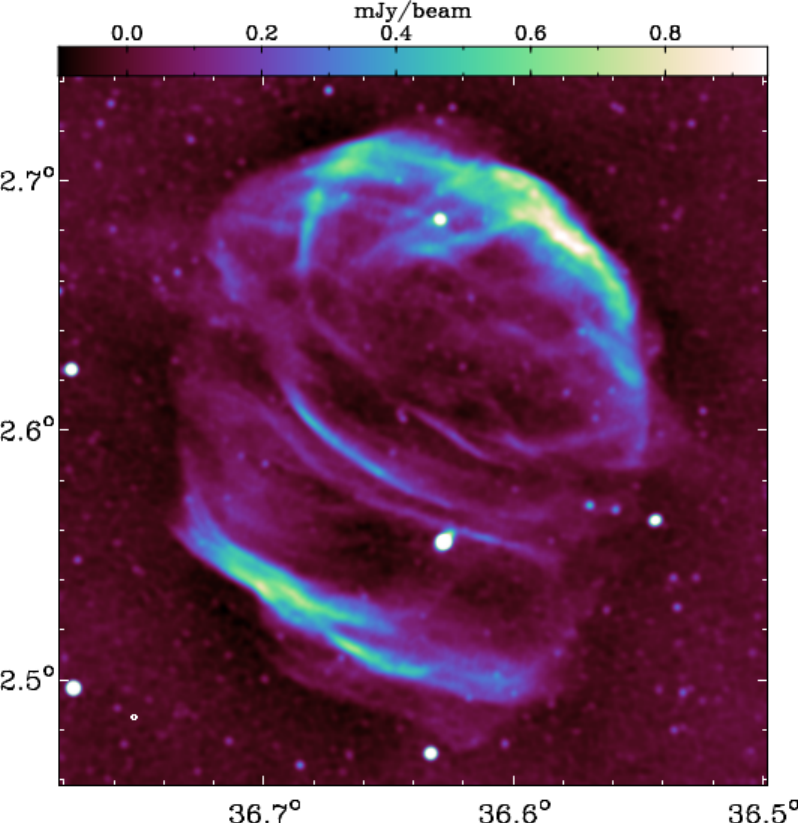}}
    \caption{Total power MeerKAT image of the SNR G36.6$+$2.6 at 1335~MHz in Galactic coordinates. The resolution of the image is 10\asec\ as indicated by the white circle in the lower left corner. \label{fig:Ig36.6}}
\end{figure}
The supernova remnant G36.6$+$2.6 was first mentioned in the catalog of new SNR candidates by \citet{Reich1988}. There have not been any follow-up observations at any wavelength. \citet{Reich1988} found a spectral index of $\alpha \approx -0.5$. The best to-date image is the 11~cm image from the Effelsberg survey at a resolution of 4.4\amin\ \citep{Reich1990-11}. This shows an extended blob of emission with a peak in the north and emission plateau in the south. 

\begin{figure}[h]
    \centerline{\includegraphics[height=0.40\textwidth]{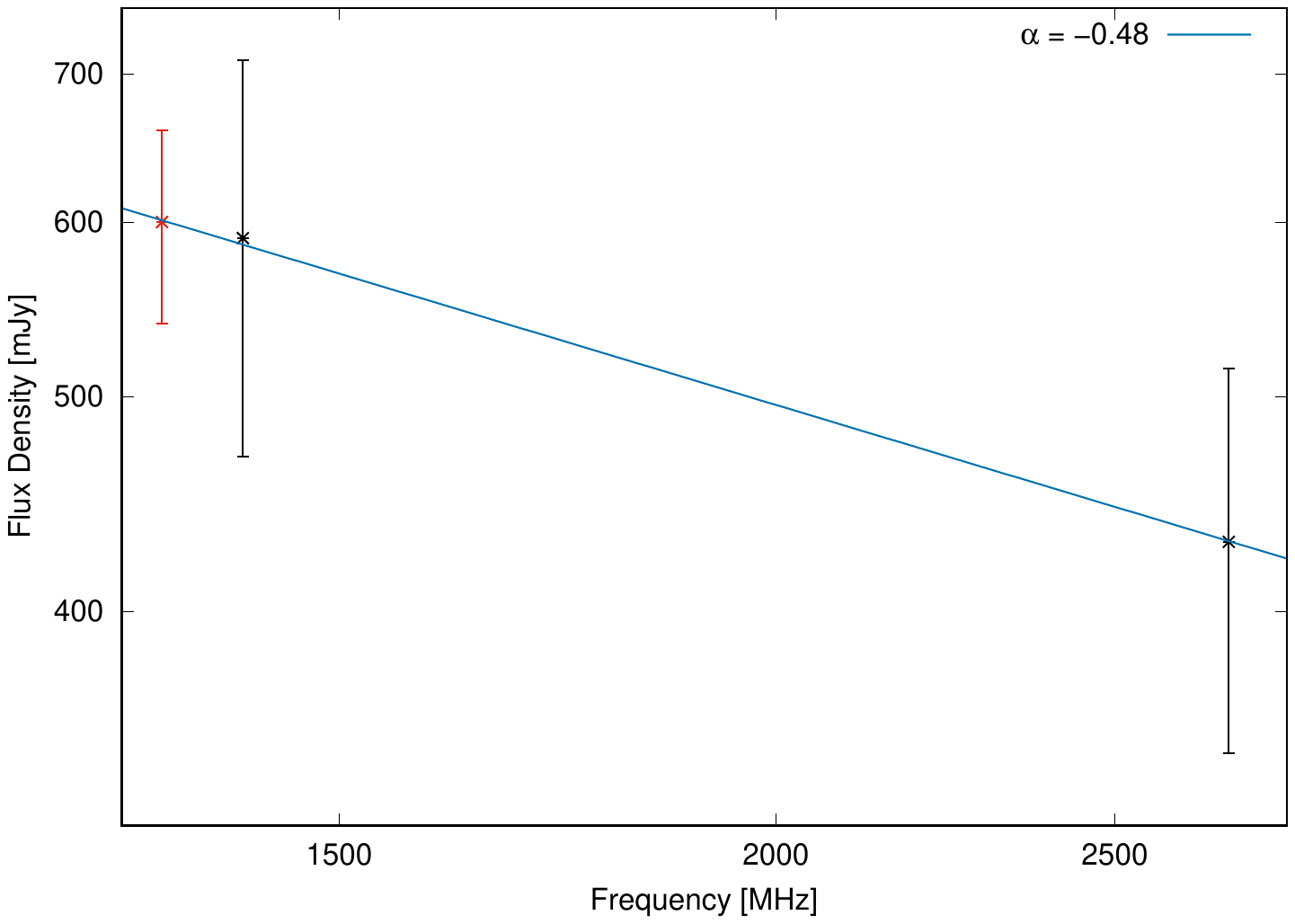}}
    \caption{Radio continuum spectrum of the SNR G36.6$+$2.6. The flux densities at 21~cm and 11~cm (black) were taken from \citet{Reich1988}. Our MeerKAT flux density is indicated in red. \label{fig:g36.6spec}}
\end{figure}

Our new MeerKAT image at 1335~MHz with a resolution of 10\asec\ is shown in Figure~\ref{fig:Ig36.6}. This SNR shows a bilateral structure with a shell in the south and a more complex and brighter shell in the north. It looks very similar to G8.7$-$5.0 as it also has a shell in the center, indicating this feature is moving towards us or away from us. Again similar to G8.7$-$5.0, the integrated flux density of $S_{1335} = 0.6 \pm 0.08$~Jy is exactly where it is expected to be, despite there being no obvious diffuse emission in the center from the shells moving towards and away from us. We updated the radio spectrum and fitted the spectral index to be $\alpha = -0.48 \pm 0.03$ (Fig.~\ref{fig:g36.6spec}).
The diameter of G36.6$+$2.6 is about 14\amin\ and we did not find any ears or blowouts.

\subsubsection{G53.6$-$2.2}
\begin{figure}[h]
    \centerline{\includegraphics[width=0.50\textwidth]{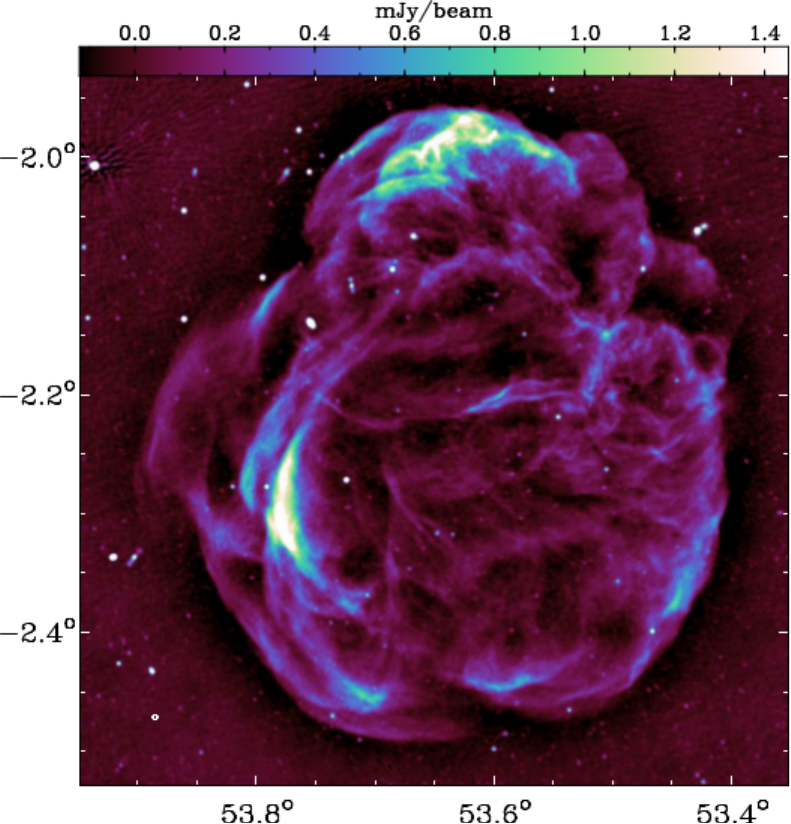}}
    \caption{Total power MeerKAT image of the SNR G53.6$-$2.2 at 1335~MHz in Galactic coordinates. The resolution of the image is 10\asec\ as indicated by the white circle in the lower left corner. \label{fig:Ig53.6}}
\end{figure}
The SNR G53.6$-$2.2, also known as NRAO~611 and 3C\,400.2, was discovered as a supernova remnant by \citet{Milne1970} from a comparison of Galactic radio surveys. The latest most up-to-date spectral analysis of this source \citep{Sun2011} reveals a radio spectral index of $\alpha \approx -0.50$, which is the canonical value for non-thermal radio emission from a supernova remnant.

Our new MeerKAT image at 1335~MHz with a resolution of 10\asec\ is shown in Figure~\ref{fig:Ig53.6}. This is the most sensitive and highest resolution image of this SNR. \citet{Dubner1994} suggest that G53.6$-$2.2 may actually consists of two separate overlapping supernova remnants, based on VLA observations at 1465~MHz. Our new higher resolution MeerKAT may actually support this assessment. The first SNR would be the bottom, almost circular, part of this radio source. The second SNR would be dominated by the bright shell at the top, which crosses with two shells on the left and right down into the other SNR. However, such a complex structure can also easily be explained by a very complex environment for a single supernova.

The integrated flux density from our data is $S_{1335} = 3.8 \pm 0.3$~Jy, which is a little more than 50\% of the value expected \citep{Sun2011}, which indicates missing large-scale emission.

\subsubsection{G55.7$+$3.4}
\begin{figure}[h]
    \centerline{\includegraphics[width=0.50\textwidth]{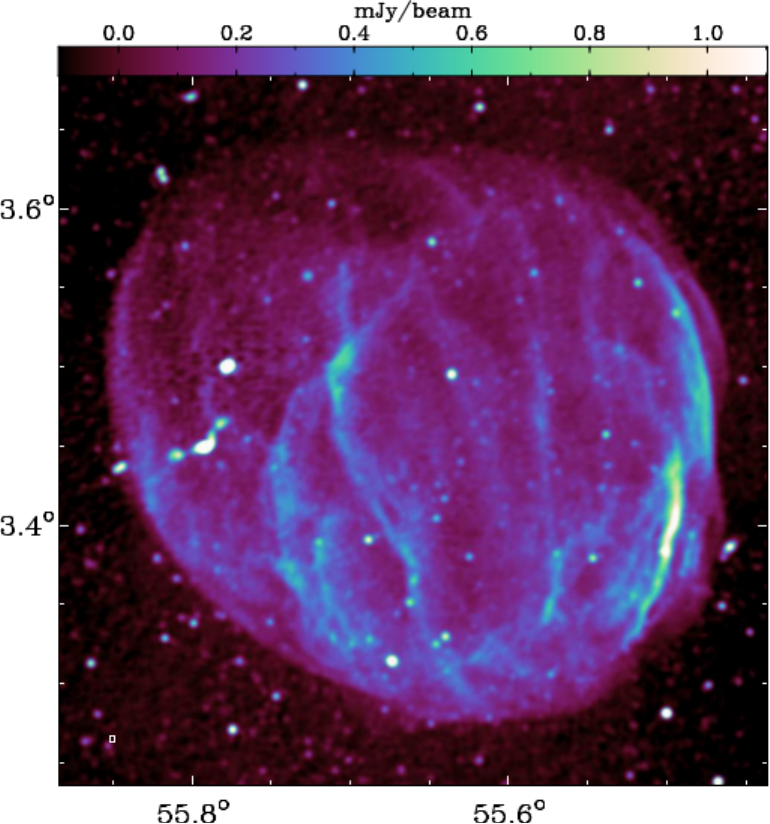}}
    \caption{Total power MeerKAT image of the SNR G55.7$+$3.4 at 1335~MHz in Galactic coordinates. The resolution of the image is 15\amin\ as indicated by the white circle in the lower left corner. \label{fig:Ig55.7}}
\end{figure}
The SNR G55.7$+$3.4 was discovered as a supernova remnant by \citet{Caswell1970} from observations at 178~MHz and 2700~MHz targeting the area around the old pulsar CP~1919, which is not related to the SNR. The latest spectral analysis of this source \citep{Sun2011} reveals a radio spectral index of $\alpha \approx -0.34$, but they actually did not take all archived flux density measurements into account.

Our new MeerKAT image at 1335~MHz with a resolution of 15\asec\ is shown in Figure~\ref{fig:Ig55.7}. This is the most sensitive and highest resolution image of this SNR. It looks very similar to the one published by \citet{Bhatnagar2011} at 1500 MHz with a resolution of 20\asec. The SNR clearly shows a bilateral structure, with the symmetry axis almost perpendicular to the plane of the Galaxy. There are a few more vertical filaments projected onto the interior of the SNR and it shows smooth background emission. The diameter of G55.7$+$3.4 is about 23\amin.
\begin{figure}[h]
    \centerline{\includegraphics[height=0.40\textwidth]{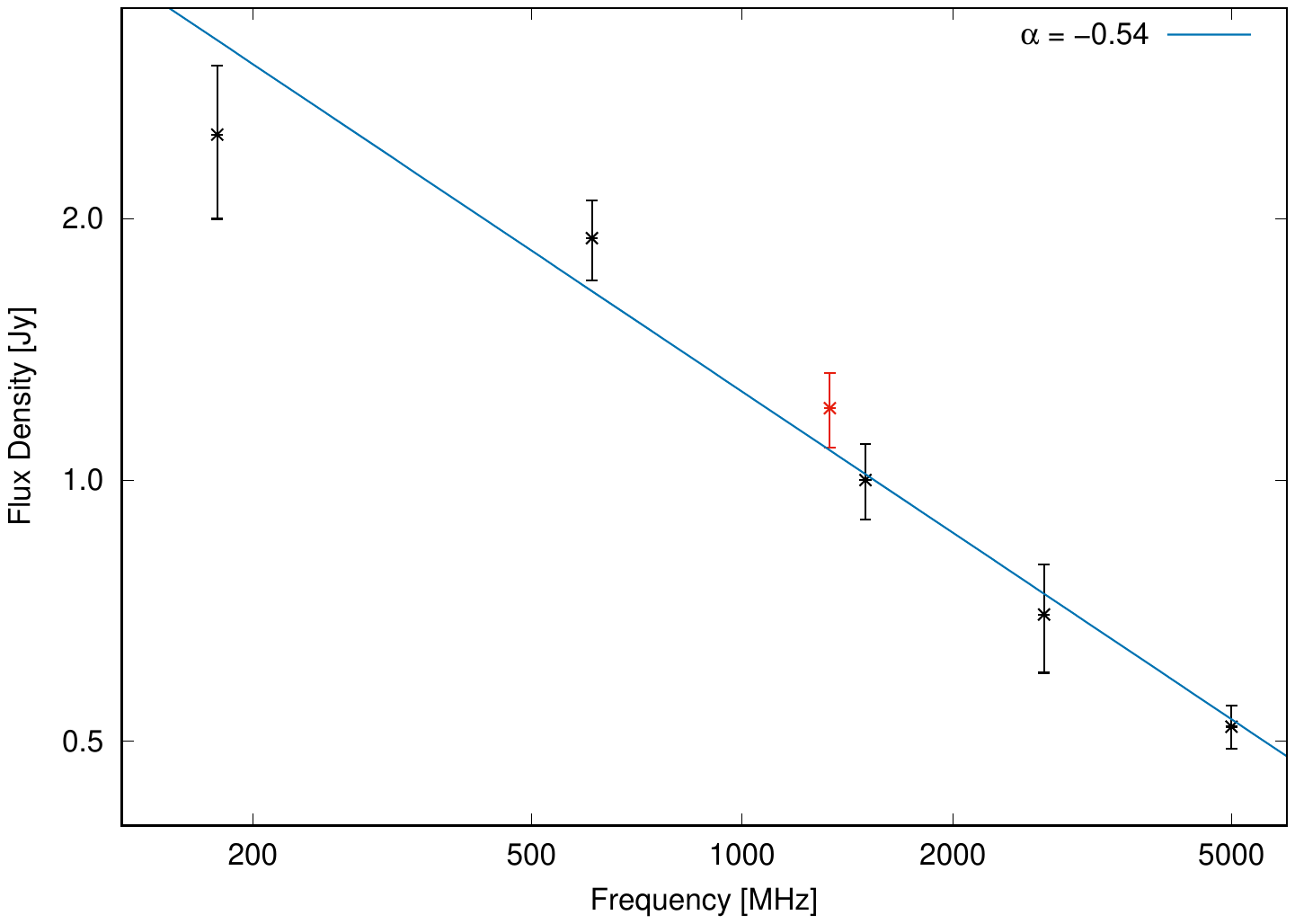}}
    \caption{Radio continuum spectrum of the SNR G55.7$+$3.4. Our flux density measurement is shown at 1335~MHz (red). Other flux density values (black) were taken from: \citet{Caswell1970}, \citet{Goss1977}, \citet{Bhatnagar2011}, and \citet{Sun2011}. \label{fig:g55.7spec}}
\end{figure}
The integrated flux density from our data is $S_{1335} = 1.25 \pm 0.12$~Jy, which is a little more than the value expected from \citet{Sun2011}. We updated the radio continuum spectrum and fitted the spectral index to be $\alpha = -0.54 \pm 0.05$ (Fig.~\ref{fig:g55.7spec}), taking all flux densities available in the literature into account. This radio spectral index is typical of the non-thermal emission coming from a mature SNR.

\subsubsection{G57.2$+$0.8}
\begin{figure}[h]
    \centerline{\includegraphics[width=0.50\textwidth]{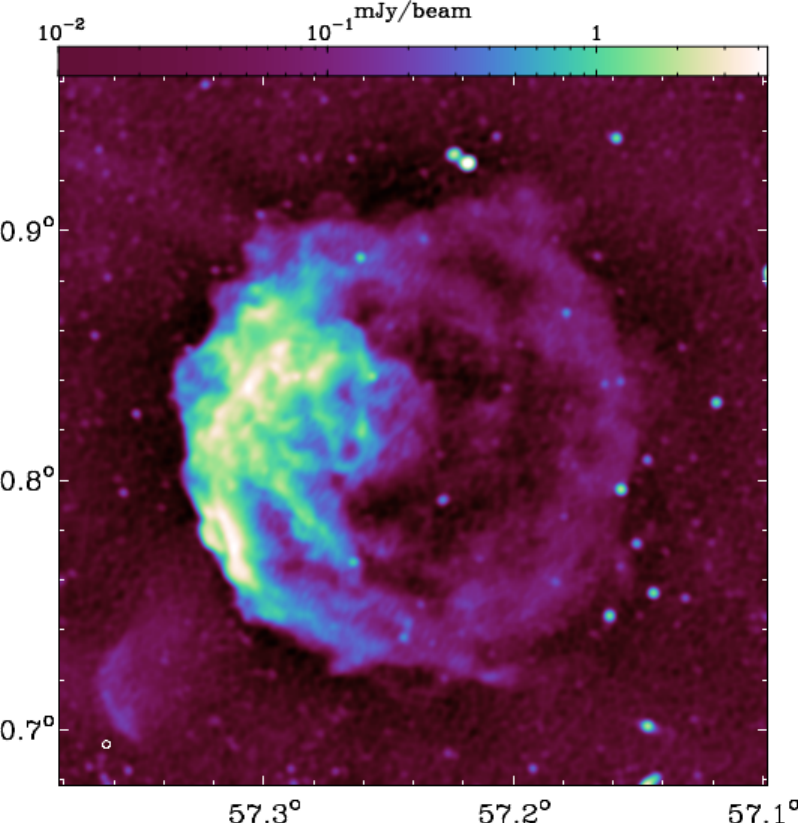}}
    \caption{Total power MeerKAT image of the SNR G57.2$+$0.8 at 1335~MHz in Galactic coordinates. The resolution of the image is 15\asec\ as indicated by the white circle in the lower left corner. \label{fig:Ig57.2}}
\end{figure}
The SNR G57.2$+$0.8 was discovered by \citet{Sieber1984} in observations of the area in the neighbourhood of millisecond pulsar PSR~1937+214 with the Effelsberg 100-m radio telescope. The millisecond pulsar is not related to the SNR. However, G57.2+0.8 is host to a magnetar that is related to the only known Galactic Fast Radio Burst observed to date \citep{Scholz2020}. The latest most up-to-date spectral analysis of this source by \citet{Kothes2018} reveals a radio spectral index of $\alpha = -0.55$, which is typical for non-thermal radio emission from a mature supernova remnant. The best previous image was also observed with MeerKAT by \citet{Bailes2021} at a resolution of 8.4\asec $\times$ 5.8\asec, which looks very similar to ours.

Our new MeerKAT image at 1335~MHz with a resolution of 15\asec\ is shown in Figure~\ref{fig:Ig57.2}. The SNR is almost circular with the brightest shell to the east. This shell is not the typical smooth shell with a sharp outer edge, but shows a lot of filamentary and cloudy substructure, while the fainter shell to the west is more smooth.
In the lower left corner of the image in Figure~\ref{fig:Ig57.2} is another small shell-like feature of unknown origin. It could be a foreground object such as a planetary nebula. 

The integrated flux density from our data is $S_{1335} = 1.25 \pm 0.1$~Jy, which is a little less than the value expected from \citet{Kothes2018}. The study by \citet{Kothes2018} indicates that the prominent shell is sitting on top of a faint emission plateau, which seems to be missing here. That would account for the missing flux.

\subsubsection{G261.9$+$5.5}
\begin{figure}[h]
    \centerline{\includegraphics[width=0.50\textwidth]{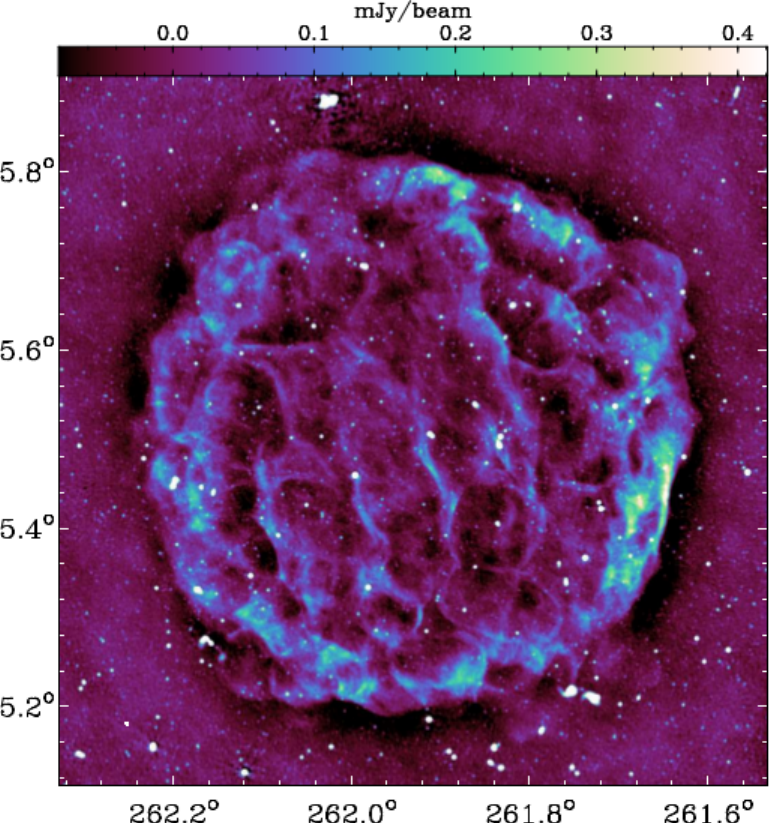}}
    \caption{Total power MeerKAT image of the SNR G261.9$+$5.5 at 1335~MHz in Galactic coordinates. The resolution of the image is 10\asec\ as indicated by the white circle in the lower left corner. \label{fig:Ig261.9}}
\end{figure}
G261.9$+$5.5 was first identified as a SNR by \citet{Hill1967} in Parkes observations at 1410~MHz and 2650~MHz. In follow-up observations with the Molonglo Observatory Synthesis Telescope (MOST) it looks like an extended, smooth and low surface brightness source with little or no limb brightening and a radio spectral index of $\alpha \approx -0.4$ \citep{Kesteven1987}.

Our new MeerKAT image at 1335~MHz with a resolution of 10\asec\ is shown in Figure~\ref{fig:Ig261.9}. It does not represent the typical characteristics of a supernova remnant. It looks like a large web of little bubbles that are confined by a mostly limb-brightened outer shell. There are little blowouts visible almost all around the object. There does not seem to be a clear symmetry axis like that for a bilateral or barrel-shaped SNR. 

The integrated flux density from our data is $S_{1335} = 2.7 \pm 0.4$~Jy, which is only about 35\% of the value expected from \citet{Kesteven1987}. Even though previous observations are very uncertain, this clearly indicates that there is large-scale emission missing for this object. This is not surprising at an average diameter of almost 40\amin.

\subsubsection{G272.2$-$3.2}
\begin{figure}[h]
    \centerline{\includegraphics[width=0.50\textwidth]{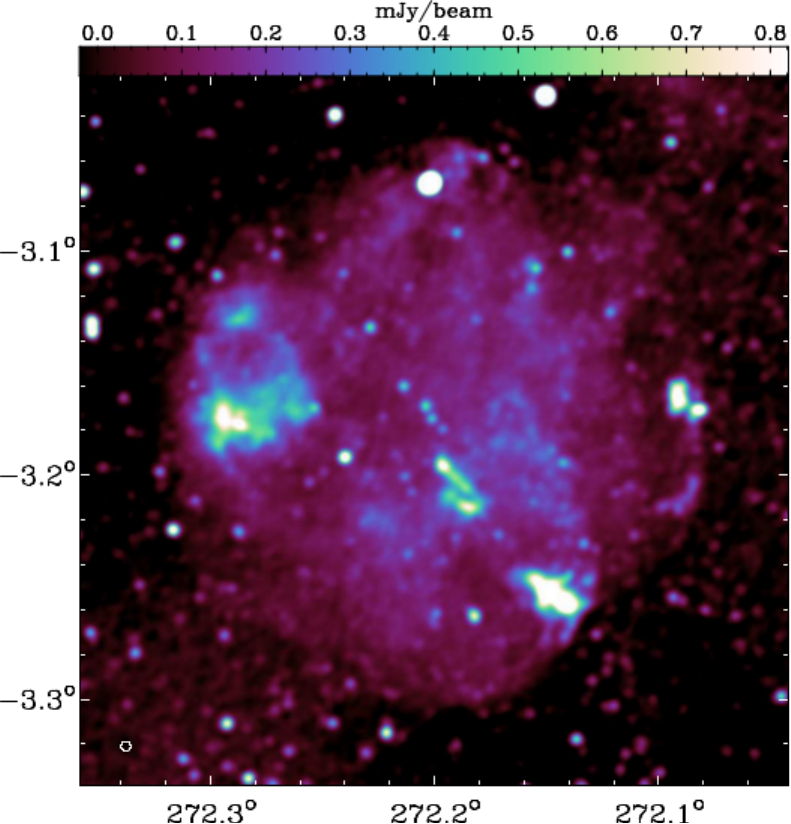}}
    \caption{Total power MeerKAT image of the SNR G272.2$-$3.2 at 1335~MHz in Galactic coordinates. The resolution of the image is 15\asec\ as indicated by the white circle in the lower left corner. \label{fig:Ig272.2}}
\end{figure}
G272.2$-$3.2 was discovered as a supernova remnant in data from the ROSAT All-Sky Survey through its X-ray emission by \citet{Greiner1994}. In the radio this source looks like a smooth, centrally brightened object with a few radio enhancements that may or may not be compact background sources \citep{Duncan1997b}. Most of those radio blobs correlate with optical emission. \citet{Duncan1997b} found a typical spectral index of $\alpha = -0.55$.
\begin{figure}[h]
    \centerline{\includegraphics[height=0.40\textwidth]{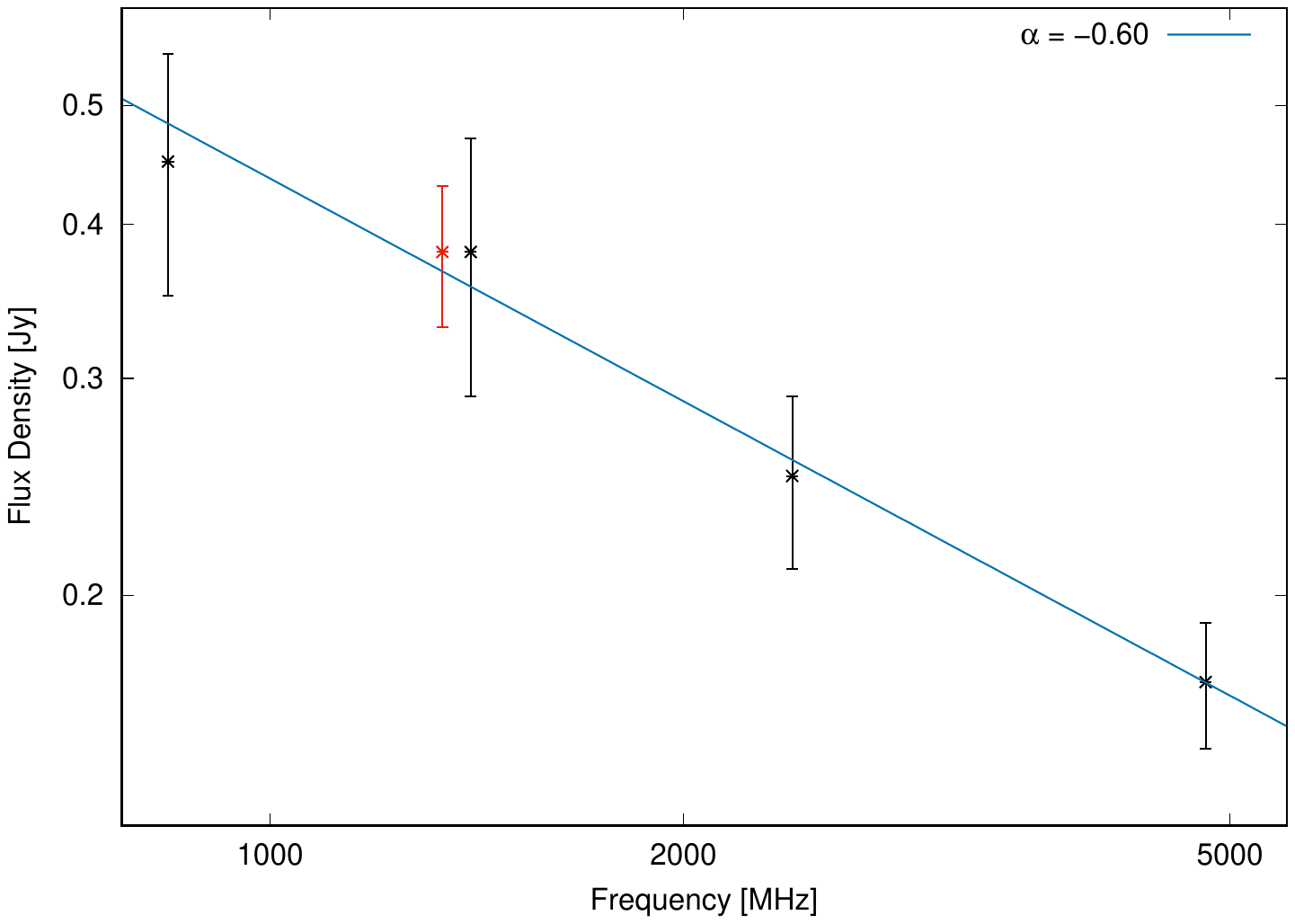}}
    \caption{Radio continuum spectrum of the SNR G272.2$-$3.2. Our flux density measurement is shown at 1335~MHz in red. Other flux density values (black) were taken from \citet{Duncan1997b}. \label{fig:g272.2spec}}
\end{figure}

Our new MeerKAT image at 1335~MHz with a resolution of 15\asec\ is shown in Figure~\ref{fig:Ig272.2}. Our image has much higher resolution than previous radio observations and therefore we can look at features in much more detail. The radio emission blobs found by \citet{Duncan1997b} to correlate with optical emission are very compact, possibly extra-galactic sources. The radio emission does not appear to be limb-brightened as would be expected from a shell-type SNR, but it is centrally filled like a pulsar wind nebula (PWN).

The integrated flux density from our data is $S_{1335} = 0.38 \pm 0.05$~Jy, with no evidence of missing short spacings. The average diameter of this source is about 15\amin. We plotted a radio continuum spectrum including our flux density and the literature values in Figure~\ref{fig:g272.2spec}. The resulting spectral index is $\alpha = -0.60 \pm 0.03$. This would be a typical value for a shell-type SNR, but unusually steep for a pulsar wind nebula for which we expect a flat radio spectrum with $\alpha \ge -0.3$. There are \chg{however} a few PWNe displaying steep radio spectra, such as G76.9+1.0 with $\alpha \approx -0.6$ \citep{Landecker1997} and the recently discovered PWN G141.2+5.0 with $\alpha \approx -0.7$
\citep{Kothes2014}. Those steep radio spectra are believed to be the result of the re-acceleration of the relativistic particles inside the PWN through the SN reverse shock \citep{Kothes2017b}.

\subsubsection{G284.3$-$1.8}
\begin{figure}[h]
    \centerline{\includegraphics[width=0.50\textwidth]{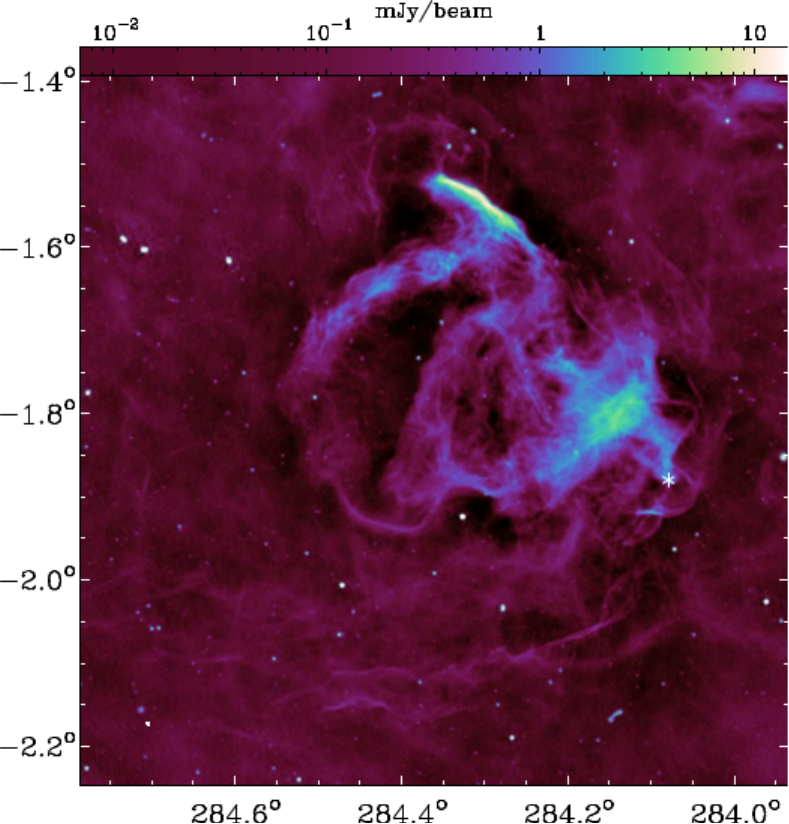}}
    \caption{Total power MeerKAT image of the SNR G284.3$-$1.8 at 1335~MHz in Galactic coordinates. The resolution of the image is 15\asec\ as indicated by the white circle in the lower left corner. The location of the young pulsar J1016$-$5857 is indicated by the white asterisk. \label{fig:Ig284.3}}
\end{figure}
G284.3$-$1.8, also know as MSH~10$-$53, was first listed as a supernova remnant by \citet{Milne1971}. There have not been many studies of this SNR in radio and therefore its spectral index is somewhat uncertain at $\alpha \approx -0.32$ \citep{Milne1989}, which is somewhat flat for a shell-type SNR. 

The full extent of this SNR is seen for the first time in our new MeerKAT image displayed in Figure~\ref{fig:Ig284.3}. The SNR shows a very complex structure. The two brightest features and the shell to the northeast were the only characteristics visible in the old radio observations \citep{Milne1989}. In addition we do see a very complex network of filaments in the central area that expand out to a diameter of about 45\amin. In particular the filaments and shells in the north and the south are responsible for doubling this SNR in size. 

\citet{Camilo2001} found a young pulsar just outside the edge of this SNR as it was known at this time. They argue that the pulsar and the SNR are related based on the positional coincidence and the presence of a X-ray PWN at its location. Later an extended radio source at the position of the pulsar was found \citep{Camilo2004} and a radio bridge between the pulsar and the shell of the SNR, indicating interaction. In our new MeerKAT image we can confirm those features and in addition we find faint SNR filaments outside the pulsar projected to the plane of the sky. There is also a faint tail behind the pulsar pointing back to the interior of the SNR. We cannot say if this tail points back to the SNR's cent\chg{er} as the determination of the geometric cent\chg{er} is not possible due to the SNR's complexity.

The integrated flux density from our data is $S_{1335} = 8.9 \pm 0.8$~Jy, which is only about 90\% of the value expected from \citet{Milne1989}. The central area clearly shows missing short spacings, while the outer shells would not have been covered in older measurements. 

\subsubsection{G292.0$+$1.8}
\begin{figure}[h]
    \centerline{\includegraphics[width=0.50\textwidth]{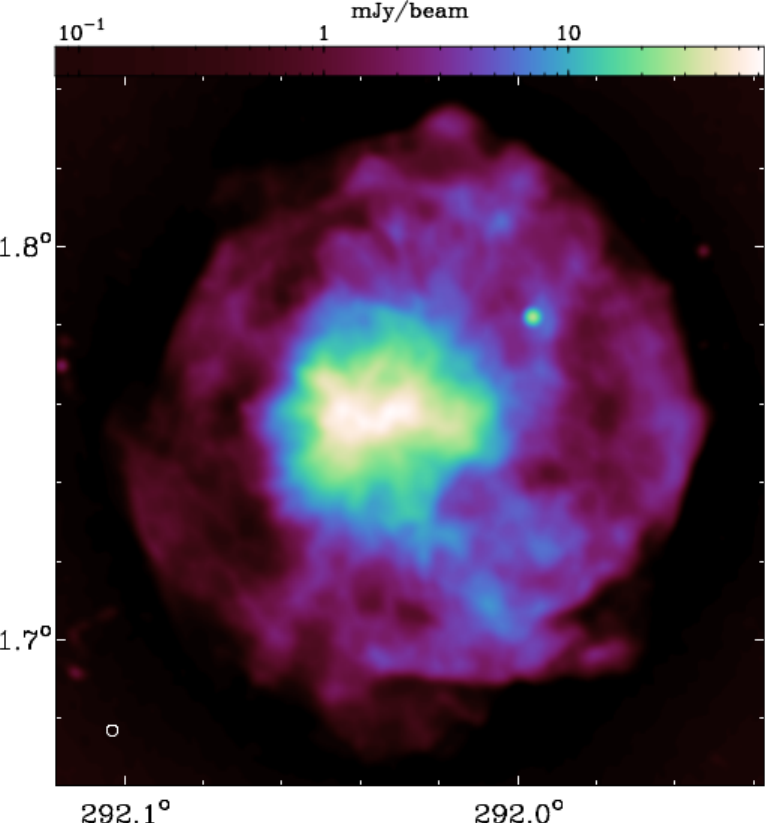}}
    \caption{Total power MeerKAT image of the SNR G292.0$+$1.8 at 1335~MHz in Galactic coordinates. The resolution of the image is 15\asec\ as indicated by the white circle in the lower left corner. \label{fig:Ig292.0}}
\end{figure}
The SNR G292.0$+$1.8 was discovered by \citet{Shaver1970} in an investigation of radio sources with the Parkes telescope at 5000~MHz and the Molonglo radio telescope at 408~MHz.
In the most recent radio study by \citet{Gaensler2003} it is argued that the radio source consists of two major components: a bright pulsar wind nebula powered by PSR~J1124$-$5916 surrounded by a plateau of radio emission representing the surrounding shell SNR. They found spectral indices of $\alpha = -0.05$ and $\alpha = -0.5$ for the PWN and the shell, respectively.
\begin{figure}[h]
    \centerline{\includegraphics[height=0.40\textwidth]{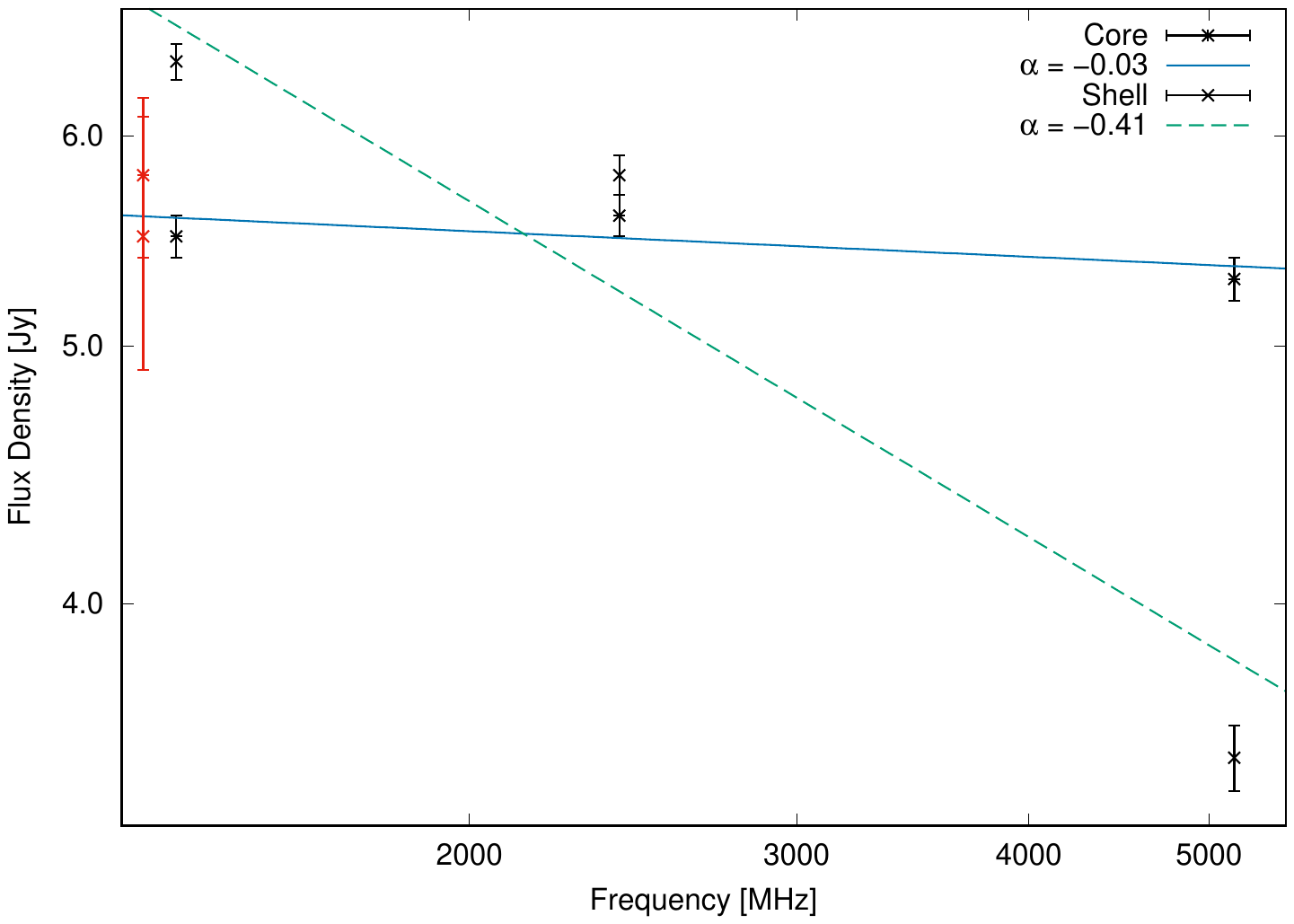}}
    \caption{Radio continuum spectrum of the SNR G292.0$+$1.8's core and plateau. Our flux density measurements are shown at 1335~MHz in red. Other flux density values were taken from \citet{Gaensler2003} and are shown in black. \label{fig:g292.0spec}}
\end{figure}
Our new MeerKAT image at 1335~MHz with a resolution of 15\asec\ is shown in Figure~\ref{fig:Ig292.0}. The PWN shows a central horizontal bar with spokes going outwards. The underlaying circular plateau shows some cloudy sub-structure, a sharp outer edge and not much limb-brightening. 

The integrated flux density from our data is $S_{1335} = 5.5 \pm 0.6$~Jy and $S_{1335} = 5.8 \pm 0.4$~Jy for the plateau and the core, respectively. The large errors are the result of the uncertainty in determining the outer radius of the core. It actually looks like some of the PWN emission features continue into the plateau. The combined spactra of the plateau and the core are displayed in Figure~\ref{fig:g292.0spec}. Considering the large uncertainties in separating the plateau from the core, our flux densities agree with those of \citet{Gaensler2003}. However, the plateau may have some missing flux.

\subsubsection{G296.5+10.0}
\begin{figure}[h]
    \centerline{\includegraphics[width=0.50\textwidth]{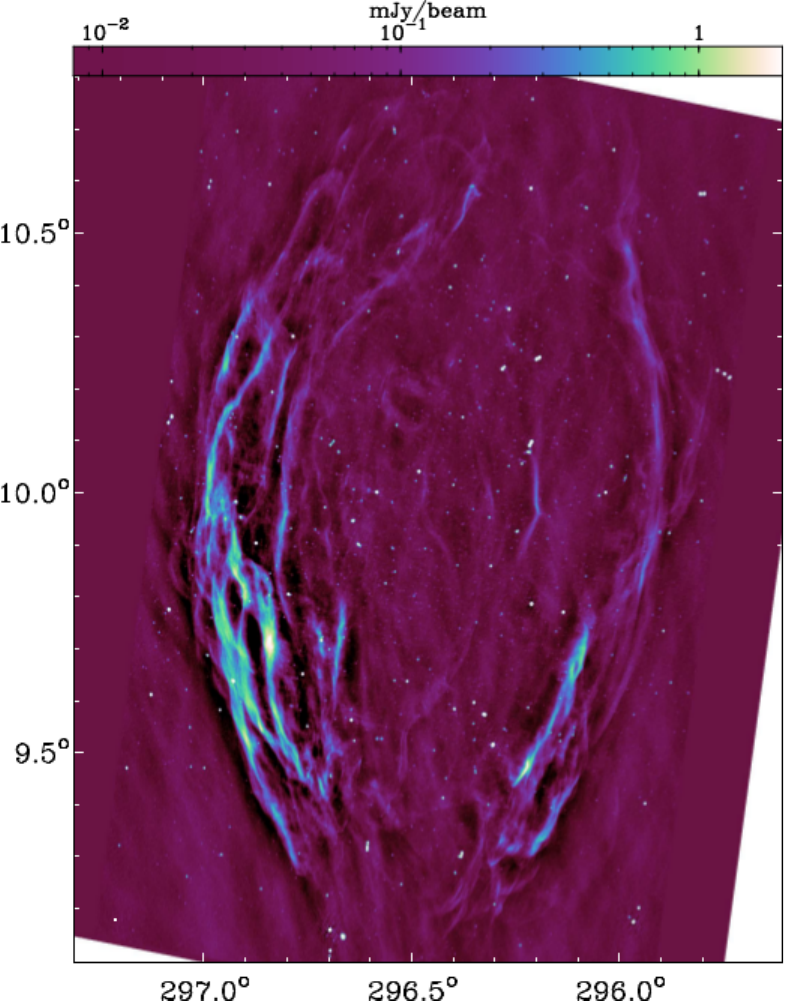}}
    \caption{Total power MeerKAT image of the SNR G296.5$+$10.0 at 1335~MHz in Galactic coordinates. The resolution of the image is 12\asec\ as indicated by the white circle in the lower left corner. \label{fig:Ig296.5}}
\end{figure}
G296.5+10.0 was identified as a supernova remnant by \citet{Whiteoak1968}, based on radio polarimetric observations with the Parkes radio telescope at 629~MHz, 1410~MHz, and 2650~MHz. In the most recent radio study, which also discussed the total power spectrum, by \citet{Milne1994}, an overall spectral index of $\alpha = -0.5$ was found, typical for a mature SNR.

Our new MeerKAT image at 1335~MHz with a resolution of 12\asec\ is shown in Figure~\ref{fig:Ig296.5}. It shows the well known bilateral structure with the symmetry axis almost perpendicular to the Galactic plane. While older Parkes observations show two smooth homogenous limbs, our high resolution observations split these two shells up into many narrow filaments. We can also see many low-level blowouts almost all around the perimeter, marked by the brightest filaments. From earlier high resolution maps \citep[e.g.][]{Roger1988,Harvey-Smith2010} we know that there are faint filaments going beyond the brighter part of the SNR in the north and the south, which were not captured in our observations. Therefore the extent of this SNR listed in Table~\ref{tab:snrchar} should be taken with a grain of salt.

The integrated flux density from our data is $S_{1335} = 3.3 \pm 0.6$~Jy. This is less than 10\% of the expected flux density at this frequency \citep{Milne1994}. As this is the largest SNR in angular dimension of our sample this large amount of missing flux is not surprising. In our observations this SNR has a width of about 68\amin\ and the height is about 97\amin, but as already mentioned above, this SNR is known to have faint filamentary extensions to the north and south. Therefore, 97\amin\ is a lower limit.

\subsubsection{G299.2$-$2.9}
\begin{figure}[h]
    \centerline{\includegraphics[width=0.50\textwidth]{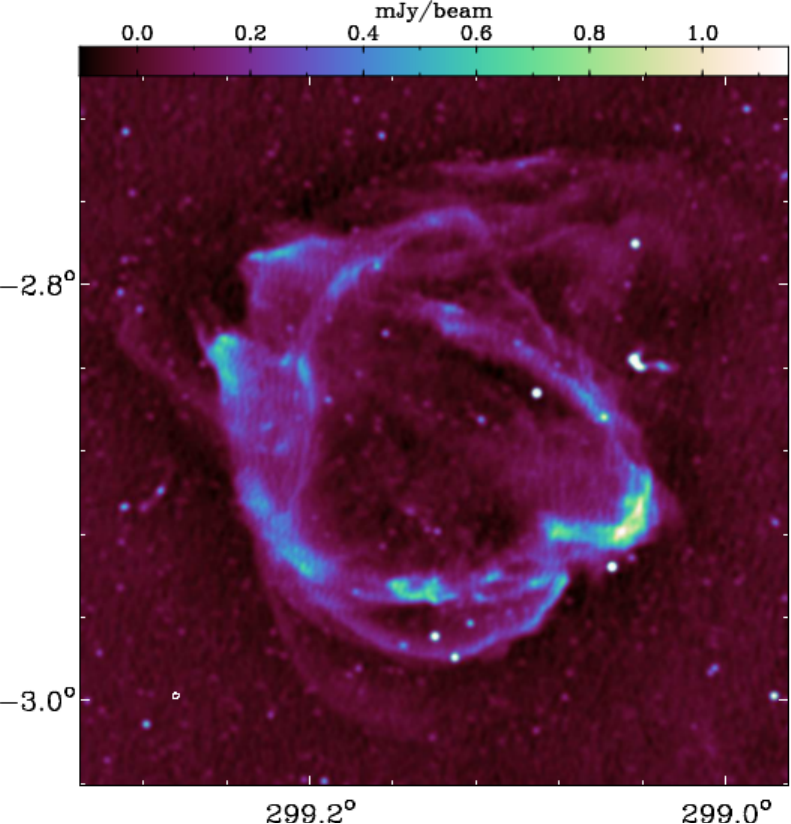}}
    \caption{Total power MeerKAT image of the SNR G299.2$-$2.9 at 1335~MHz in Galactic coordinates. The resolution of the image is 15\asec\ as indicated by the white circle in the lower left corner. \label{fig:Ig299.2}}
\end{figure}
G299.2$-$2.9 was discovered as a supernova remnant in data from the ROSAT All-Sky Survey through its X-ray emission by \citet{Busser1996}. There has not been any dedicated radio study of this SNR and no integrated flux density has been published.

Our MeerKAT image displayed in Figure~\ref{fig:Ig299.2} is the first map ever published for this SNR. It shows multiple shells in a complex pattern with the brightest to the southwest. No symmetry axis for a bilateral structure is clearly visible. There seem to be a few blowouts in particular to the south and the east. The integrated flux density from our data is $S_{1335} = 0.54 \pm 0.05$~Jy. The background in our image looks relatively flat, which may indicate that we are not dealing with significant missing large-scale emission. In our observations this SNR has a diameter of about 18\amin.

\subsubsection{G312.5$-$3.0}
\begin{figure}[h]
    \centerline{\includegraphics[width=0.50\textwidth]{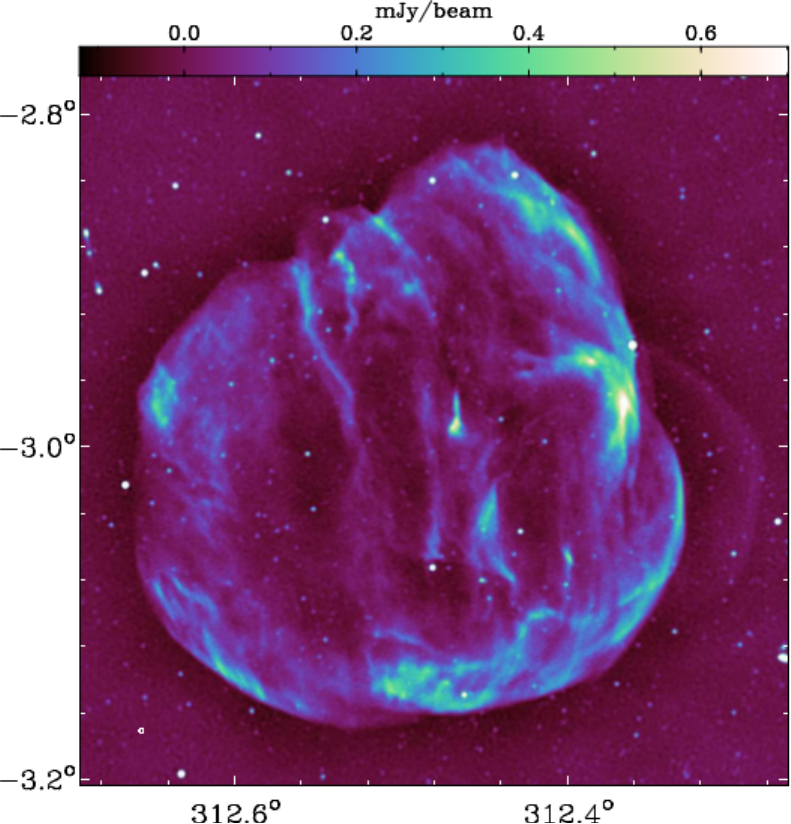}}
    \caption{Total power MeerKAT image of the SNR G312.5$-$3.0 at 1335~MHz in Galactic coordinates. The resolution of the image is 8\asec\ as indicated by the white circle in the lower left corner. \label{fig:Ig312.5}}
\end{figure}
G312.5$-$3.0 was first listed as a SNR candidate in a Parkes 2.4~GHz southern Galactic plane survey by \citet{Duncan1995}. Follow-up observations by \citet{Kane2003} with the Australia Telescope Compact Array (ATCA) confirmed this source as a SNR. As all of the previous observations have missing large scale emission, a spectral index could not be determined.

Our new MeerKAT image at 1335~MHz with a resolution of 8\asec\ is shown in Figure~\ref{fig:Ig312.5}. It resolves the smooth shells seen in the images of \citet{Kane2003} into multiple filaments all around the edge of the supernova remnant. The SNR does not show a clear bilateral structure although the filaments in the interior may indicate an almost vertical symmetry axis. There is a large blowout of a very faint filament seen in the west.

The integrated flux density from our data is $S_{1335} = 1.0 \pm 0.1$~Jy, which is less than half the value published by \citet{Kane2003} at 1380~MHz. The diameter of G312.5$-$3.0 is about 22\amin.

\subsubsection{G315.4$-$2.3}
\begin{figure}[h]
    \centerline{\includegraphics[width=0.50\textwidth]{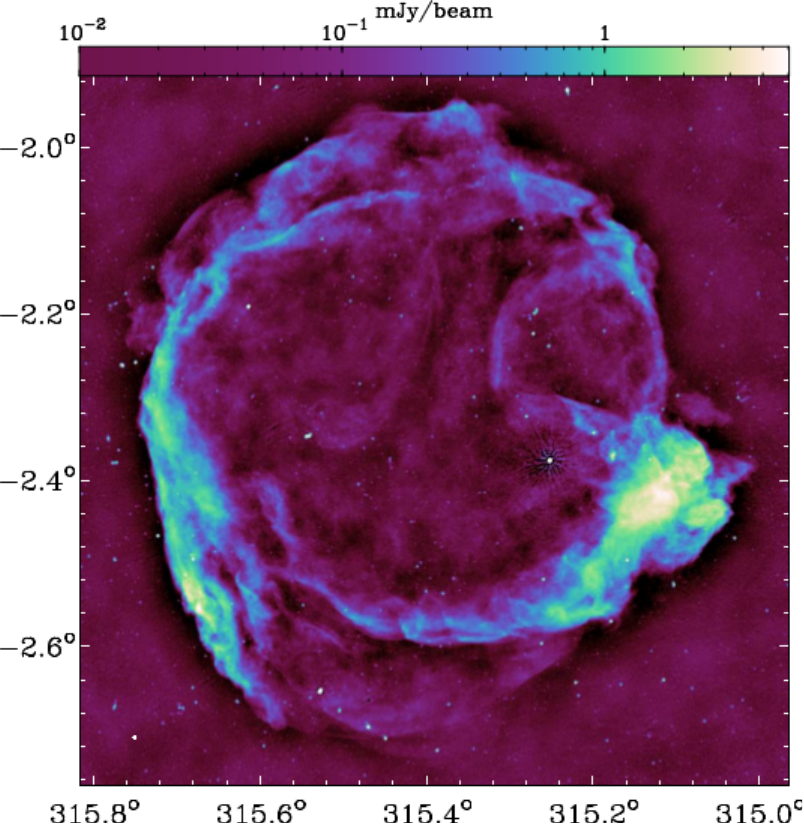}}
    \caption{Total power MeerKAT image of the SNR G315.4$-$2.3 at 1335~MHz in Galactic coordinates. The resolution of the image is 8\asec\ as indicated by the white circle in the lower left corner. \label{fig:Ig315.4}}
\end{figure}
G315.4$-$2.3, also known as RCW~86 and MSH~$14-63$, was first identified as a SNR by \citet{Hill1967} in Parkes observations at 1410~MHz and 2650~MHz. As this is a very large southern SNR, Parkes radio observations are necessary to capture all of its emission. The latest reliable spectral study was done by \citet{Caswell1975} and resulted in a spectral index of $\alpha = -0.62$. This SNR is well studied at all wavelengths.

Our new MeerKAT image at 1335~MHz with a resolution of 8\asec\ is shown in Figure~\ref{fig:Ig315.4}. This image has the same characteristics as the observation published by \citet{Dickel2001}, observed with ATCA at 1340~MHz at a resolution of 8\asec. But the improvement of the MeerKAT image in sensitivity and image fidelity is quite remarkable, thanks to 2000 baselines with MeerKAT \chg{compared to only} 15 with ATCA. G315.4$-$2.3 is almost circular with shells all around and no favoured bilateral structure. This would indicate either a young SNR or that the ambient magnetic field is almost parallel to the line of sight. It shows a few protrusions all around its perimeter and a bright extended non-filamentary region of radio emission to the west.

The integrated flux density from our data is $S_{1335} = 15.4 \pm 1.0$~Jy, which is about a third of the expected value \citep{Caswell1975}. The diameter of about 45\amin\ makes G315.4$-$2.3 one of the largest SNRs in our sample.

\subsubsection{G326.3$-$1.8}
\begin{figure}[h]
    \centerline{\includegraphics[width=0.50\textwidth]{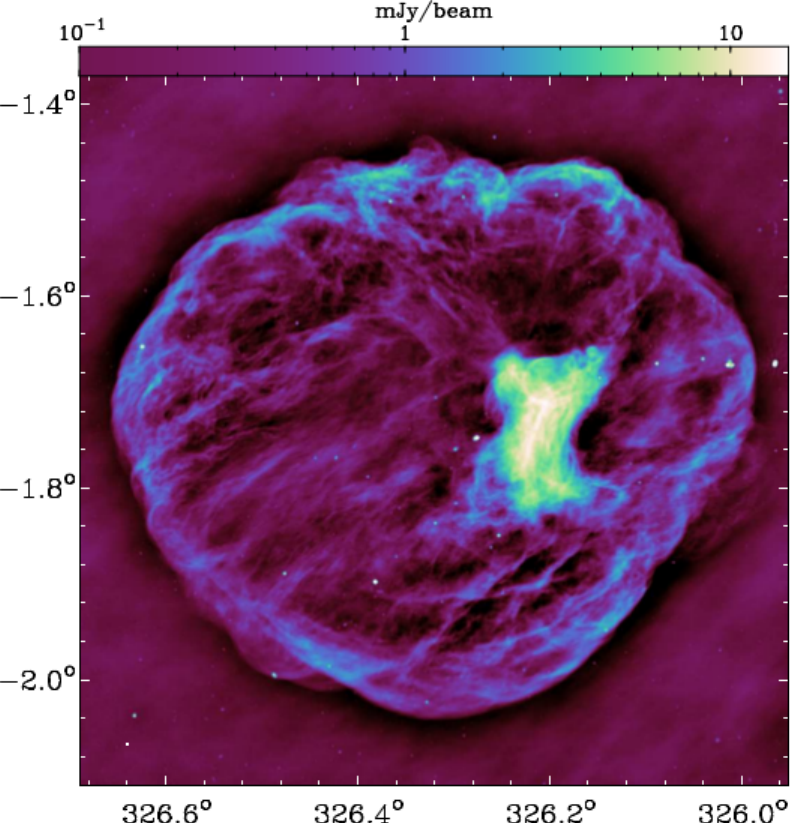}}
    \caption{Total power MeerKAT image of the SNR G326.3$-$1.8 at 1335~MHz in Galactic coordinates. The resolution of the image is 8\asec\ as indicated by the white circle in the lower left corner. \label{fig:Ig326.3}}
\end{figure}
G326.3$-$1.8, also known as MSH~$15-56$, was first identified as a SNR by \citet{Hill1967} in Parkes observations at 1410~MHz and 2650~MHz. This SNR is well studied with many telescopes at all available wavelengths. The most recent spectrum was published by \citet{Ball2023} resulting in an overal spectral index of $\alpha = -0.32 \pm 0.04$, which also includes a flux density at 935~MHz observed with the Australian SKA Pathfinder (ASKAP).

Our new MeerKAT image at 1335~MHz with a resolution of 8\asec\ is shown in Figure~\ref{fig:Ig326.3}. This is a very large SNR in angular extent, consisting of a shell-type remnant and a bright pulsar wind nebula to west of the central area. The shell-type SNR shows limb-brightening all around its perimeter with no clear symmetry axis or a bilateral structure. A few protrusions are visible to the north and in the south-eastern area. There are also many filaments in its interior. The PWN has a bar-like structure with a lot of filaments compressed in the center of the bar, spreading out to either end of it.
\begin{figure}[h]
    \centerline{\includegraphics[height=0.40\textwidth]{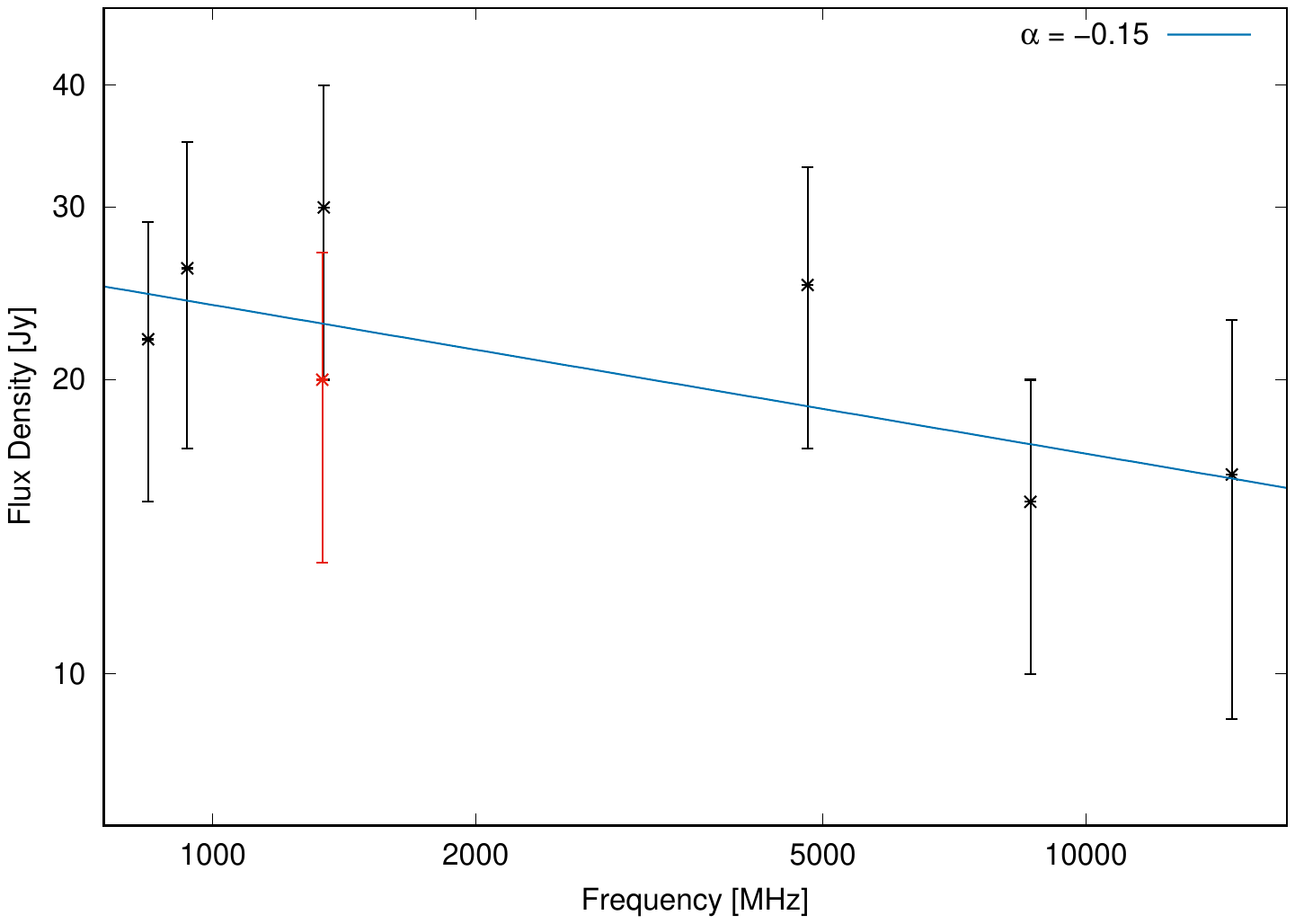}}
    \caption{Radio continuum spectrum of the PWN inside SNR G326.3$-$1.8. Our flux density measurement is shown at 1335~MHz in red. We also integrated the PWN emission at 935~MHz from the observations of \citet{Ball2023}. Other flux density values were taken from \citet{Dickel2000}. \label{fig:g326.3spec}}
\end{figure}

The integrated flux density from our data is $S_{1335} = 40 \pm 4$~Jy for the whole SNR at a diameter of about 40\amin\ and about 20~Jy for the PWN. The uncertainty for the PWN flux density is rather high since its background is highly fluctuating and the edge is somewhat uncertain as well. We plotted the updated spectrum of the PWN in Figure~\ref{fig:g326.3spec}. The uncertainties were taken to be one third, following the recommendation by \citet{Dickel2000}, because of the high uncertainty in the background estimate for the PWN. The resulting radio spectral index for the PWN is $\alpha = -0.15 \pm 0.07$, confirming the typical flat radio continuum spectrum of a PWN and earlier results.

\subsubsection{G327.6$+$14.6}
\begin{figure}[h]
    \centerline{\includegraphics[width=0.50\textwidth]{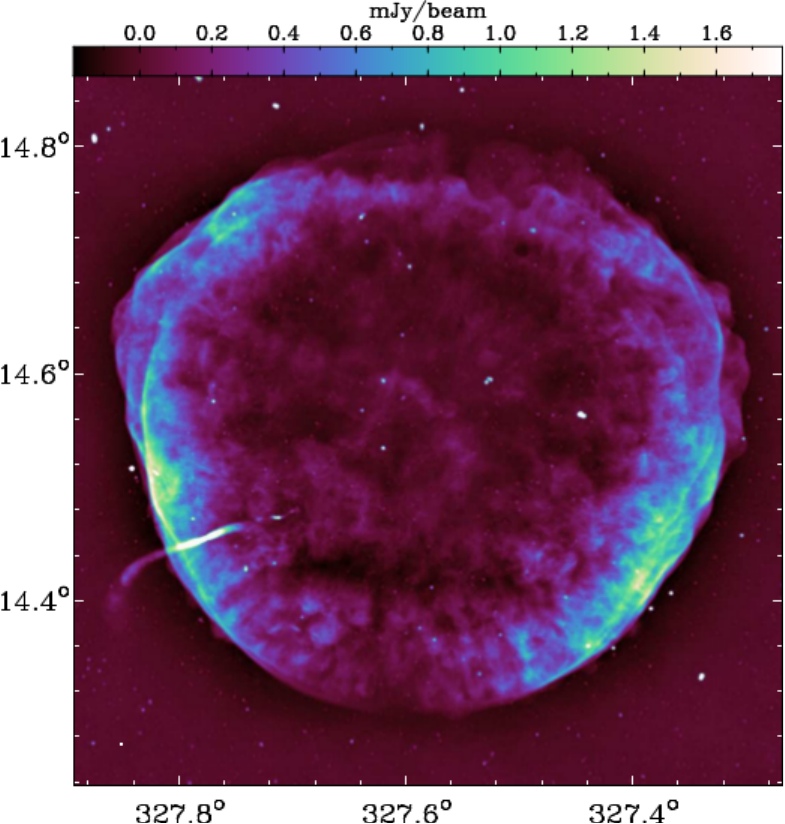}}
    \caption{Total power MeerKAT image of the SNR G327.6$+$14.6 at 1335~MHz in Galactic coordinates. The resolution of the image is 8\asec\ as indicated by the white circle in the lower left corner. \label{fig:Ig327.6}}
\end{figure}
The radio SNR G327.6$+$14.6, also known as PKS~1459--41 or SN1006, is the result of a supernova that was observed in 1006 \citep{Gardner1965}. This supernova remnant has been studied at all wavelengths many times. As this is, at more than 30\amin, a very large object, typically single antenna observations are required to study the full spectrum. The radio spectral index of the full source is, at $\alpha \approx -0.6$, typical for a young supernova remnant. Our new MeerKAT image at 1335~MHz with a resolution of 8\asec\ is shown in Figure~\ref{fig:Ig327.6}. It shows the well-known structure of this SNR with a bilateral structure around an almost vertical symmetry axis and an overall fluffy interior and many Rayleigh-Taylor instabilities on the inside of the shells. The integrated flux density from our data is $S_{1335} = 8.2 \pm 0.7$~Jy for the whole SNR at a diameter of about 33\amin. This is about 50\% of the flux we expect to see at this frequency.

\subsubsection{G332.5$-$5.6}
\begin{figure}[h]
    \centerline{\includegraphics[width=0.50\textwidth]{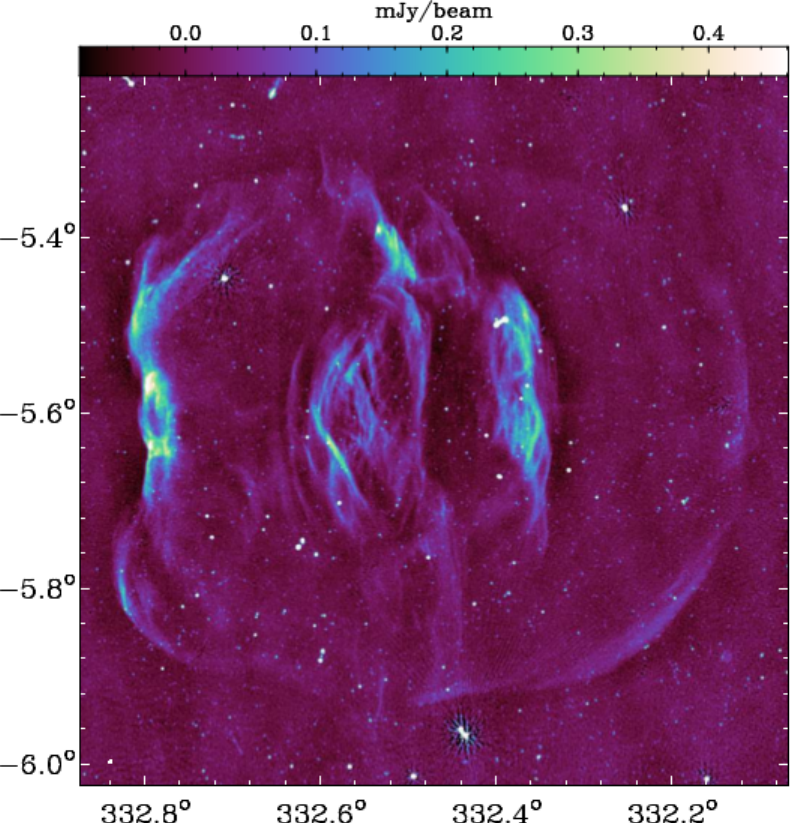}}
    \caption{Total power MeerKAT image of the SNR G332.5$-$5.6 at 1335~MHz in Galactic coordinates. The resolution of the image is 8\asec\ as indicated by the white circle in the lower left corner. \label{fig:Ig332.5}}
\end{figure}
G332.5$-$5.6 was first listed as a SNR candidate in a Parkes 2.4~GHz southern Galactic plane survey by \citet{Duncan1995}. Follow-up observations by \citet{Reynoso2007} with ATCA confirmed this source as a SNR. As all of the reliable previous observations have missing large scale emission, a full spectral index could not be determined. \citet{Reynoso2007} found this object to consist of three components which have spectral indices between $-0.6$ and $-0.7$.

Our new MeerKAT image at 1335~MHz with a resolution of 8\asec\ is shown in Figure~\ref{fig:Ig332.5}. The three components described by \citet{Reynoso2007} are the relatively bright vertical features at longitudes of about 332.35\degr\ and 332.8\degr\ described as the outer shell and between 332.5\degr\ and 332.6\degr\ described as a central component. A look at our new MeerKAT image (Fig.~\ref{fig:Ig332.5}) shows that it is not quite as simple as that. The vertical filament at 332.8\degr\ seems to be the eastern shell of a bilateral shaped SNR with the western shell at about 332.1\degr\ being much fainter. This western shell has not been seen before and increases the size of this SNR quite dramatically. The two other vertical components discovered by \citet{Reynoso2007} seem to be in the central area of the SNR. 

The integrated flux density from our data is $S_{1335} = 1.6 \pm 0.1$~Jy for the whole SNR at a diameter of about 40\amin. This is a little less than the combined flux density at 1384~MHz of the three components \citep{Reynoso2007}. 

\subsubsection{G343.1$-$2.3}
\begin{figure}[h]
    \centerline{\includegraphics[width=0.50\textwidth]{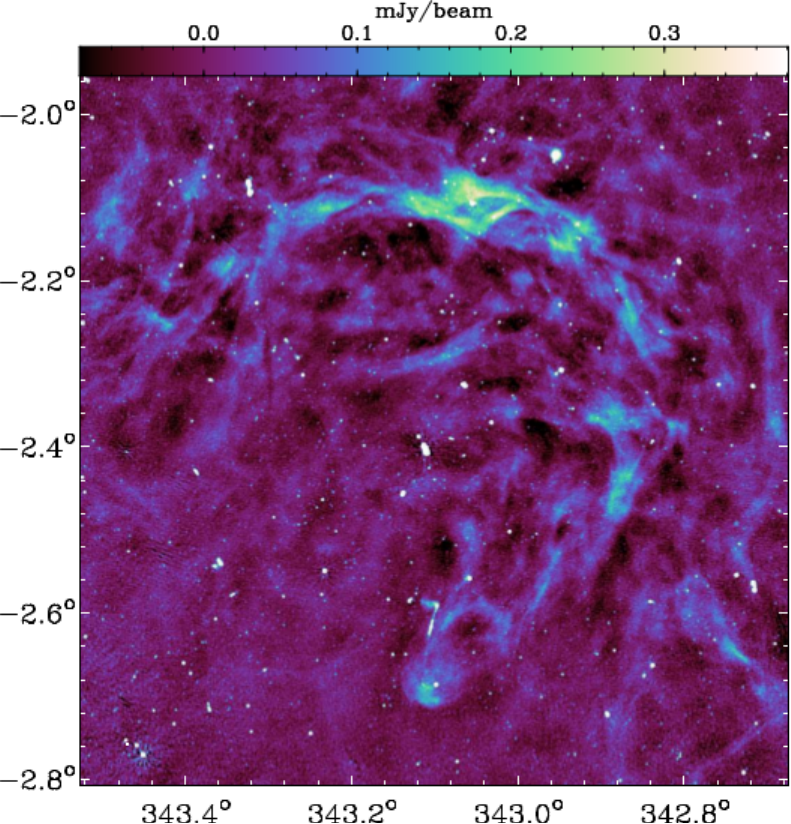}}
    \caption{Total power MeerKAT image of the SNR G343.1$-$2.3 at 1335~MHz in Galactic coordinates. The resolution of the image is 8\asec\ as indicated by the white circle in the lower left corner. \label{fig:Ig343.1}}
\end{figure}
The radio supernova remnant G343.1$-$2.3 was discovered in a radio survey of the area around the $\gamma$-ray pulsar B1706$-$44 by \citet{McAdam1993}. Its radio spectral index is uncertain, as there are no reliable radio observations of the whole source available due to its low radio surface brightness and its full extent is still unknown. PSR~B1706$-$44 shows a faint wind nebula around it of diameter 4\amin\ and seems to drag a tail behind it \citep{Frail1994}. A proper motion study of the pulsar by \citet{deVries2021} makes an association of the pulsar with the SNR unlikely.

\begin{figure}[h]
    \centerline{\includegraphics[width=0.50\textwidth]{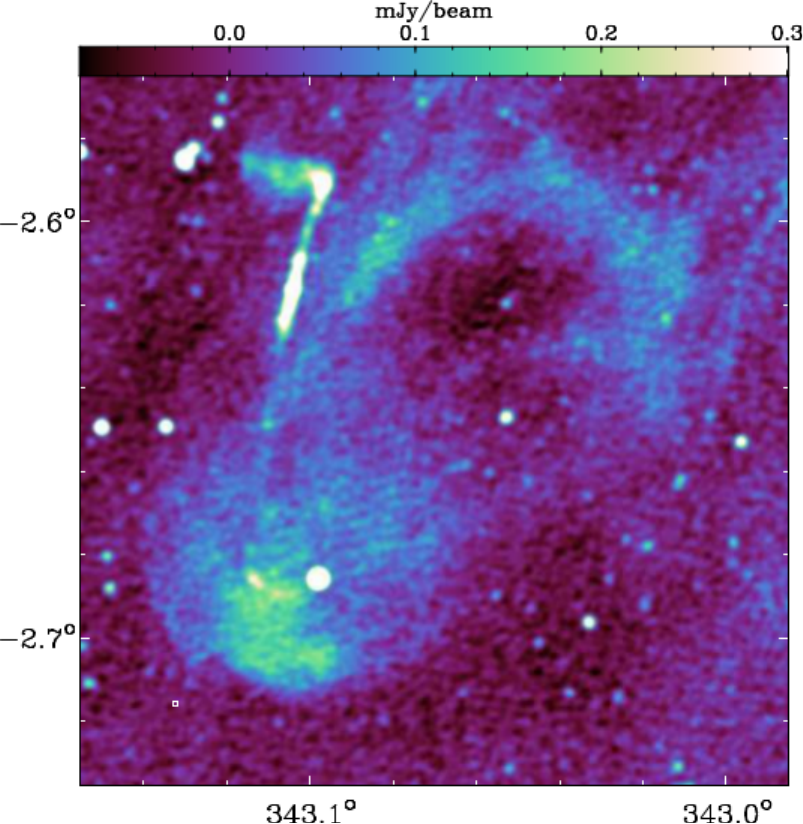}}
    \caption{Total power MeerKAT image of the PWN of SNR G343.1$-$2.3 at 1335~MHz in Galactic coordinates. The resolution of the image is 8\asec\ as indicated by the white circle in the lower left corner. The pulsar is seen as a bright point source in the cent\chg{er} of the PWN.\label{fig:Ig343pwn}}
\end{figure}
\begin{figure}[h]
    \centerline{\includegraphics[height=0.40\textwidth]{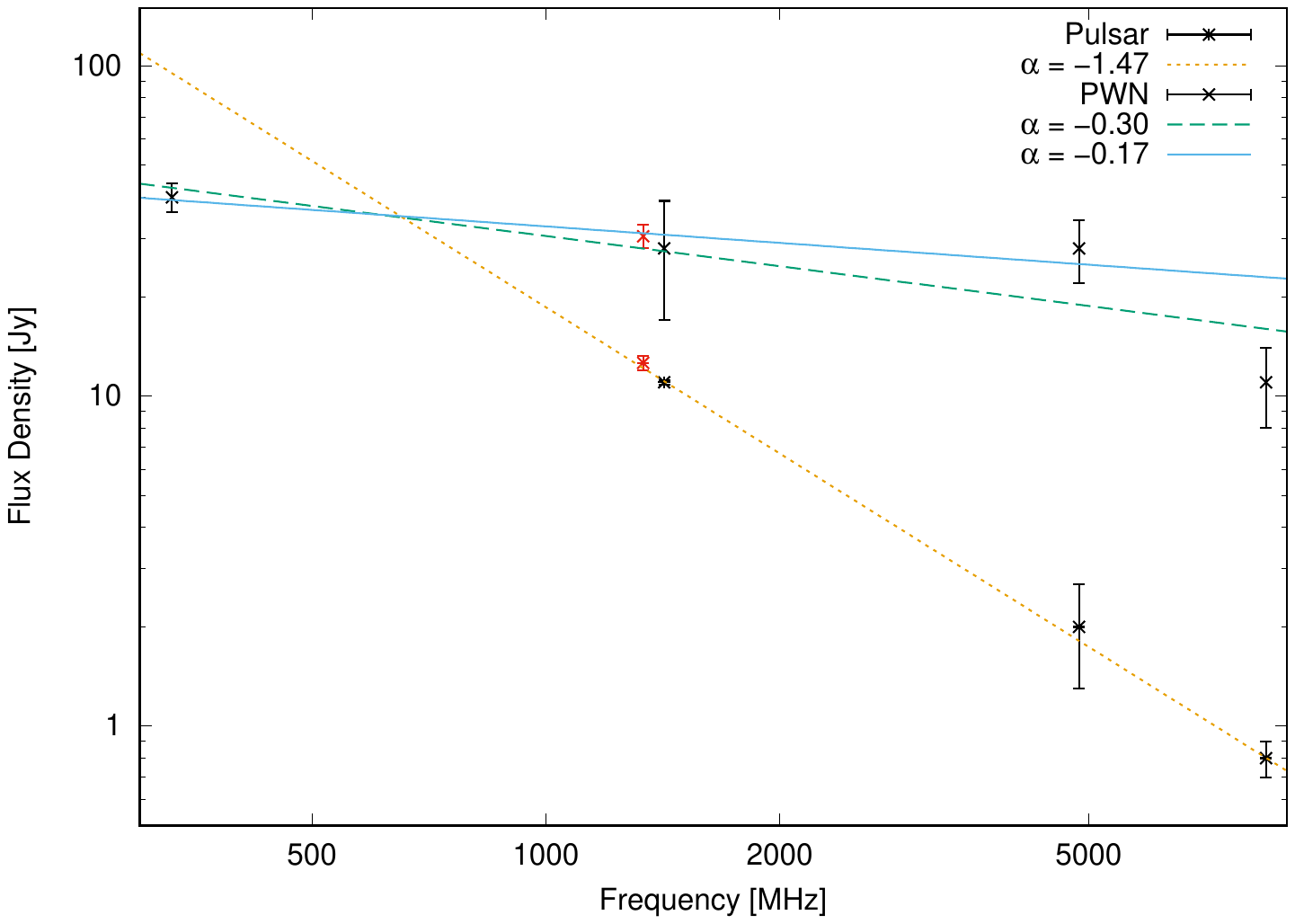}}
    \caption{Radio continuum spectrum of the PWN and its pulsar near SNR G343.1$-$2.3. Our flux density measurements are shown at 1335~MHz in red. Other flux density values were taken from \citet{Giacani2001}. \label{fig:g343spec}}
\end{figure}
Our new MeerKAT image of the SNR G343.1$-$2.3 at 1335~MHz with a resolution of 8\asec\ is shown in Figure~\ref{fig:Ig343.1} and a zoomed in image of the PWN around the pulsar B1706$-$44 is shown in Figure~\ref{fig:Ig343pwn}. We were not able to get a reliable flux density value for the SNR because of its low surface brightness, the background fluctuations, and we simply do not know where the SNR ends as there are so many filaments and shells overlapping in this area.

However, we determined the flux density of the pulsar and its wind nebula to be $S_{1335} = 43 \pm 5$~Jy. We plotted a spectrum of the pulsar and the nebula including flux values from the literature in Figure~\ref{fig:g343spec}. The resulting spectral index for the pulsar is $\alpha = -1.47 \pm 0.04$, which is a typical steep radio spectrum for a pulsar. The PWN's spectrum fit results in a spectral index of $\alpha = -0.30 \pm 0.08$, including all available flux values and $\alpha = -0.17 \pm 0.03$ if we leave out the highest frequency value, which seems to be below the others. This could also indicate a synchrotron cooling break between 5 and 8.5~GHz. This would warrant follow-up observations of the PWN at higher radio frequencies. The flatter spectral index is more typical of a PWN.

\subsubsection{G350.0$-$2.0}
\begin{figure}[h]
    \centerline{\includegraphics[width=0.50\textwidth]{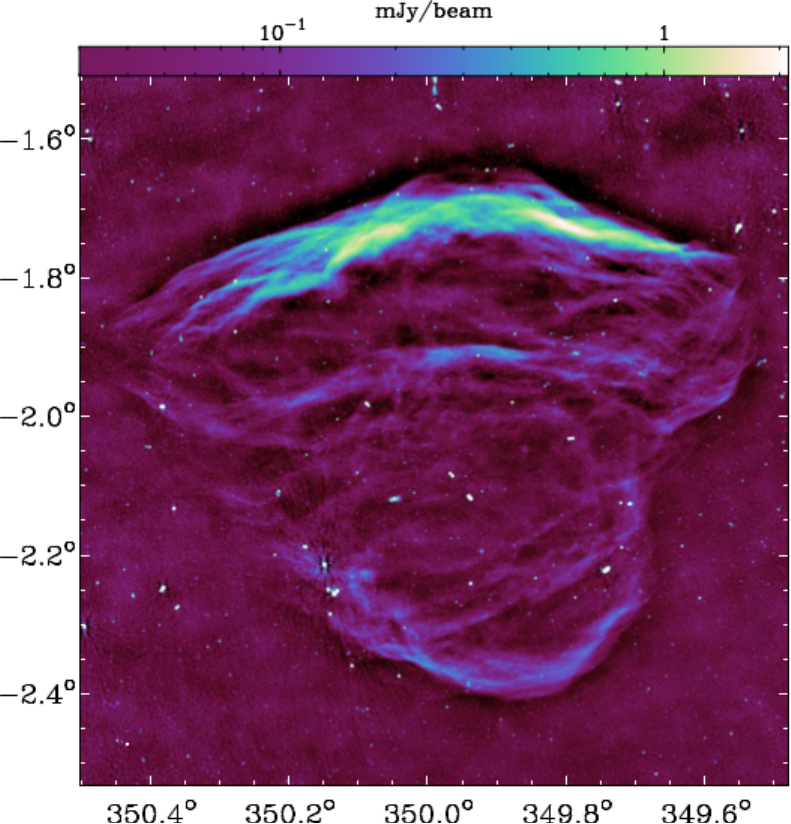}}
    \caption{Total power MeerKAT image of the SNR G350.0$-$2.0 at 1335~MHz in Galactic coordinates. The resolution of the image is 8\asec\ as indicated by the white circle in the lower left corner. \label{fig:Ig350.0}}
\end{figure}
G350.0$-$2.0 was first identified as a SNR by \citet{Clark1973} with a comparison of 408~MHz observations with the Molonglo radio telescope and 5000~MHz Parkes observations. It shows an overall spectral index of $\alpha = -0.51$ \citep{Clark1975}. Our new MeerKAT image of the SNR G350.0$-$2.0 at 1335~MHz with a resolution of 8\asec\ is shown in Figure~\ref{fig:Ig350.0}. It shows a clear bilateral structure with a symmetry axis almost parallel to the Galactic plane. It is bounded by a bright shell to the north and a fainter shell to the south. A third shell above the central region shows the same surface brightness as the southern shell and this lead some astronomers to assume that this could be two supernova remnants \citep{Gaensler1998}. But this third shell projected on to the interior can be easily explained by the 3 dimensional nature of these objects with a structured environment.

The integrated flux density from our data is $S_{1335} = 5.8 \pm 0.5$~Jy for the whole SNR at a diameter of about 50\amin. This is only about 20\% of the expected value from \citet{Clark1975}. Given the huge extent of this source there may actually be a lot of  flux in smooth emission that is not captured in our observations.

\subsubsection{G351.0$-$5.4}
\begin{figure}[h]
    \centerline{\includegraphics[width=0.50\textwidth]{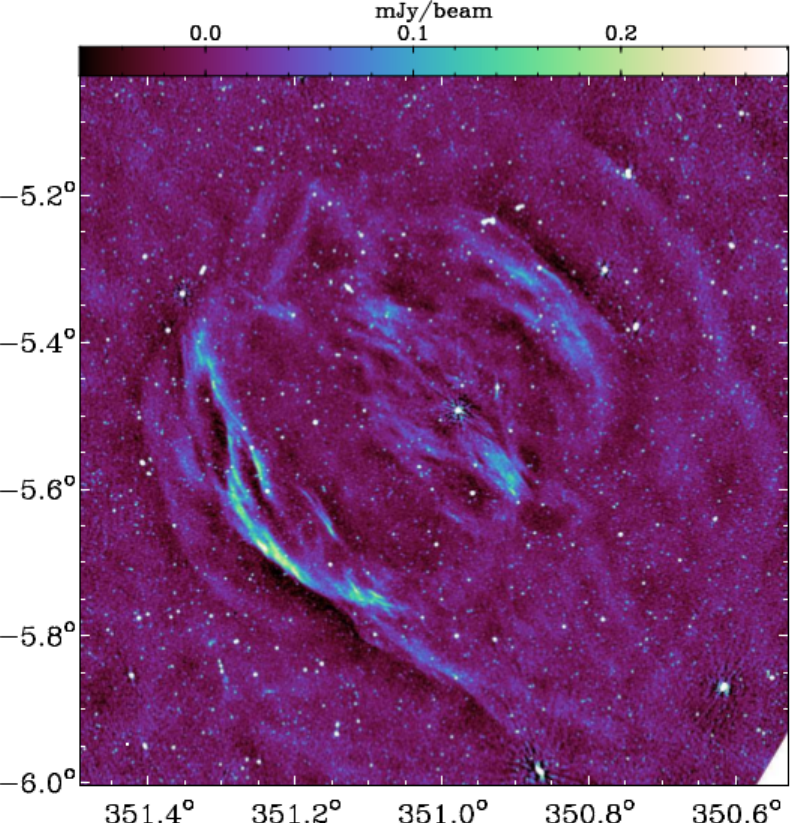}}
    \caption{Total power MeerKAT image of the SNR G351.0$-$5.4 at 1335~MHz in Galactic coordinates. The resolution of the image is 8\asec\ as indicated by the white circle in the lower left corner. \label{fig:Ig351.0}}
\end{figure}
G351.0$-$5.4 was first identified as a SNR candidate by \citet{deGasperin2014} with GMRT follow-up observations of diffuse radio emission discovered in the NVSS \citep{NVSS}. There have not been any later studies at any wavelength and no radio flux density is known. Our new MeerKAT image of this SNR at 1335~MHz with a resolution of 8\asec\ is shown in Figure~\ref{fig:Ig351.0}. It shows two opposing shells forming a bilateral SNR with some fainter filamentary emission coming from the center. Both shells have another very faint shell of similar curvature on their outside, very close for the south eastern shell and a bit further away for the north western shell. We were not able to determine an integrated flux density as the SNR surface brightness is very low and its full extent is not clear. The SNR has a diameter of about 50\amin, however some emission seems to disappear towards the edge due to lower sensitivity.

\subsubsection{G353.9$-$2.0}
\begin{figure}[h]
    \centerline{\includegraphics[width=0.50\textwidth]{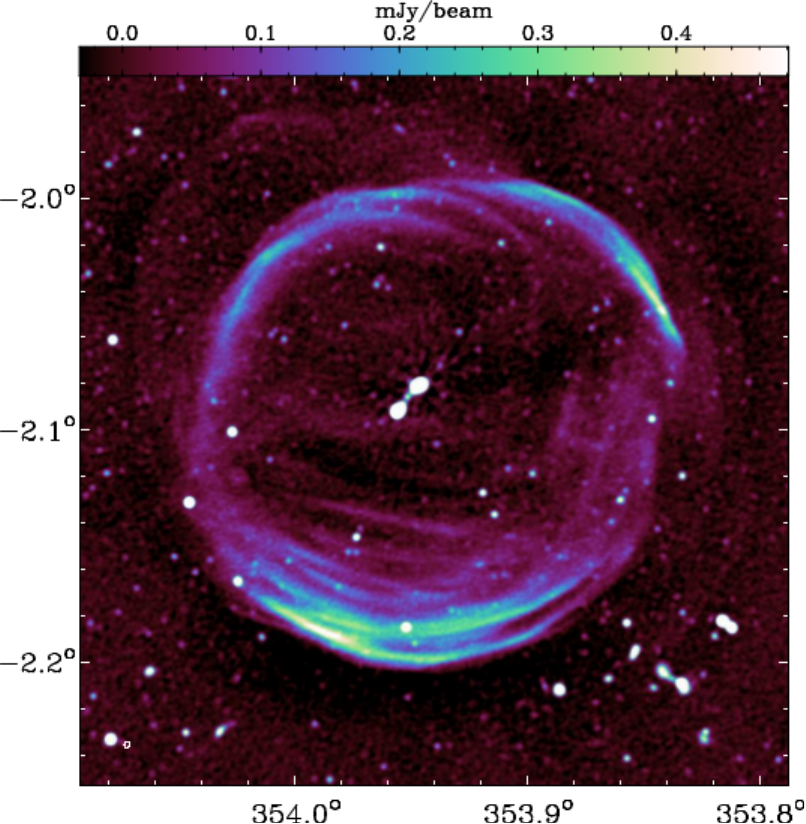}}
    \caption{Total power MeerKAT image of the SNR G353.9$-$2.0 at 1335~MHz in Galactic coordinates. The resolution of the image is 8\asec\ as indicated by the white circle in the lower left corner. \label{fig:Ig353.9}}
\end{figure}
G353.9$-$2.0 was first identified as a SNR candidate by \citet{Green2001} with VLA follow-up observations of shell-like emission found in the NVSS \citep{NVSS}. They found a spectral index of $\alpha = -0.55 \pm 0.15$ assuming that their VLA P-band observation captures all spatial scales and comparing it with a flux density determined from the 2.4-GHz survey of \citet{Duncan1995} with the Parkes telescope. 

 Our new MeerKAT image of G353.9$-$2.0 at 1335~MHz with a resolution of 8\asec\ is shown in Figure~\ref{fig:Ig353.9}. This SNR is a beautiful circular bilateral SNR with the symmetry axis almost parallel to the Galactic plane. Both opposing shells show filamentary substructure and their ends are connected by faint smooth shells. This could indicate that this SNR is in the transition from free expansion to adiabatic expansion. No emission is coming from the interior. But there are large faint protrusions on the outside of the SNR to the north-east, the west and the south-west. 

 The integrated flux density from our data is $S_{1335} = 430 \pm 40$~mJy for the whole SNR at a diameter of about 13\amin. This is almost identical to the flux density determined by \citet{Green2001} from their 1420 MHz observations with the VLA.
 
\subsubsection{G355.9$-$2.5}
\begin{figure}[h]
    \centerline{\includegraphics[width=0.50\textwidth]{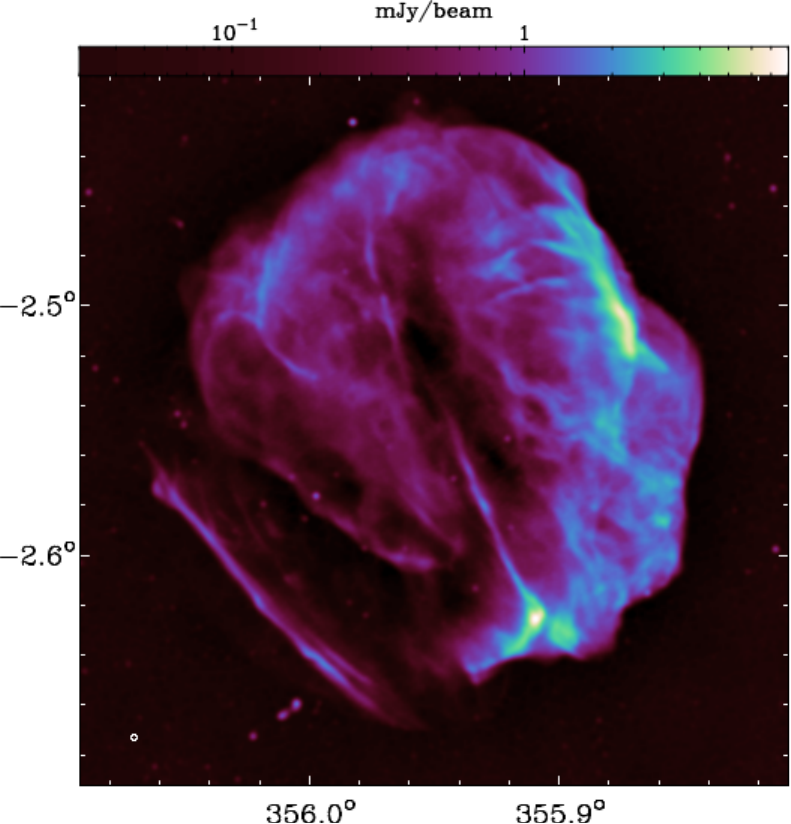}}
    \caption{Total power MeerKAT image of the SNR G355.9$-$2.5 at 1335~MHz in Galactic coordinates. The resolution of the image is 8\asec\ as indicated by the white circle in the lower left corner. \label{fig:Ig355.9}}
\end{figure}
G355.9$-$2.5 was first identified as a SNR by \citet{Clark1973} with a comparison of 408~MHz observations with the Molonglo radio telescope and 5000~MHz Parkes observations. It shows an overall spectral index of $\alpha = -0.51$ \citep{Clark1975}. Our new MeerKAT image of the SNR G355.9$-$2.5 at 1335~MHz with a resolution of 8\asec\ is shown in Figure~\ref{fig:Ig355.9}. This SNR looks quite peculiar. It shows unusual narrow linear features to the south-east and in the bottom central area. The main almost circular part looks like the rib X-ray of a human being, with several filaments of the same curvature. It is very difficult to explain this peculiar shape of the SNR.
\begin{figure}[h]
    \centerline{\includegraphics[height=0.40\textwidth]{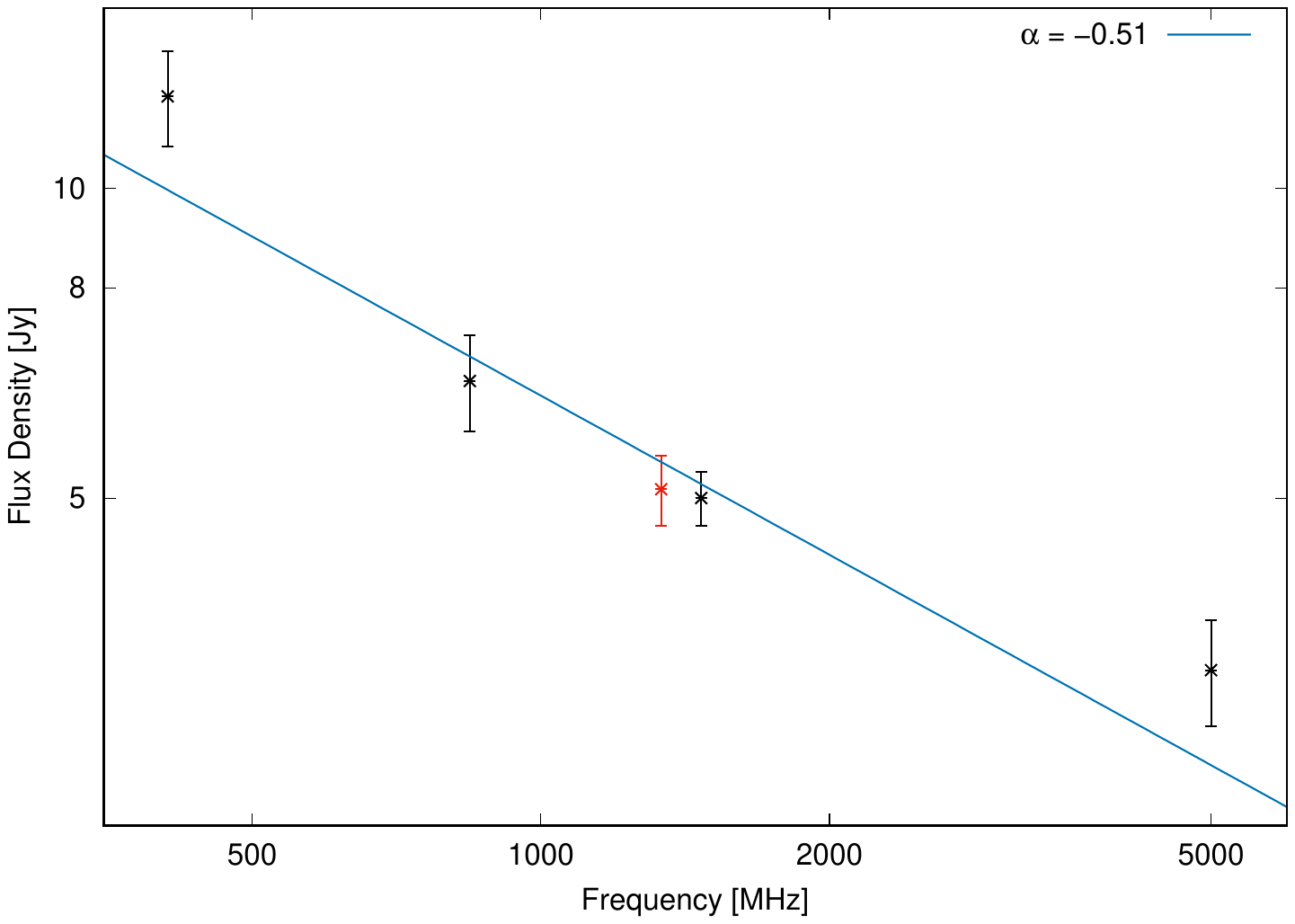}}
    \caption{Radio continuum spectrum of the SNR G355.9$-$2.5. Our flux density measurement is shown at 1335~MHz in red. Other flux density values were taken from \citet{Clark1973}, \citet{Dubner1993}, and \citet{Gray1994} and are shown in black. \label{fig:g355spec}}
\end{figure}
The integrated flux density from our data is $S_{1335} = 5.1 \pm 0.4$~Jy for the whole SNR at a diameter of about 14\amin. A combined spectrum including archival flux densities is shown in Figure~\ref{fig:g355spec}. The resulting spectral index is $\alpha = -0.51 \pm 0.11$. 

\subsubsection{G356.2$+$4.5}
\begin{figure}[h]
    \centerline{\includegraphics[width=0.50\textwidth]{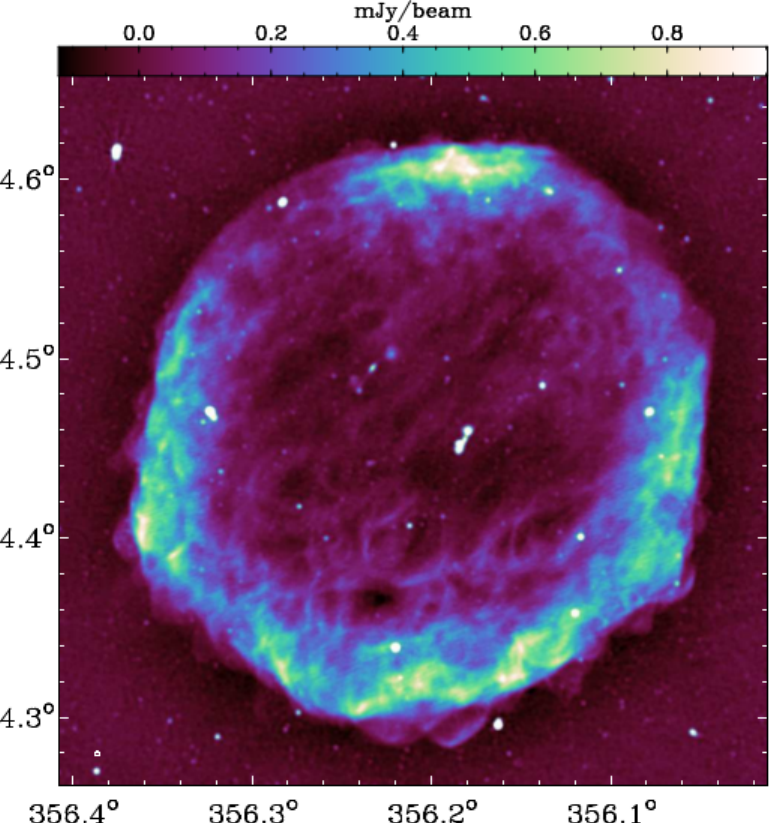}}
    \caption{Total power MeerKAT image of the SNR G356.2$+$4.5 at 1335~MHz in Galactic coordinates. The resolution of the image is 8\asec\ as indicated by the white circle in the lower left corner. \label{fig:Ig356.2}}
\end{figure}
G356.2$+$4.5 was first listed as a SNR candidate in a Parkes 2.4~GHz southern Galactic plane survey by \citet{Duncan1995}. Follow-up observations by \citet{Bhatnagar2000} with the GMRT confirmed this source as a SNR. They found an overall spectral index of $\alpha = -0.66$. Our new MeerKAT image at 1335~MHz with a resolution of 8\asec\ is shown in Figure~\ref{fig:Ig356.2}. This SNR shows somewhat \chg{of} a bilateral structure, however there is emission from an unusually thick shell almost all around. This shell does not display the typical limb-brightening, but looks more chaotic or turbulent, maybe indicating a very young SNR.

\begin{figure}[h]
    \centerline{\includegraphics[height=0.40\textwidth]{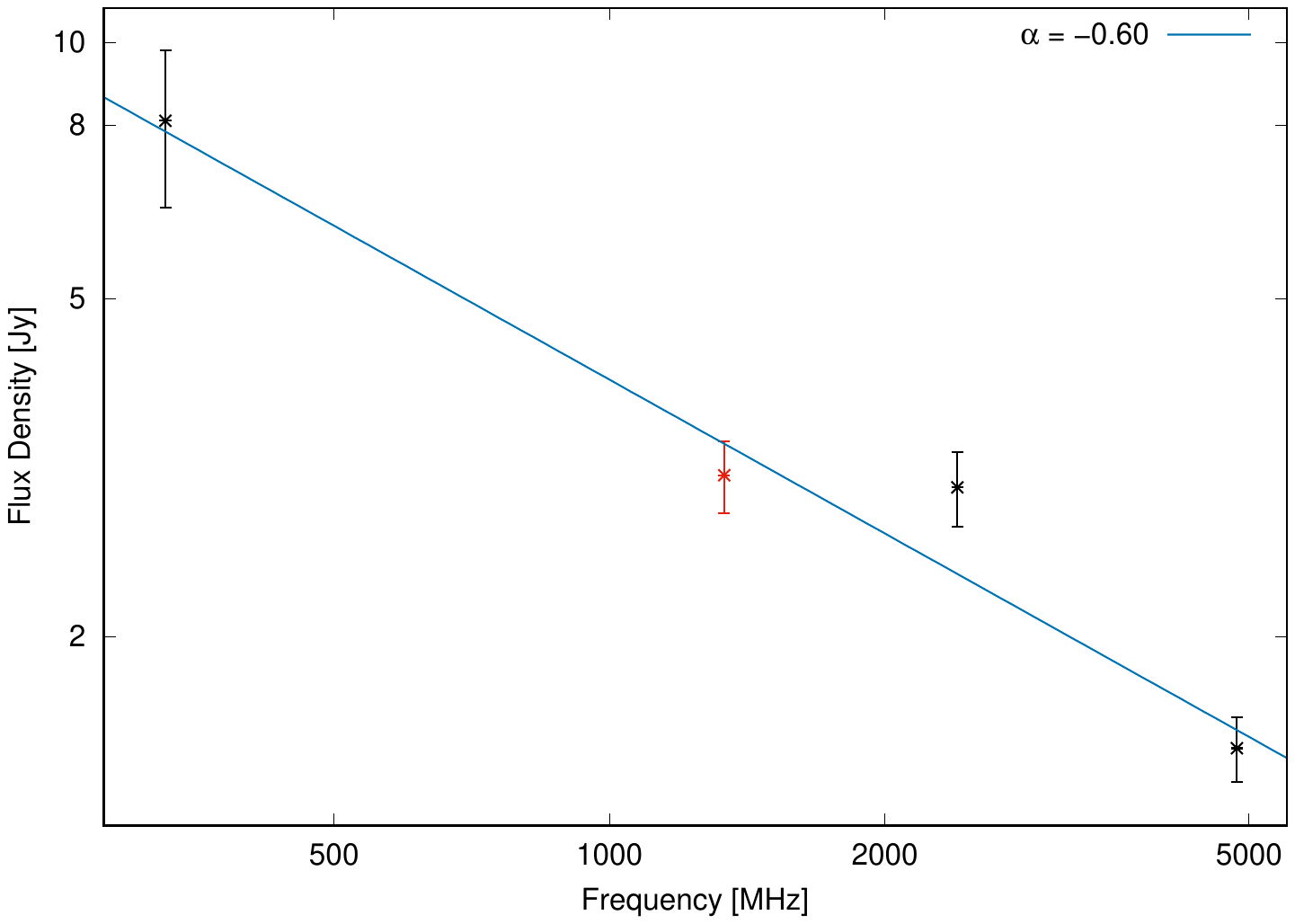}}
    \caption{Radio continuum spectrum of the SNR G356.2$+$4.5. Our flux density measurement is shown at 1335~MHz in red. Other flux density values were taken from \citet{Duncan1995} and \citet{Bhatnagar2000} and are shown in black. \label{fig:g356spec}}
\end{figure}
The integrated flux density from our data is $S_{1335} = 3.1 \pm 0.3$~Jy for the whole SNR at a diameter of about 20\amin. A combined spectrum, including archival flux densities, is shown in Figure~\ref{fig:g356spec}. The resulting spectral index is $\alpha = -0.60 \pm 0.12$. 

\subsubsection{G358.0$+$3.8}
\begin{figure}[h]
    \centerline{\includegraphics[width=0.50\textwidth]{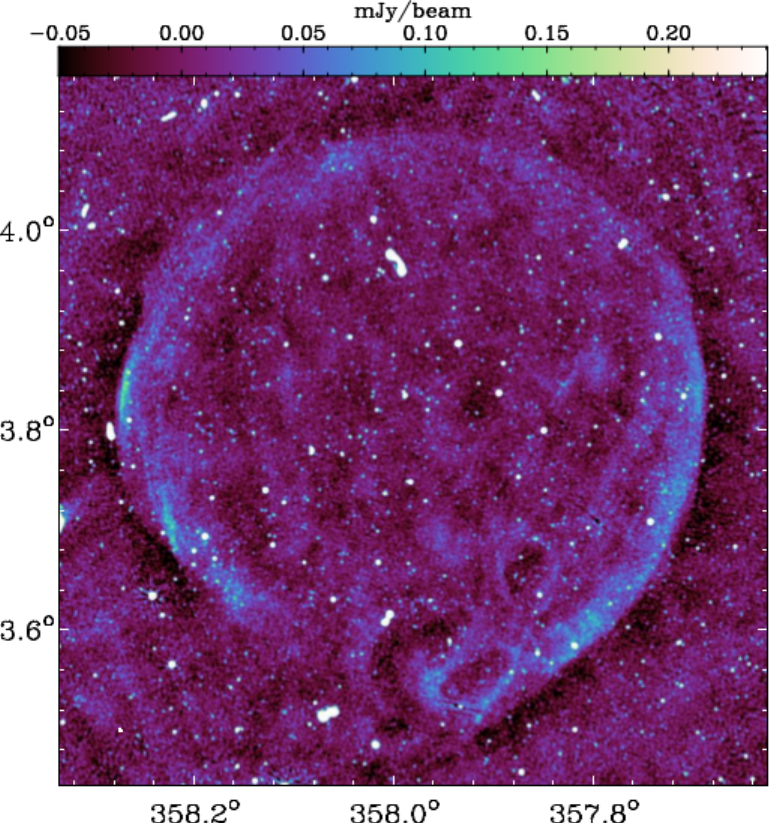}}
    \caption{Total power MeerKAT image of the SNR G358.0$+$3.8 at 1335~MHz in Galactic coordinates. The resolution of the image is 8\asec\ as indicated by the white circle in the lower left corner. \label{fig:Ig358.0}}
\end{figure}
G358.0$+$3.8 was first listed as a SNR candidate in a Parkes 2.4~GHz southern Galactic plane survey by \citet{Duncan1995}. Follow-up observations by \citet{Bhatnagar2000} with the GMRT confirmed this source as a SNR. But this SNR is so faint that no reliable flux densities could be determined. Our new MeerKAT image at 1335~MHz with a resolution of 8\asec\ is shown in Figure~\ref{fig:Ig358.0}. This SNR has the typical bilateral structure with an almost vertical symmetry axis. Besides the well-defined shells to the west and east, not much is visible.

The integrated flux density from our data is $S_{1335} = 0.8 \pm 0.15$~Jy. G358.0$+$3.8 is almost circular with a diameter of about 36\amin.

\subsection{Detailed Polarimetry\label{HiResPol}} 
For a selected subset of the SNRs, higher frequency resolution imaging was done in linear polarization; 1\% maximum fractional bandwidth giving 68 channels across the \chg{bandpass}. 
This gives good sensitivity to Faraday depth of well over $\pm$1000 rad m$^{-2}$, see \citet{Obit78}. 
A joint Q/U CLEAN deconvolution was used.
A search in Faraday depth similar to that in \cite{XGalaxy} was performed covering the range $\pm$400 rad m$^{-2}$, except as noted.
This process derives unwrapped polarized flux densities, Faraday depth of the peak and the polarization angle (EVPA) at wavelength $\lambda$=0. 
For optically thin synchrotron sources, the polarization ``B'' vectors give the projected orientation of the magnetic field.

\subsubsection{G315.4$-$2.3}
\begin{figure*}
\centerline{
  \includegraphics[width=3.0in]{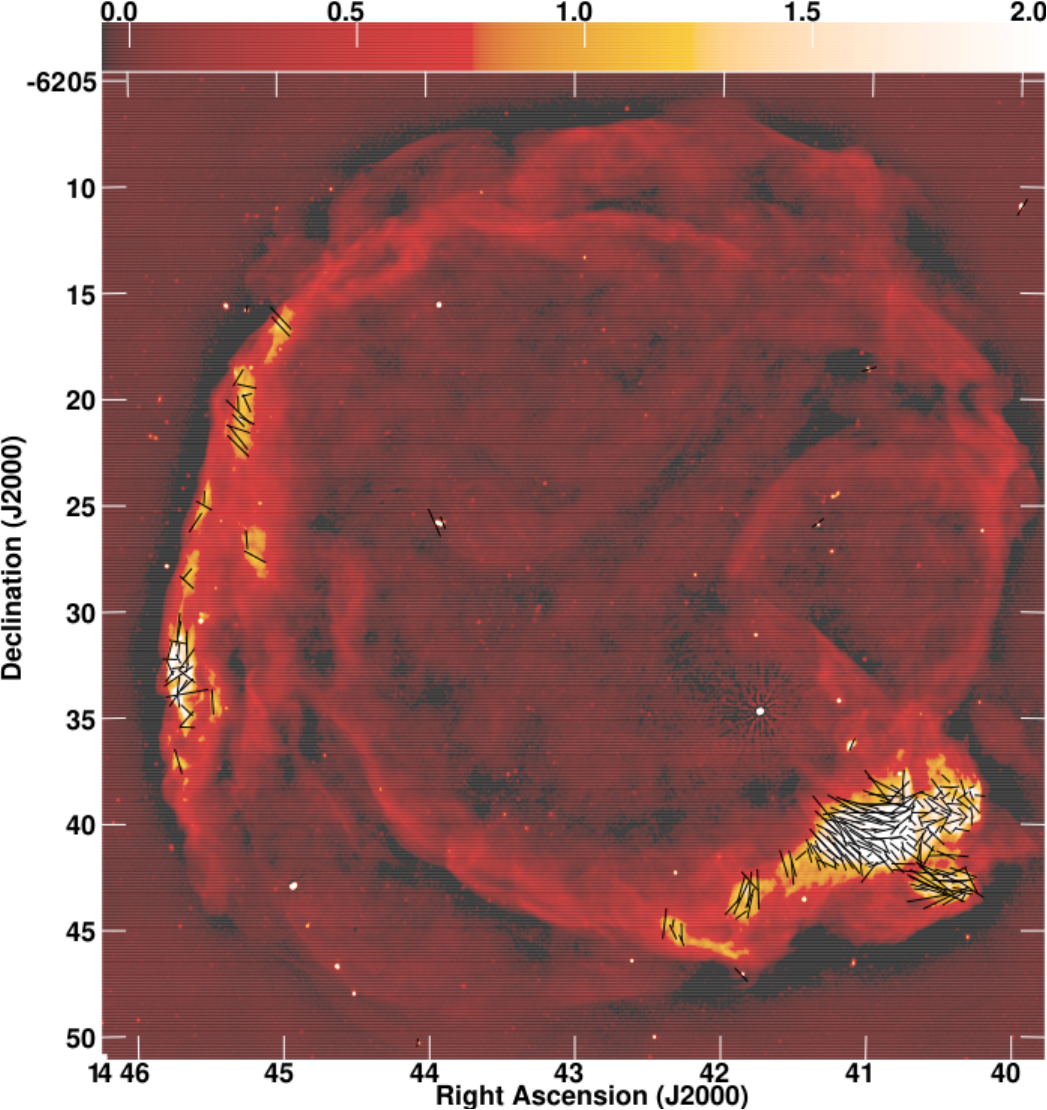}
  \includegraphics[width=3.0in]{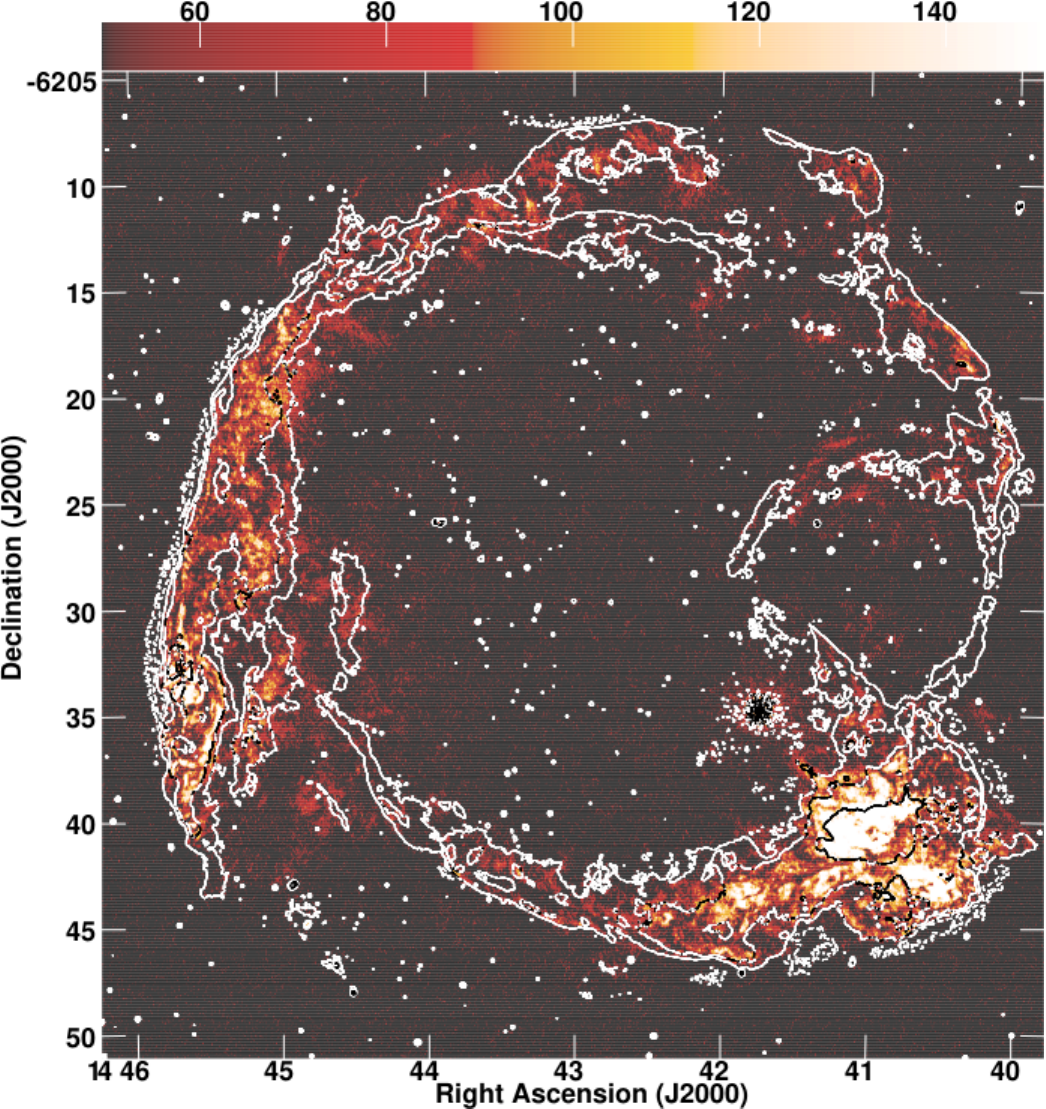}
}
\centerline{
  \includegraphics[width=3.0in]{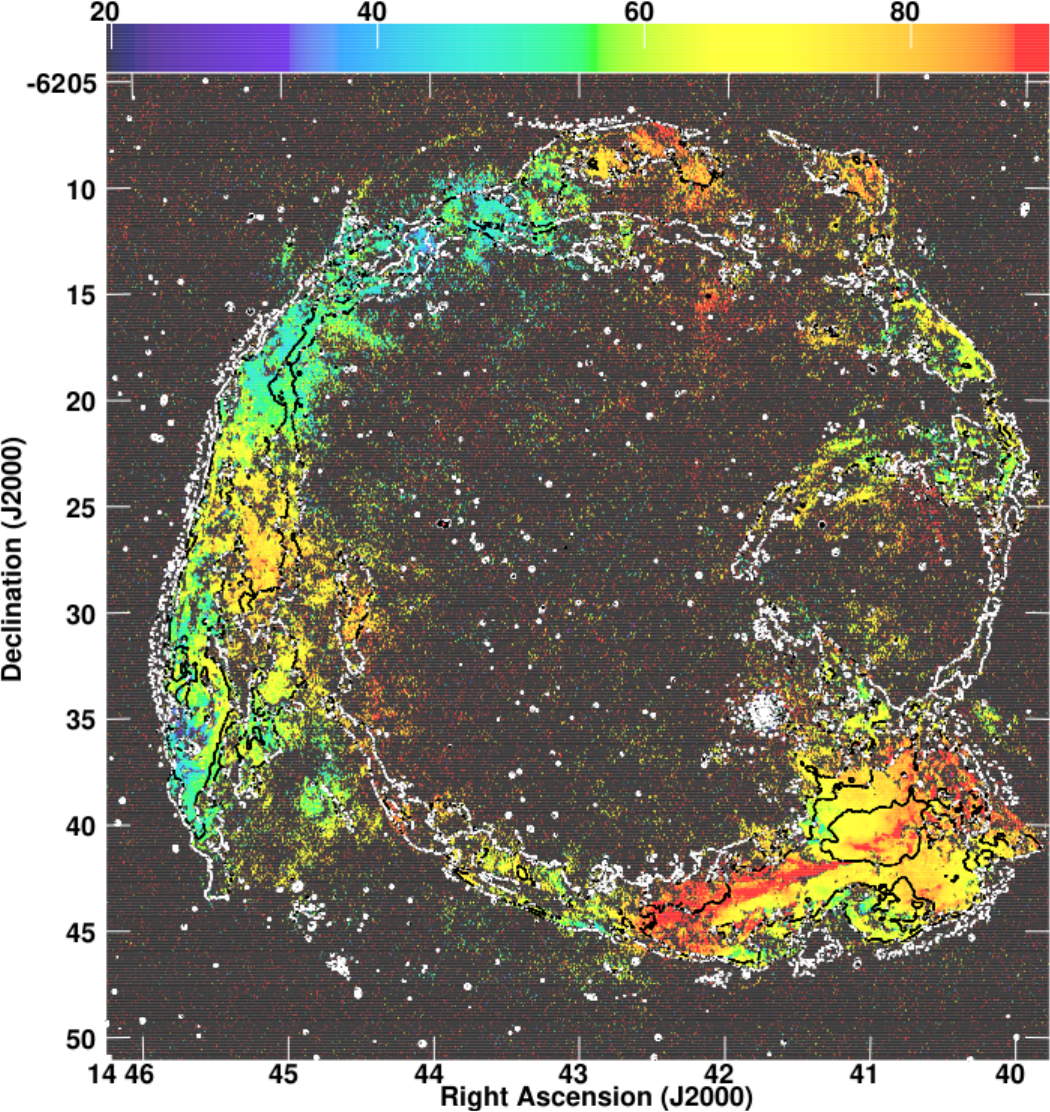}
}
\caption{\chg{G315.4$-$2.3}. \hfill\break
{\bf Top left}: Bias corrected fractional polarization B vectors on heat image of Stokes I; scale bar labeled in mJy/beam at top.
\chg{ The restoring beam is 7.8\asec $\times$ 7.4\asec at position angle = 13.9$^\circ$.}\hfill\break
{\bf Top right}: Bias corrected polarized intensity as heat image with scale bar labeled in $\mu$Jy/beam on top; Stokes I contours at $-$0.1, 0.1, 0.4, 1.6, 6.4 and 25.6 mJy/beam.\hfill\break
{\bf Bottom}: Peak Faraday depth in color given by scale bar at the top with Stokes I contours as shown in {\bf Top right}.
}
\label{fig:G315.4Poln}
\end{figure*}

Both the total intensity and polarized emission from G315.4$-$2.3 are dominated by the edge of the remnant as shown in Figure \ref{fig:G315.4Poln}.  Variations in Faraday rotation are relatively small and smooth across the remnant.

\subsubsection{G326.3$-$1.8}
\begin{figure*}
\centerline{
  \includegraphics[width=3.0in]{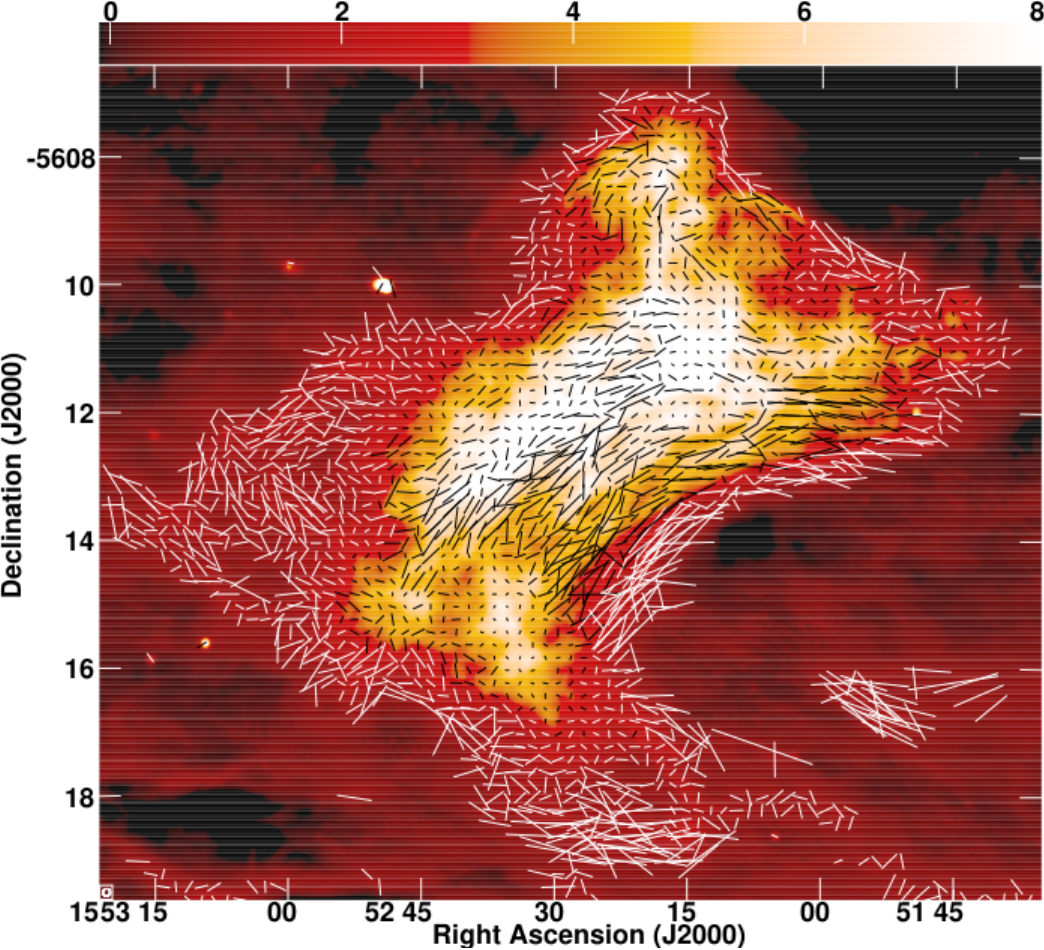}
  \includegraphics[width=3.0in]{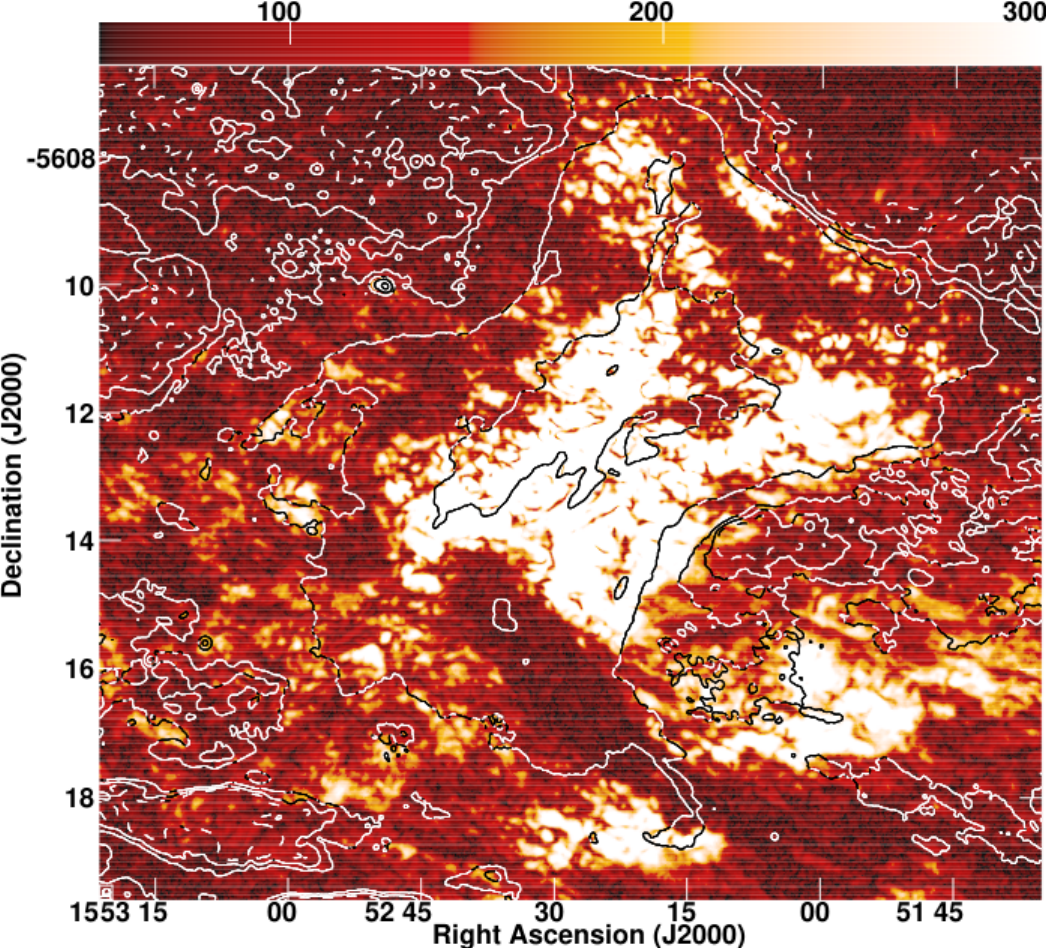}
}
\centerline{
  \includegraphics[width=3.0in]{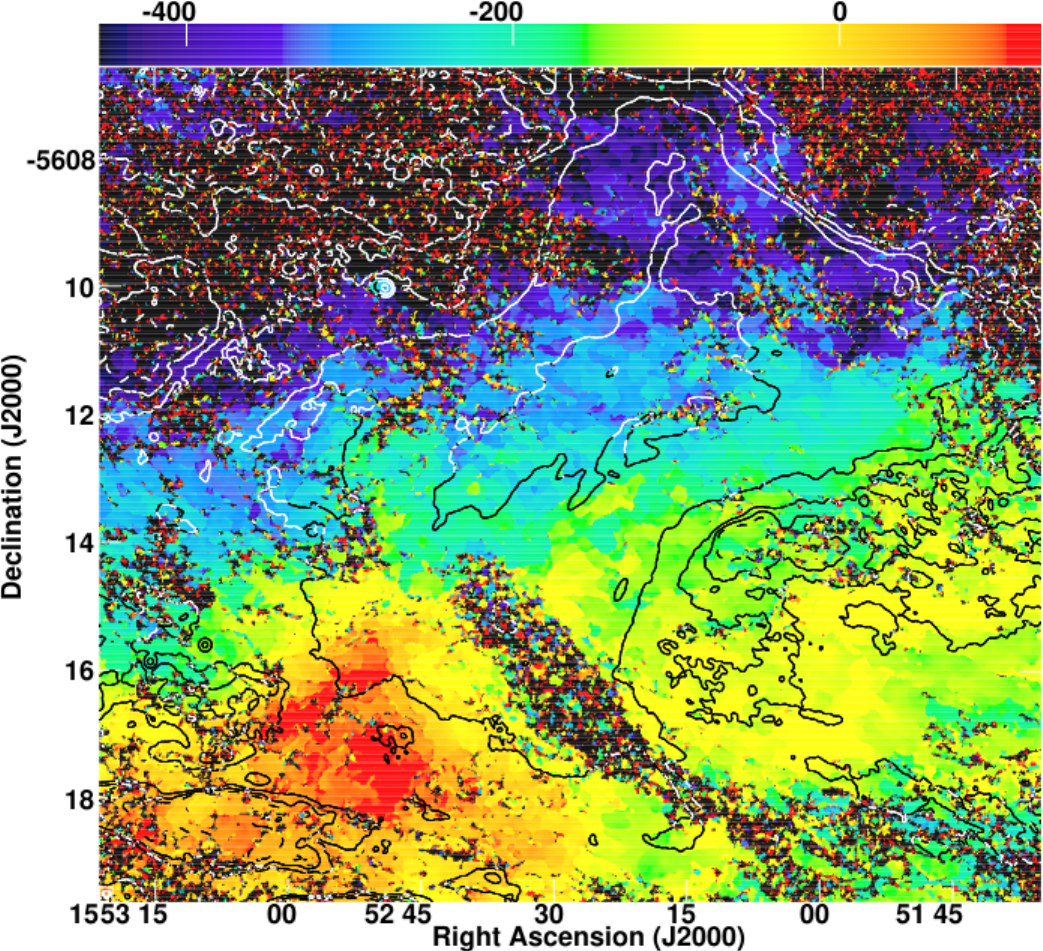}
}
\caption{\chg{Like Figure \ref{fig:G315.4Poln} but} G326.3$-$1.8 pulsar wind nebula.
\chg{The restoring beam is 7.3\asec $\times$ 7.1\asec at position angle = 5.0$^\circ$.}
}
\label{fig:G326.3Poln}
\end{figure*}

Both the total intensity and polarized emission from G326.3$-$1.8 is dominated by the pulsar wind nebula (PWN) which is shown in Figure \ref{fig:G326.3Poln}.
The Faraday depth search range was $\pm$1500 rad m$^{-2}$.
The orientation of the B vectors is predominantly along the main axis of the PWN but with considerable variation.
There  is a banana shaped feature to the south seen in the polarized intensity and RM images in which the emission is strongly depolarized.
This depolarization feature is superposed on a ridge of total intensity emission.  
There is also a strong RM gradient near this feature, +120 to $-450$ rad m$^{-2}$. 

\subsubsection{G327.6+14.6 (SN1006)}
X-ray imaging polarimetry of the north east region of this SNR is reported in \cite{Zhou2023}. The X-ray polarization shows that the magnetic field is predominantly radial with an average fractional polarization of 22 $\pm$ 4 \%.

\begin{figure*}
\centerline{
  \includegraphics[width=3.0in]{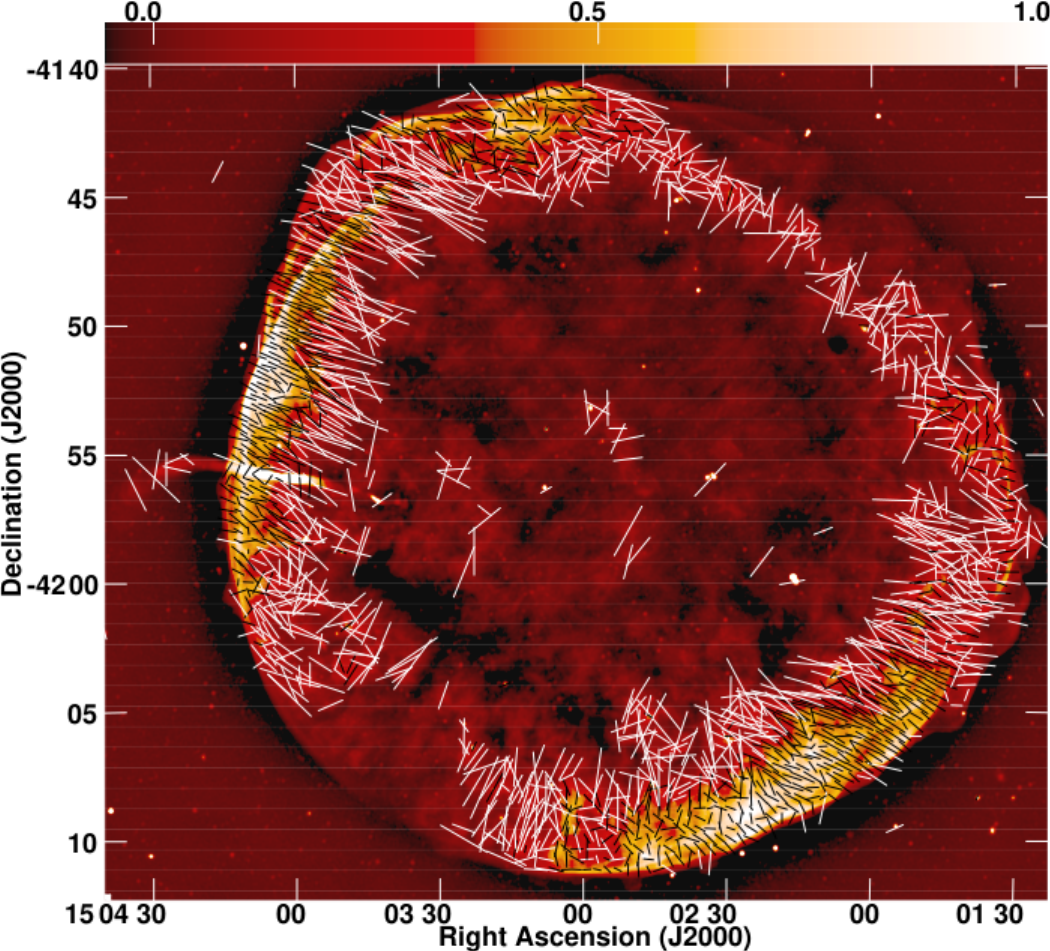}
  \includegraphics[width=3.0in]{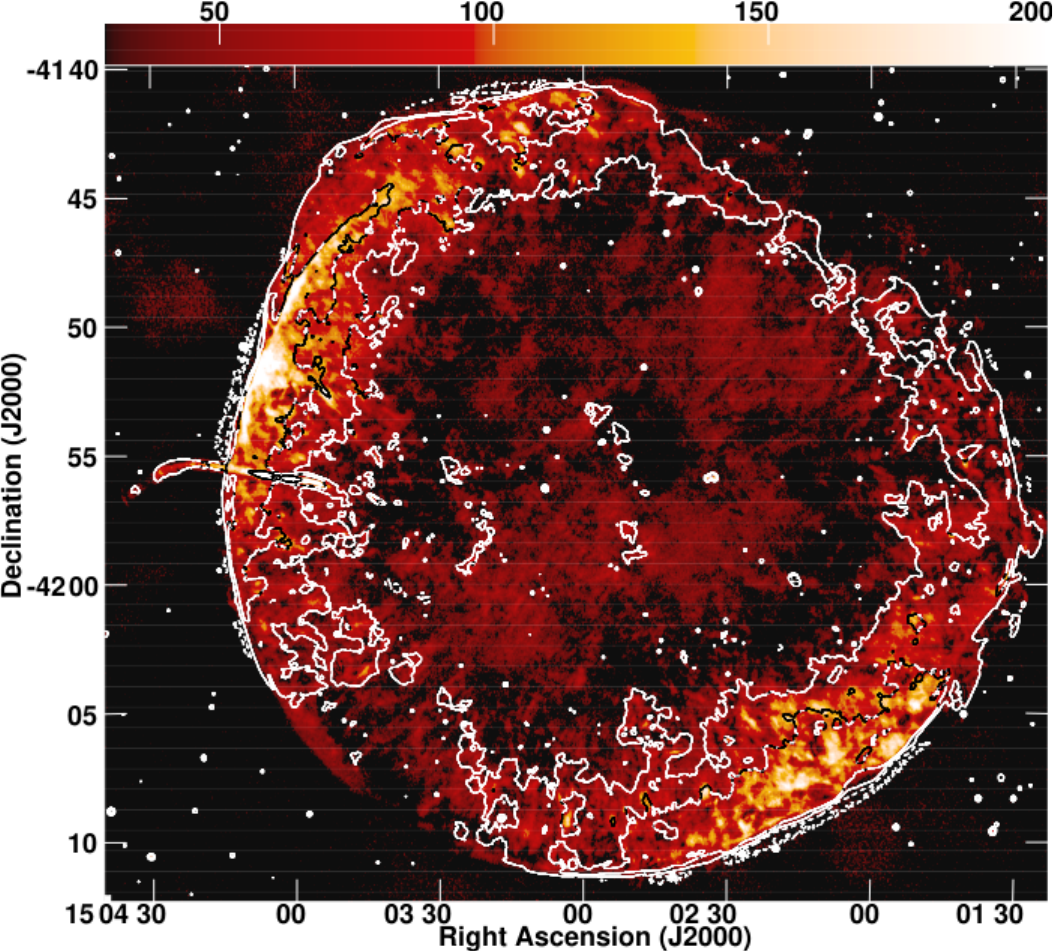}
}
\centerline{
  \includegraphics[width=3.0in]{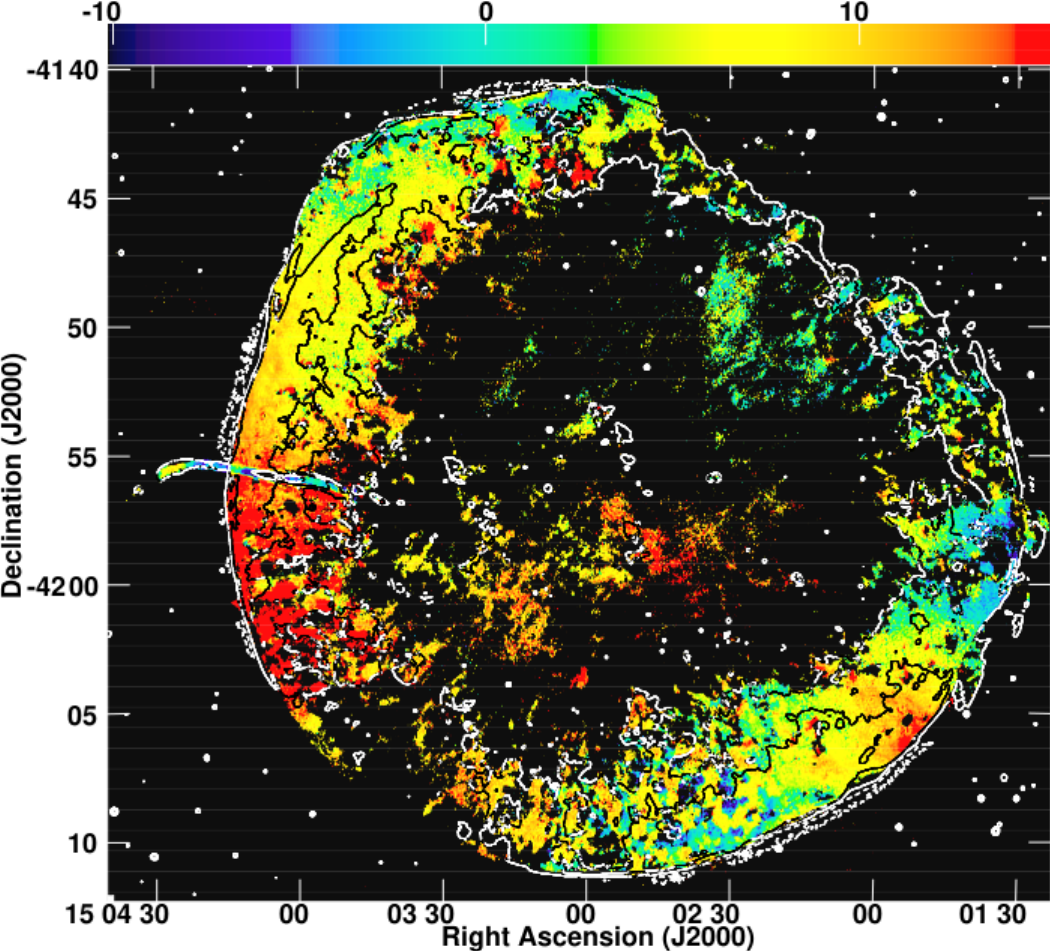}
}
\caption{Like Figure \ref{fig:G315.4Poln} but showing G327.6+14.6.
\chg{ The restoring beam is 7.4\asec $\times$ 7.1\asec at position angle = 7.9$^\circ$.}
}
\label{fig:G327.6Poln}
\end{figure*}

\begin{figure}[h]
\centerline{
  \includegraphics[width=3.5in]{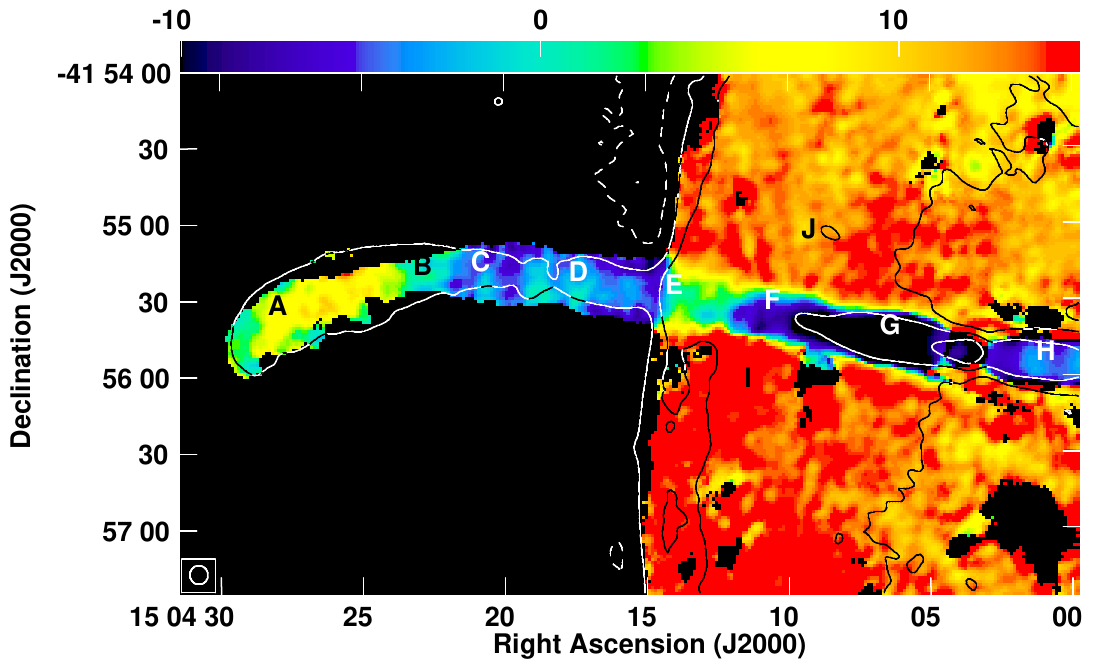}
}
\caption{G327.6+14.6: close up of section of Figure \ref{fig:G327.6Poln}, lower
  panel marking locations of RM spectra.  The resolution is given by the ellipse in the box in the lower left corner.}
\label{fig:G327.6+14.6_Close_RM}
\end{figure}

The polarized radio emission in G327.6+14.6 is also dominated by radial B vectors indicating it is still in the free expansion phase; see Figure \ref{fig:G327.6Poln}.
Resolution of the extended emission of the remnant reduces the Stokes I flux density so that a reliable estimate of the radio fractional polarization of the remnant is not possible.
A background FRI AGN crosses the limb of the remnant.
This AGN is WISEA J150403.60-415550.9 at z=0.039207 \citep{2012ApJS..199...26H} and a closeup of the peak Faraday depth around the AGN is given in Figure \ref{fig:G327.6+14.6_Close_RM}.

\begin{figure*}
\centerline{
  \includegraphics[width=6.3in]{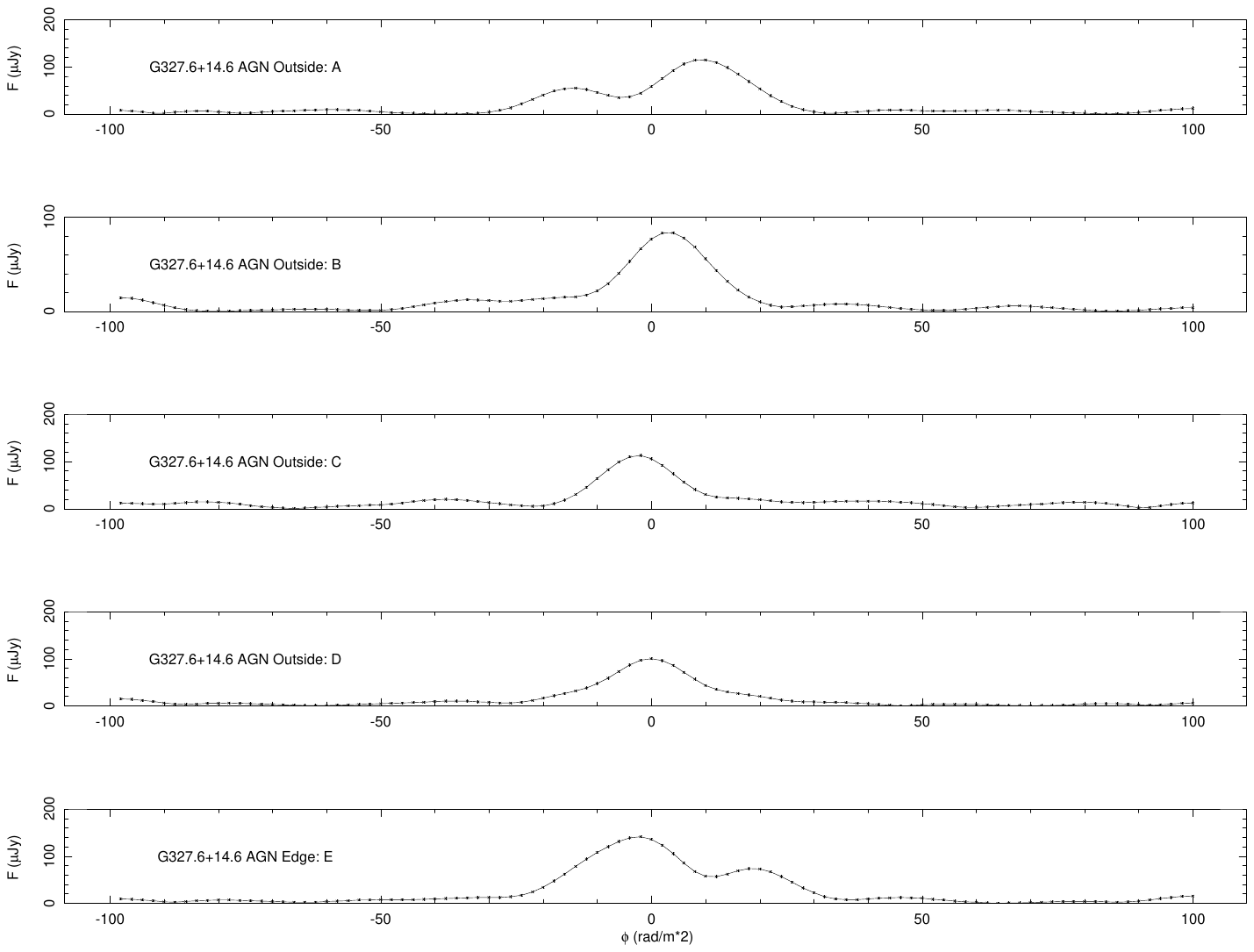}
}
\centerline{
  \includegraphics[width=6.3in]{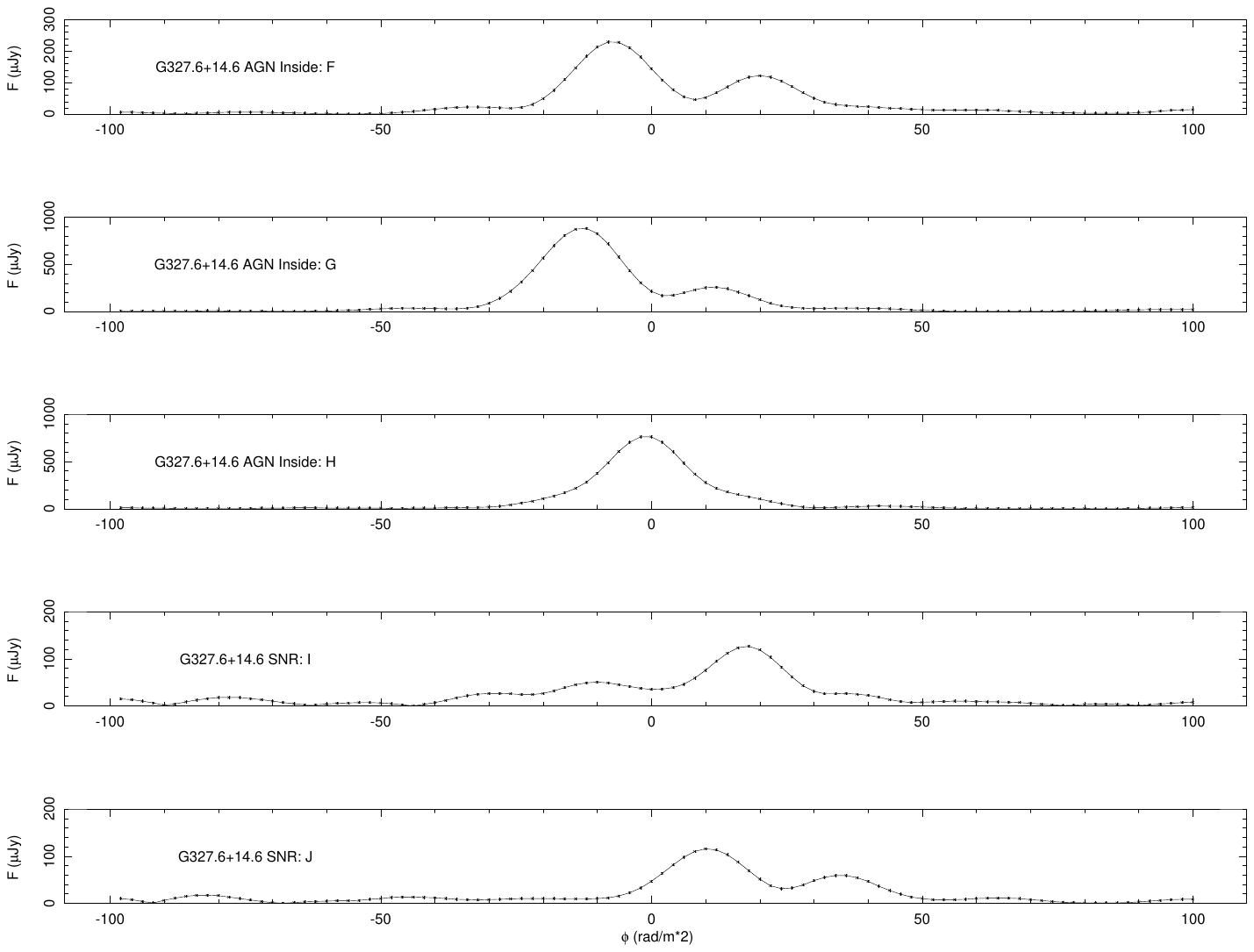}
}
\caption{G327.6+14.6: Faraday spectra of selected locations indicated in Figure
  \ref{fig:G327.6+14.6_Close_RM}.} 
\label{fig:G327.6+14.6_RMSpectra}
\end{figure*}

The  background AGN can be  a probe of the Faraday depth and depolarization through the remnant.
Since sightlines to this background AGN pass through intermittent layers of polarized emission and Faraday rotating magnetized plasma, the Faraday spectra are expected to contain multiple components showing the difference in Faraday rotation between the polarized emission layers.
A Faraday synthesis with complex CLEAN deconvolution following \cite{Rudnick2023} was performed over the range of $\pm$100 rad m$^{-2}$ and the restoring function had a Gaussian FWHM of 15 rad m$^{-2}$.
Faraday spectra at locations indicated in Figure \ref{fig:G327.6+14.6_Close_RM} are displayed in Figure \ref{fig:G327.6+14.6_RMSpectra}.
Spectra of sightlines toward the AGN jet passing through the SNR and close enough to the limb that the polarized emission from the SNR is detected (sightlines E-G) show two components.
The first appears with a peak between 0 and $-12$ rad m$^{-2}$ and is the polarized emission from the AGN and the second, weaker component, appearing between +12 and +20 rad m$^{-2}$  is the polarized emission from the SNR.

\begin{figure}
    \centerline{
    \includegraphics[width=3.5in]{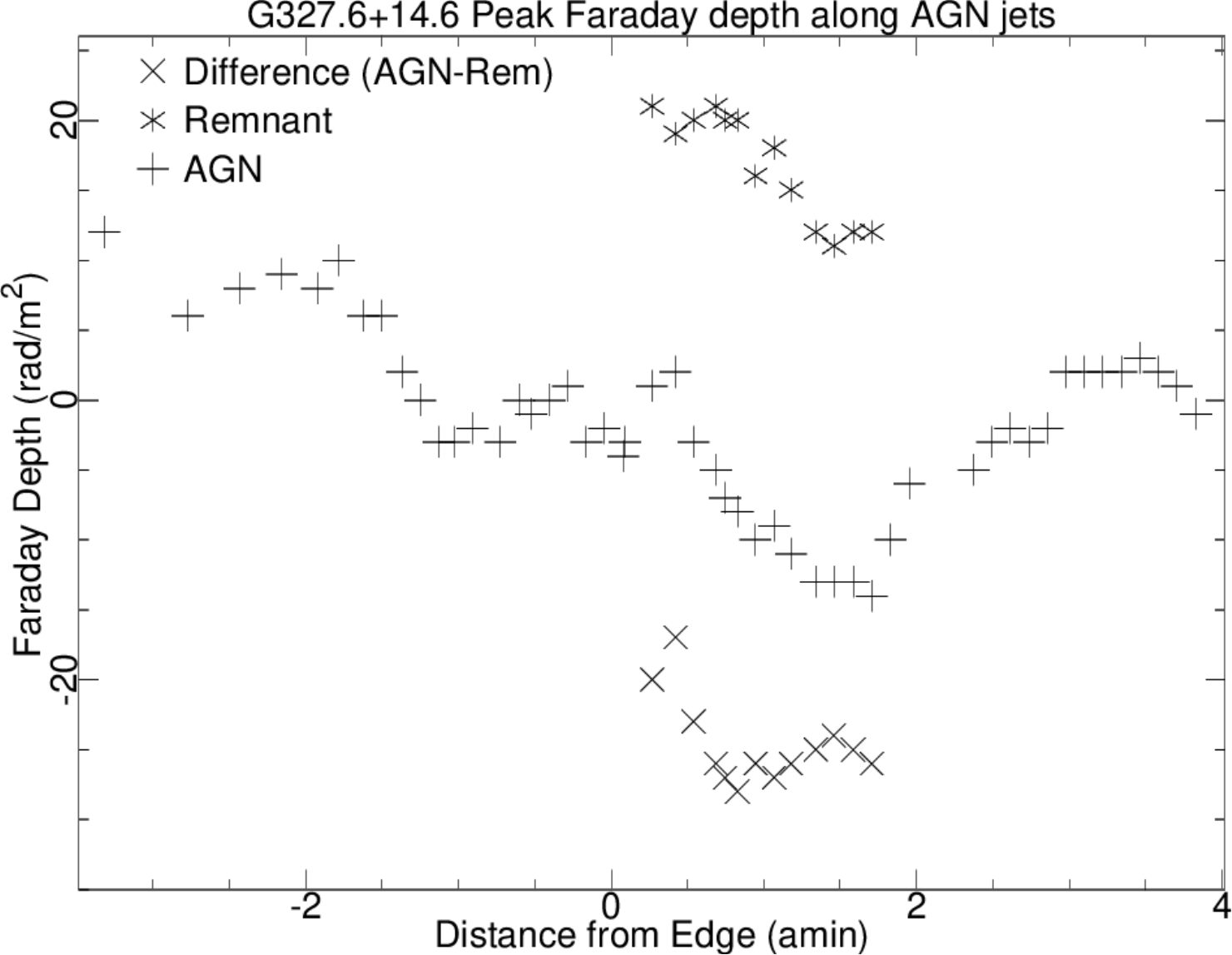}
    }
    \caption{G327.6+14.6: Peak Faraday depths of AGN and remnant polarized emission and difference at locations along the large background FRI AGN.  The AGN component is shown by pluses (``+"), the remnant emission by ``*" and the difference (AGN-Remnant) by ``x".}
    \label{fig:G327.6+14.6_AGN_Rem_RM}
\end{figure}

Figure \ref{fig:G327.6+14.6_RMSpectra} shows that the Faraday depth varies along the length of the jet for both the background AGN and the remnant emission; this could be due to some combination of the Faraday depth through the Galaxy plus some additional contribution from the interior of the remnant.  Assuming that the bulk of the Galactic variation is in material between us and the remnant, the difference between the peak Faraday depths of the AGN and remnant on the same sightline should remove this variation.  If the interior of the remnant is uniformly filled with a Faraday rotating gas, the difference between the AGN and remnant peak Faraday depths should increase with the length of the line of sight through the remnant. Figure \ref{fig:G327.6+14.6_AGN_Rem_RM} shows the various estimates of the peak Faraday depth and their difference.  

\begin{figure}[h]
    \centering
    \includegraphics[width=3.5in]{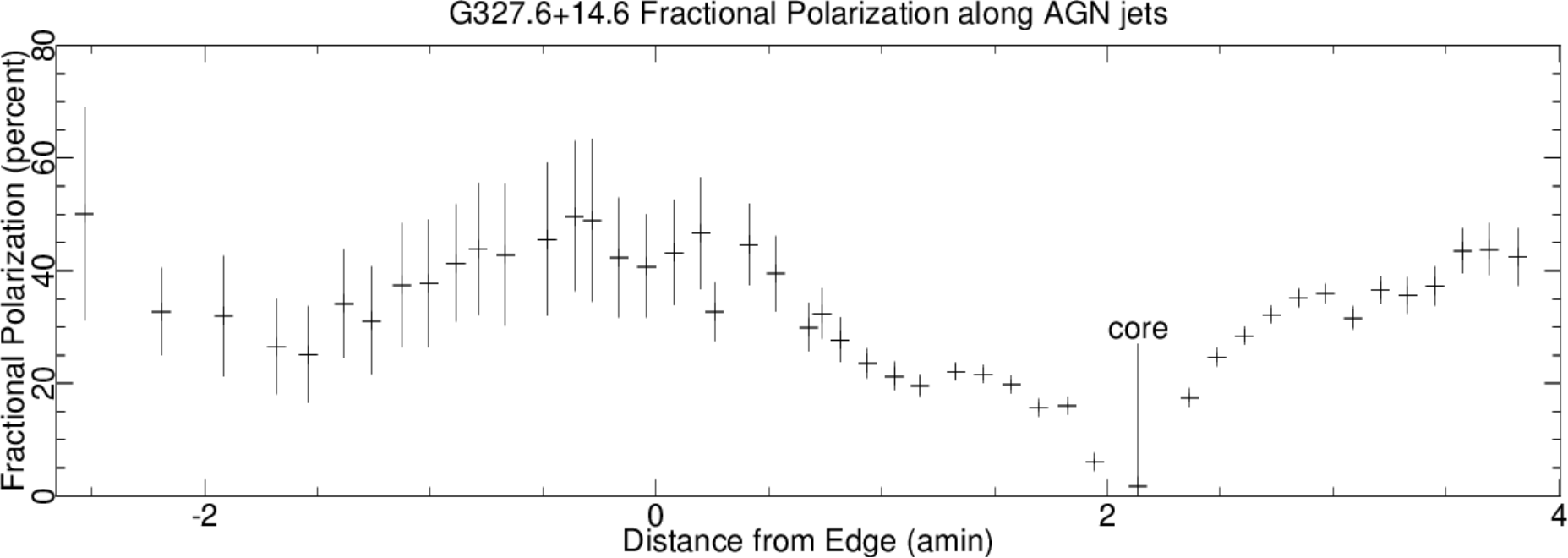}
    \caption{G327.6+14.6: Fractional polarization of the background AGN emission at locations along the FRI jets as a function of distance from the edge of the remnant.  The location of the core component of the AGN is indicated.  Only locations with polarized intensity greater than 50~$\mu$Jy~beam$^{-1}$ and total intensity greater than 80~$\mu$Jy~beam$^{-1}$ are shown.}
    \label{fig:G327.6+14.6_FPol}
\end{figure}

If instead of a uniform, constant Faraday screen filling the remnant, it contained a chaotic and perhaps filamentary magnetized plasma, the effect on the AGN emission passing through it might be to depolarize it, i.e reduce the fractional polarization.  Figure \ref{fig:G327.6+14.6_FPol} shows the fractional polarization at locations along the AGN jet after correcting the jet total intensity for the adjacent remnant emission and the negative bowl surrounding the remnant. The fractional polarization inside and outside the remnant is relatively symmetric, rising to about 40\%.  

\subsubsection{G355.9$-$2.5}
\begin{figure*}
\centerline{
  \includegraphics[width=3.0in]{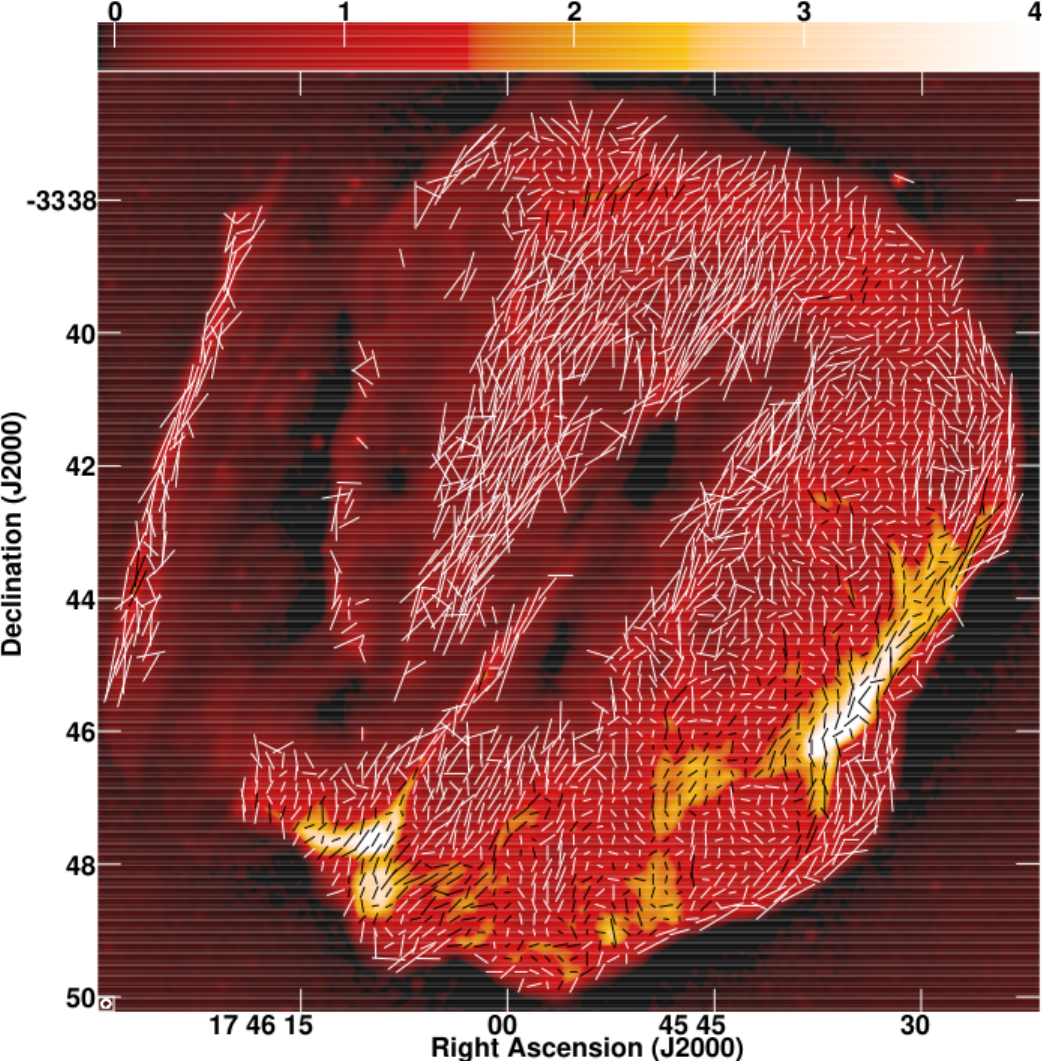}
  \includegraphics[width=3.0in]{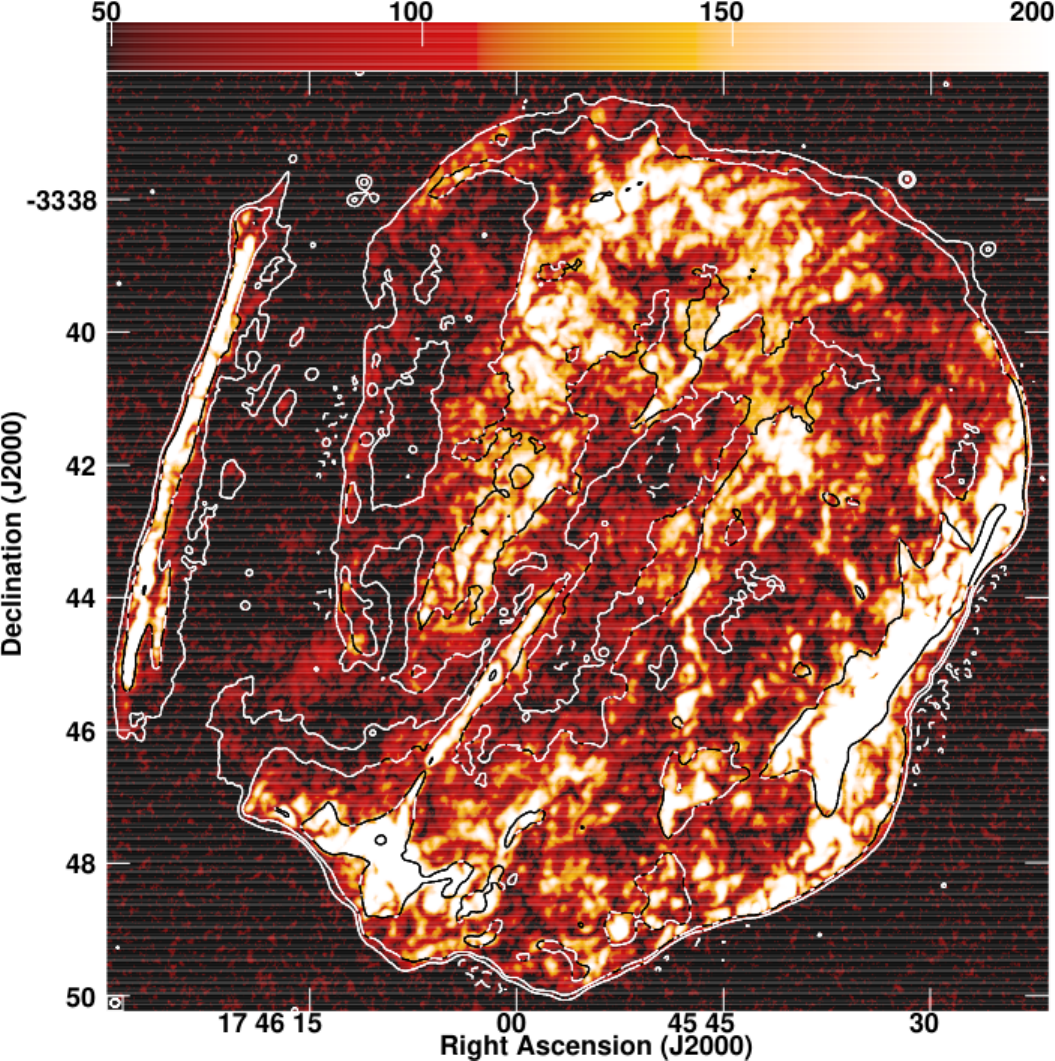}
}
\centerline{
  \includegraphics[width=3.0in]{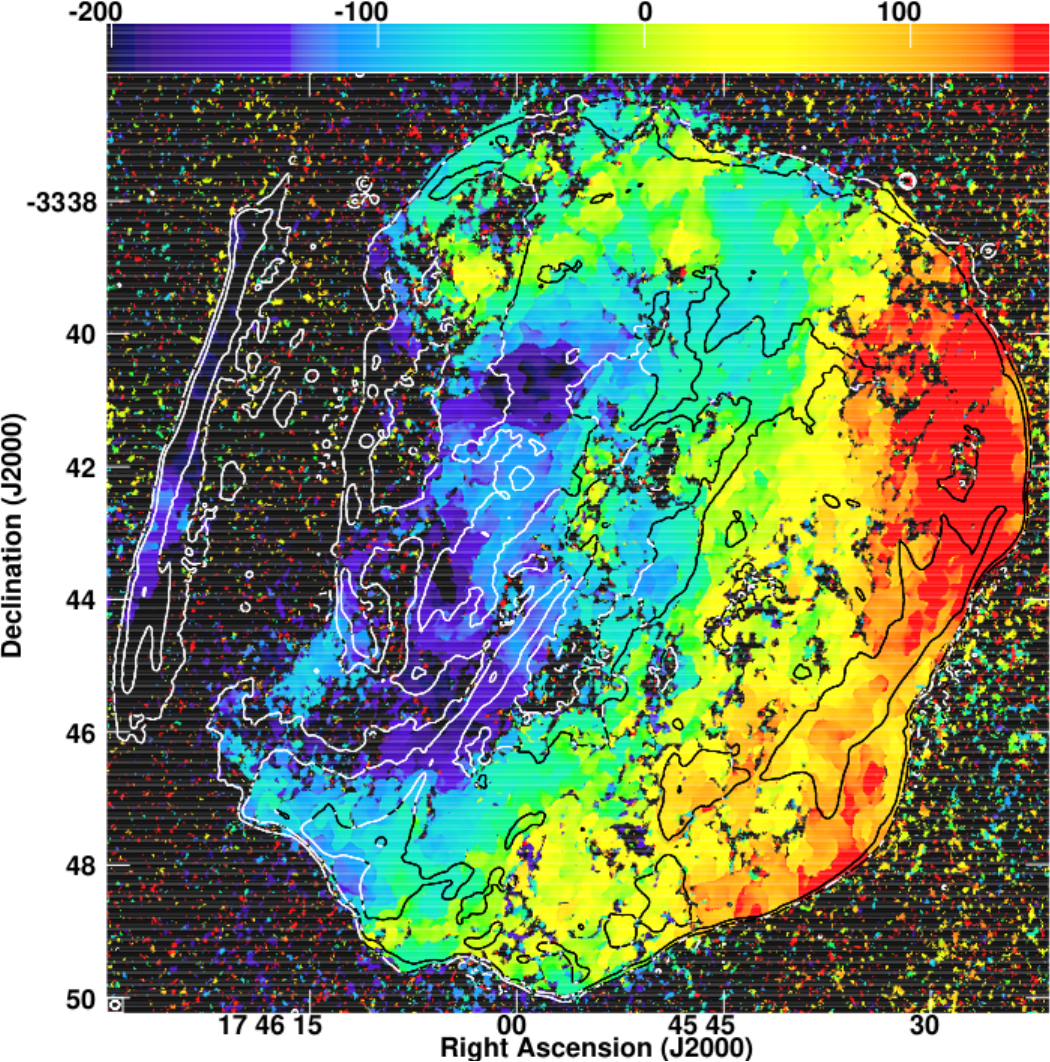}
}
\caption{Like Figure \ref{fig:G315.4Poln} but showing G355.9$-$2.5.
\chg{ The restoring beam is 7.7\asec $\times$ 7.3\asec at position angle = -66.4$^\circ$.}
}
\label{fig:G355.9Poln}
\end{figure*}

G355.9$-$2.5 has strong polarized emission from the remnant with little apparent organization of the B vectors except in the filamentary structure to the east in which the B field runs along the filament; it is shown in Figure \ref{fig:G355.9Poln}.
There is a strong gradient in RM across the remnant, +150 to $-200$ rad m$^{-2}$. 

\subsubsection{G356.2+4.5}
\begin{figure}
\centerline{
  \includegraphics[width=3.0in]{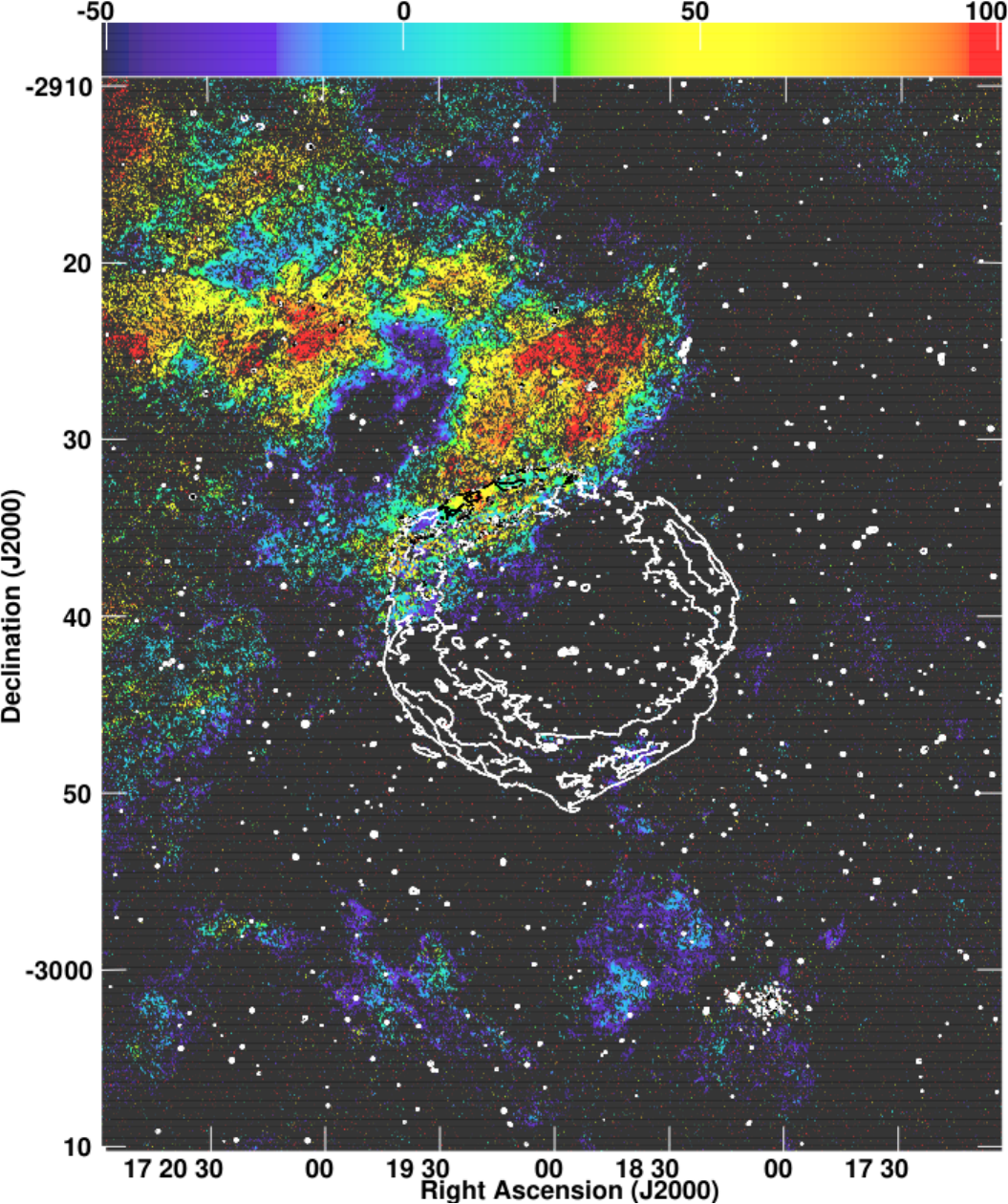}
}
\caption{G356.2+4.5. Peak Faraday depth in color given by scale bar at the top labeled in rad m$^{-2}$ with Stokes I contours at $-$0.1, 0.1, 0.4, 1.6, 6.4, 25.6, and 102.4 mJy/beam. 
\chg{ The restoring beam is 7.6\asec $\times$ 7.2\asec at position angle = -71.4$^\circ$.}
The SNR itself has only weak polarization.
}
\label{fig:G356.2Poln}
\end{figure}

The radio emission from G356.2+4.5, shown in Figure \ref{fig:G356.2Poln} is weakly polarized but there is a Faraday screen visible extending to the north east.
This screen introduces small scale structure in the Q and U images by Faraday rotating the polarized Galactic disk emission.
The corresponding Stokes I emission is smooth on a large enough scale to be completely filtered out by the interferometer array.

\subsubsection{G358.0+3.8}
\begin{figure}
\centerline{
  \includegraphics[width=4.0in]{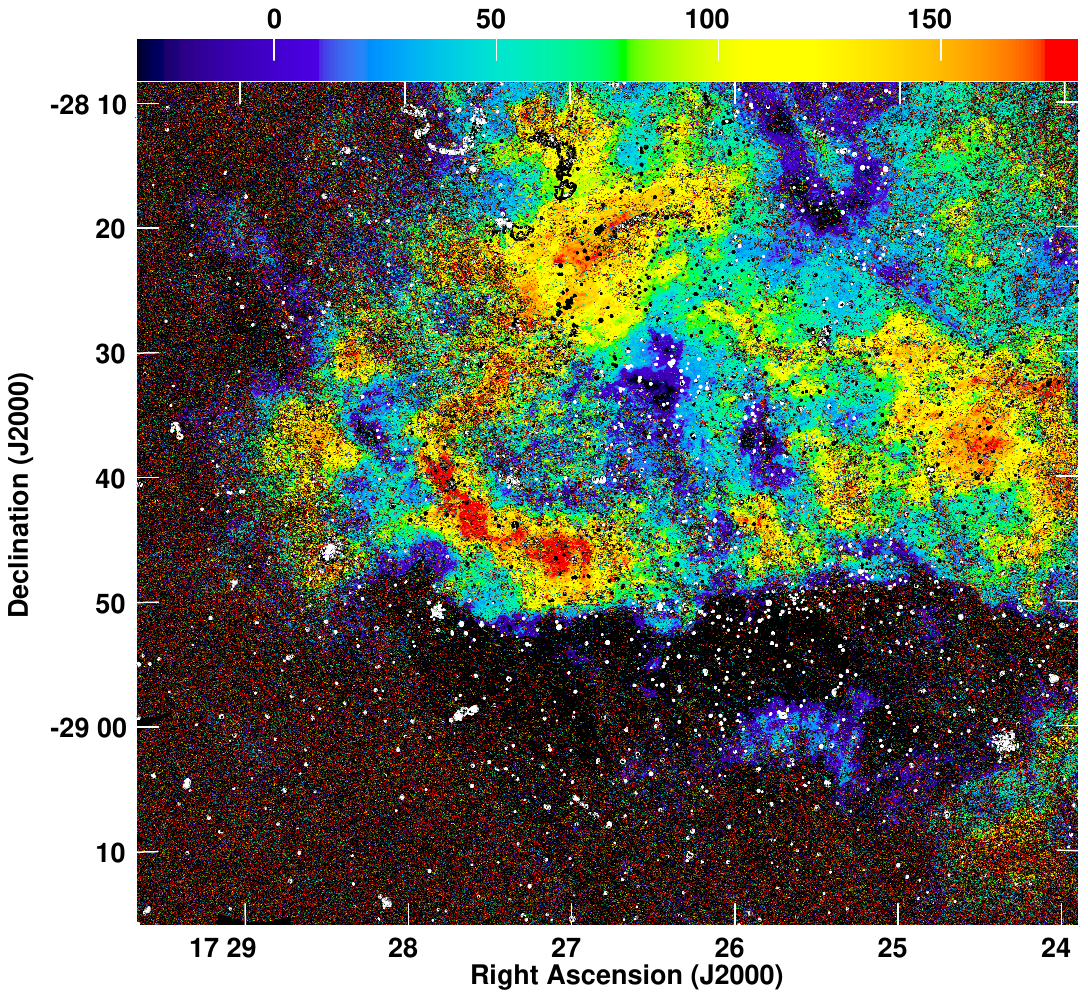}
}
\caption{G358.0+3.8. Peak Faraday depth in color given by scale bar at the top labeled in rad m$^{-2}$ with Stokes I contours (\chg{nearly invisible} white on dark background, black on bright) at $-$0.05, 0.05, 0.2, 0.8, 3.2, 12.8, and 51.2 mJy/beam. 
The SNR itself has only \chg{very} weak total and polarized intensity.
\chg{ The restoring beam is 7.4\asec $\times$ 7.3\asec at position angle = -6.8$^\circ$.}\
}
\label{fig:G358.0Poln}
\end{figure}

The radio emission from G358.0+3.8 (Figure \ref{fig:G358.0Poln}) is faint and weakly polarized but there is a Faraday screen visible extending beyond the extent of the total intensity.
As for G356.2+4.5, the polarized emission visible is from the Faraday rotation of the polarized Galactic disk emission in magnetized plasmas near the SNR.

\subsubsection{G4.8+6.2}
\begin{figure*}
\centerline{
  \includegraphics[width=3.0in]{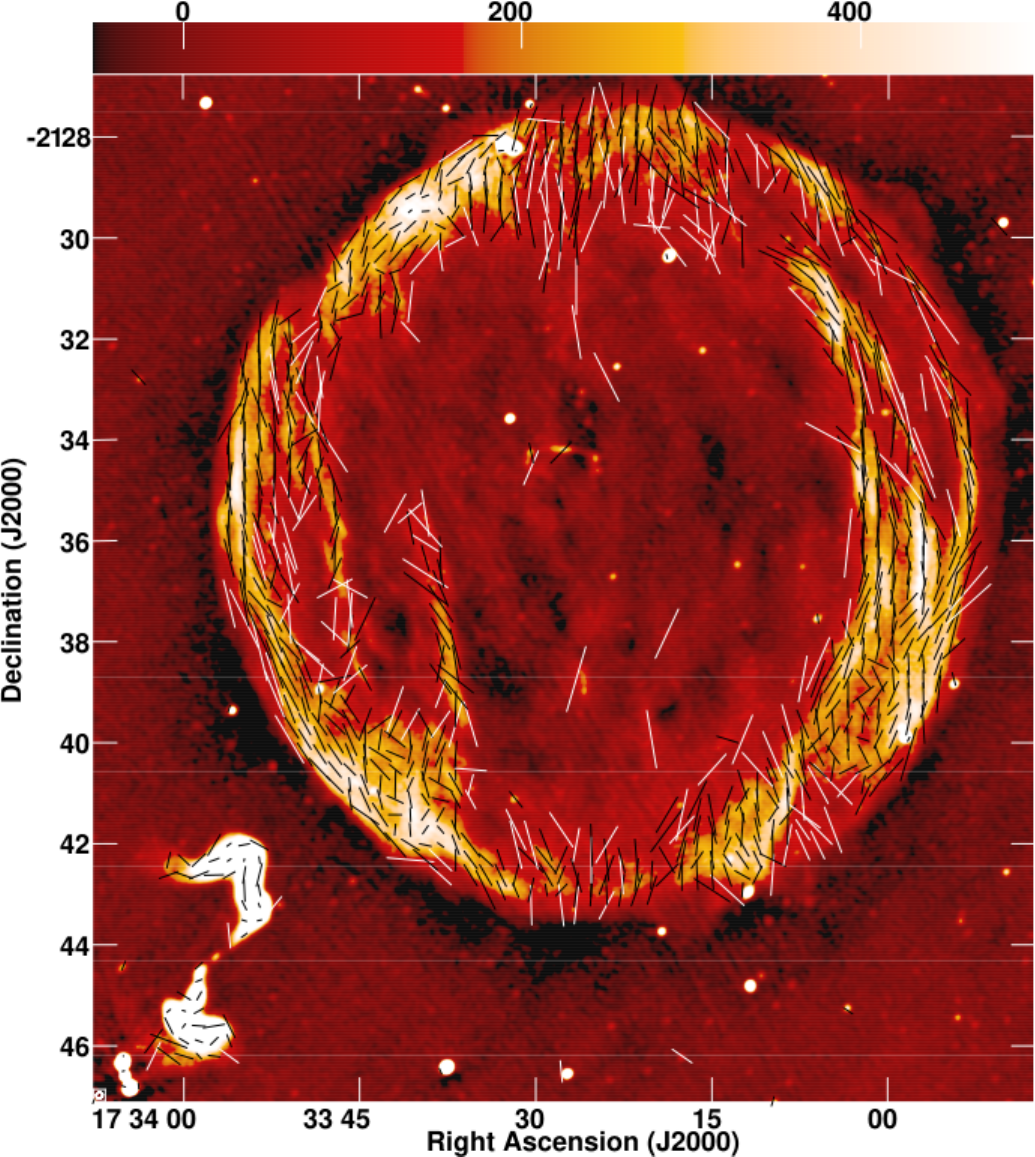}
  \includegraphics[width=3.0in]{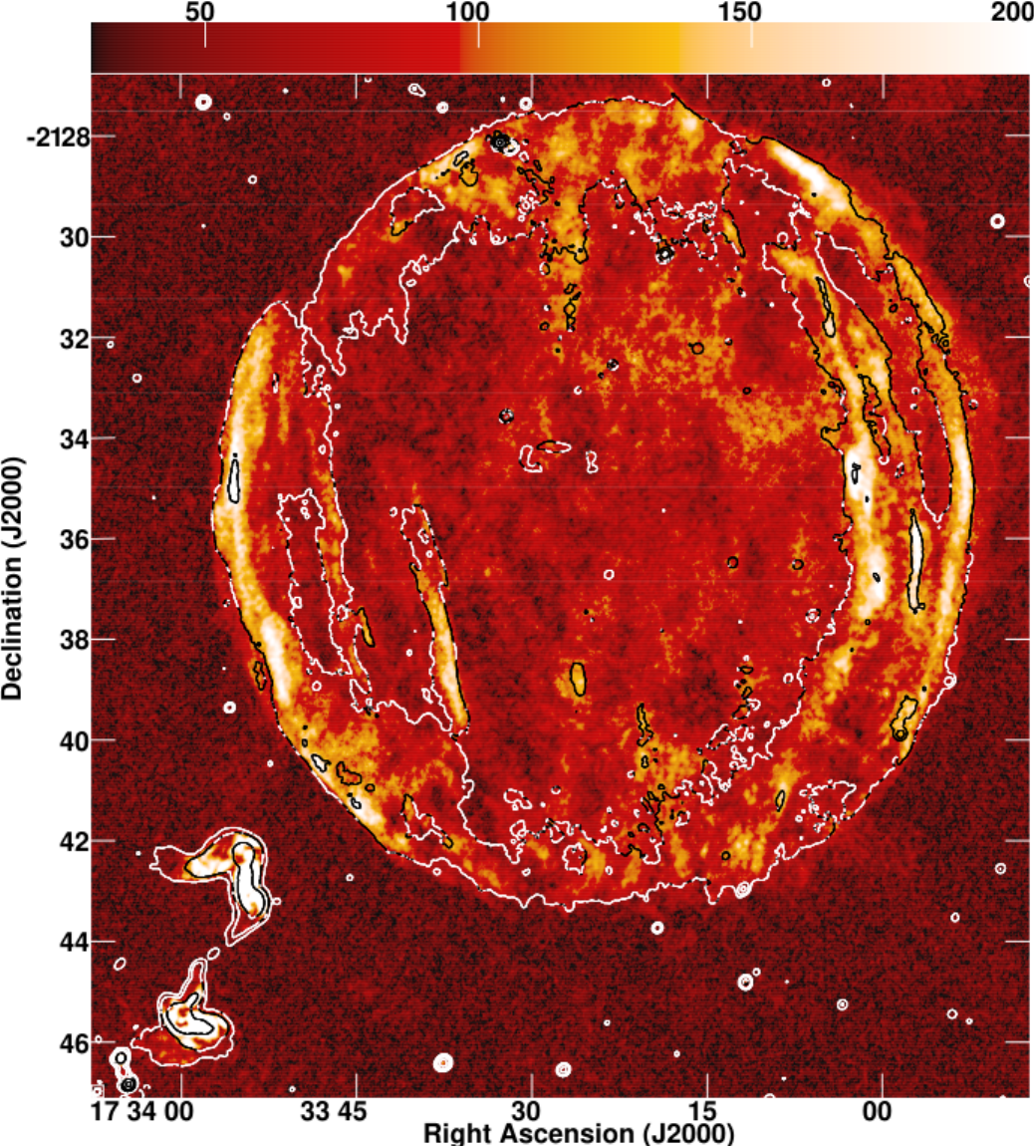}
}
\centerline{
  \includegraphics[width=3.0in]{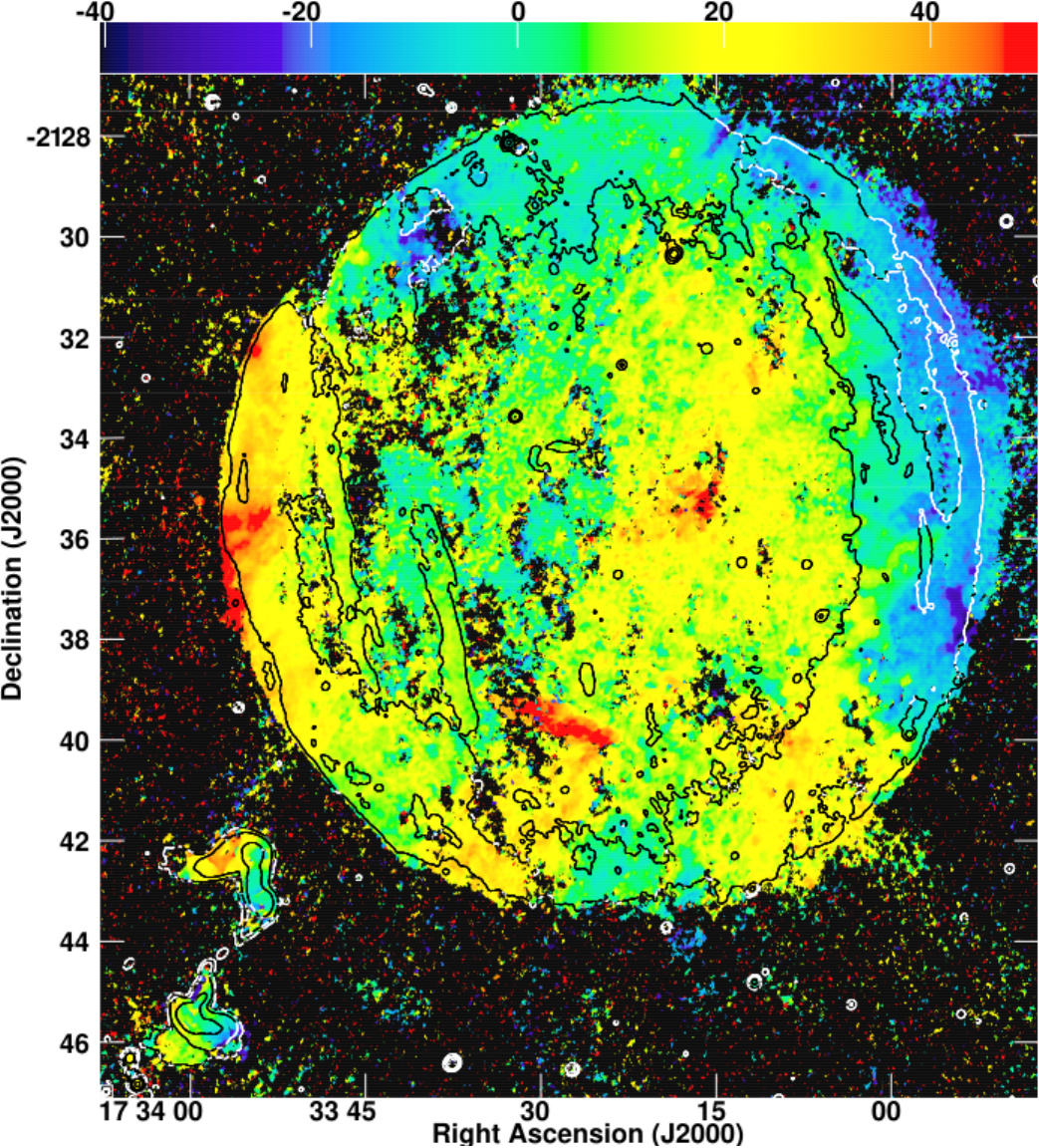}
}
\caption{Like Figure \ref{fig:G315.4Poln} but showing G4.8+6.2.
\chg{ The restoring beam is 7.9\asec $\times$ 7.3\asec at position angle = -47.5$^\circ$.}
}
\label{fig:G4.8Poln}
\end{figure*}

G4.8+6.2 (Figure \ref{fig:G4.8Poln}) has a polarization structure with the B vectors tangent to the shell except to the north and south where the polarization is predominantly radial.
The Faraday rotation structure in front of the remnant varies from $\pm$40 rad m$^{-2}$ indicating reversals in the line of sight magnetic field.
A background AGN has a similar range of Faraday depths.

\subsubsection{G7.7$-$3.7}
\begin{figure*}
\centerline{
  \includegraphics[width=3.0in]{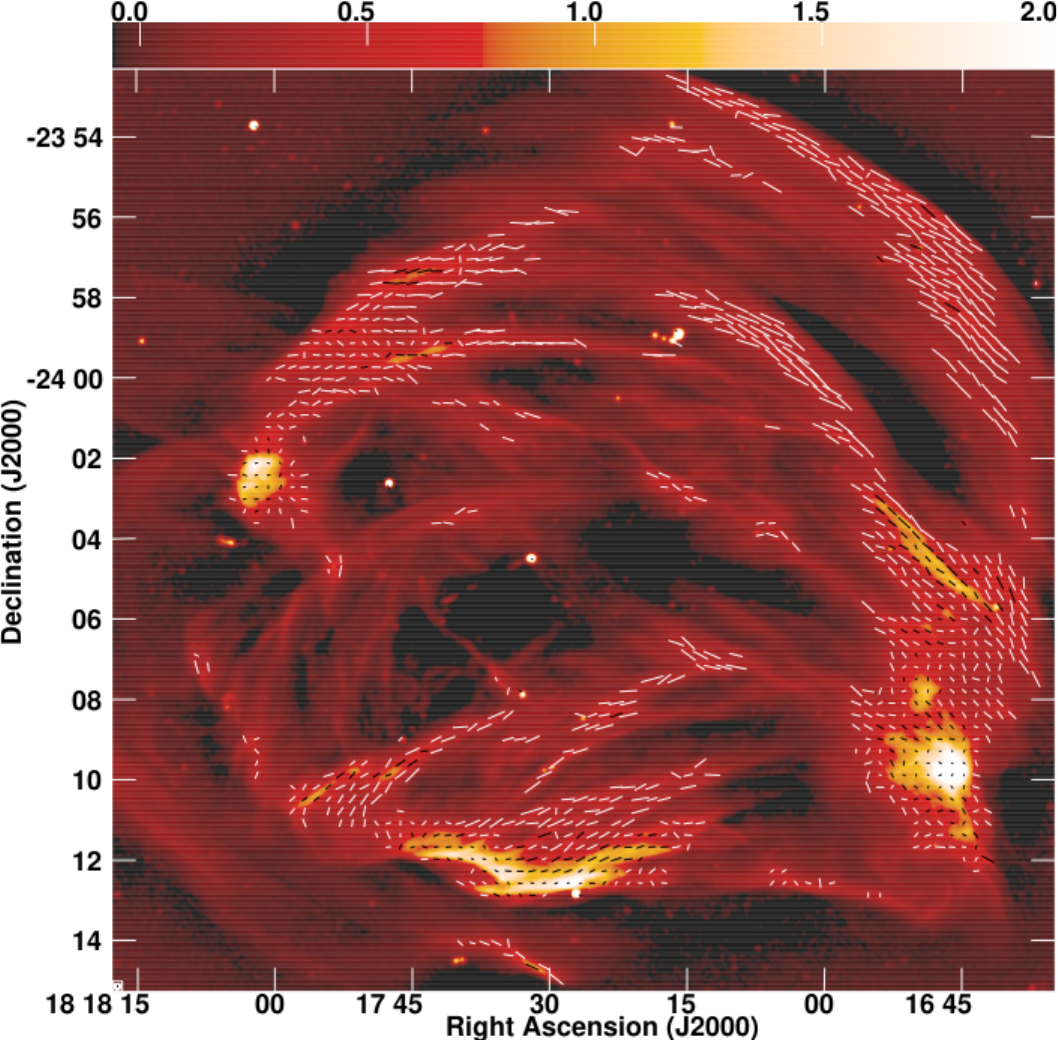}
  \includegraphics[width=3.0in]{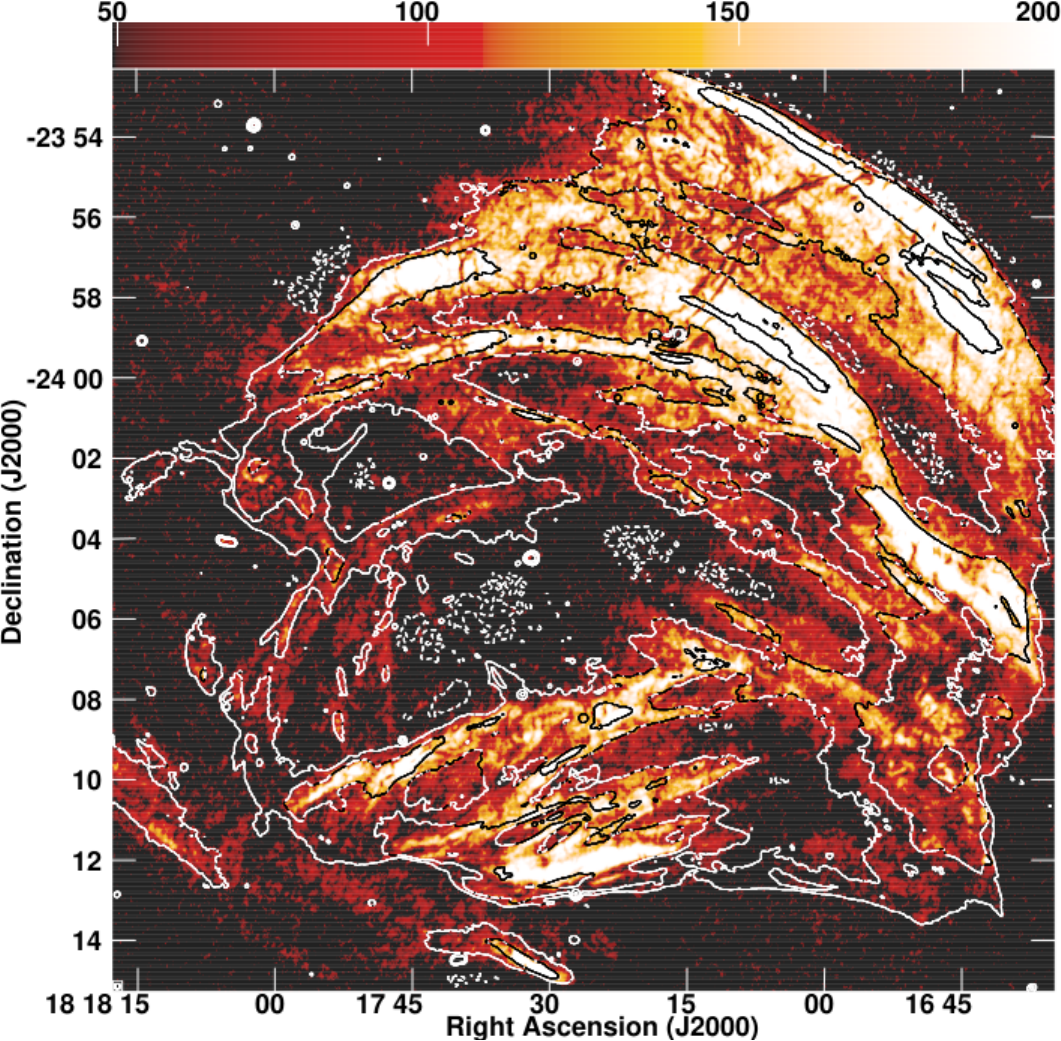}
}
\centerline{
  \includegraphics[width=3.0in]{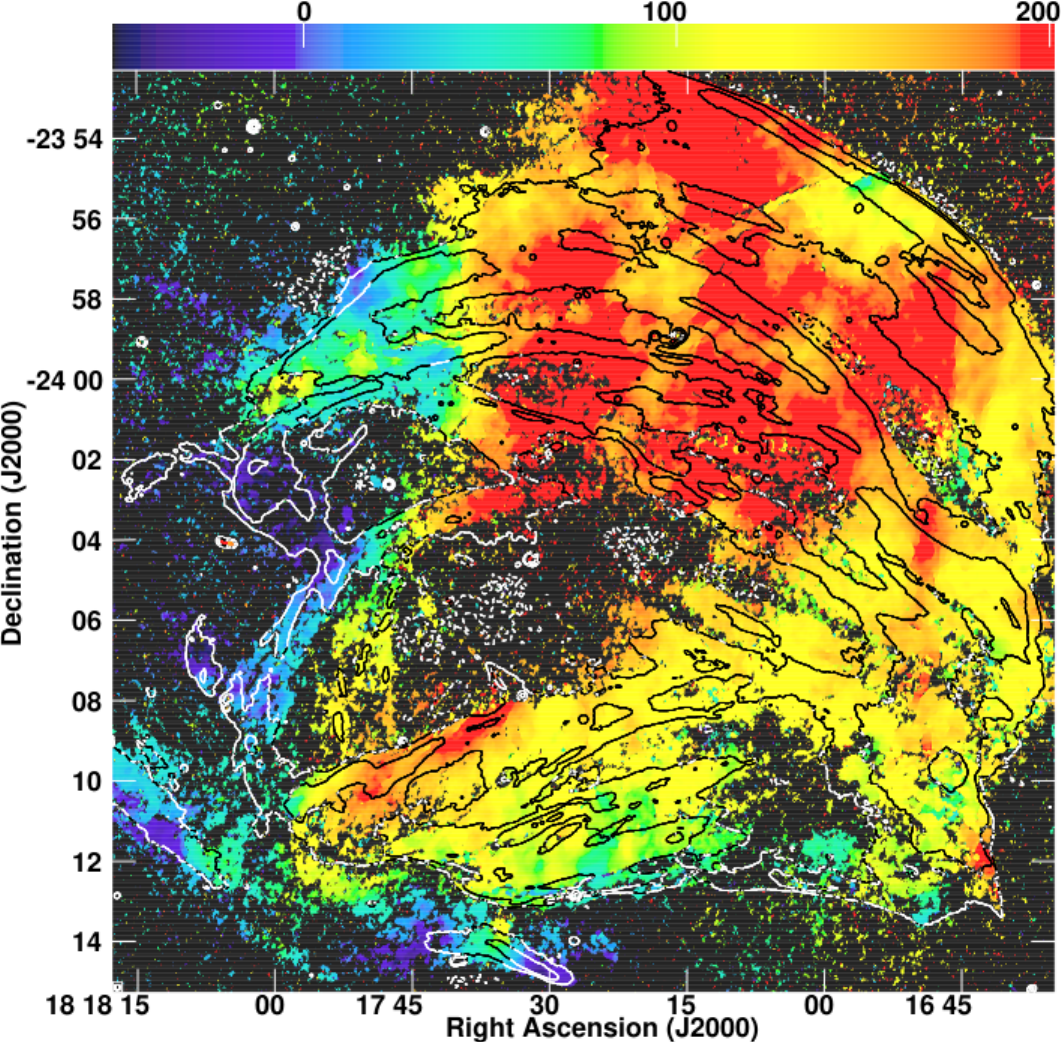}
}
\caption{Like Figure \ref{fig:G315.4Poln} but showing G7.7$-$3.7.
\chg{ The restoring beam is 7.6\asec $\times$ 7.5\asec at position angle = -44.3$^\circ$.}
}
\label{fig:G7.7Poln}
\end{figure*}

G7.7$-$3.7 (Figure \ref{fig:G7.7Poln}) has a strong filamentary structure with the filaments showing significant polarization with the B vectors predominantly aligned with the filaments.
There is a steep gradient in Faraday rotation with the largest values being in the north east.

\subsubsection{G15.1$-$1.6}
In Section \ref{IPol_G15.1} it was suggested that G15.1$-$1.6 was dominated by an HII region.  There was no linear polarization detected from this source although several filaments were bright enough that, if they were nonthermal, detectable linear polarization would be present.  The nondetection of polarized emission supports the conclusion that this object is not a SNR.

\subsubsection{G53.6$-$2.2}
\begin{figure*}
\centerline{
  \includegraphics[width=3.0in]{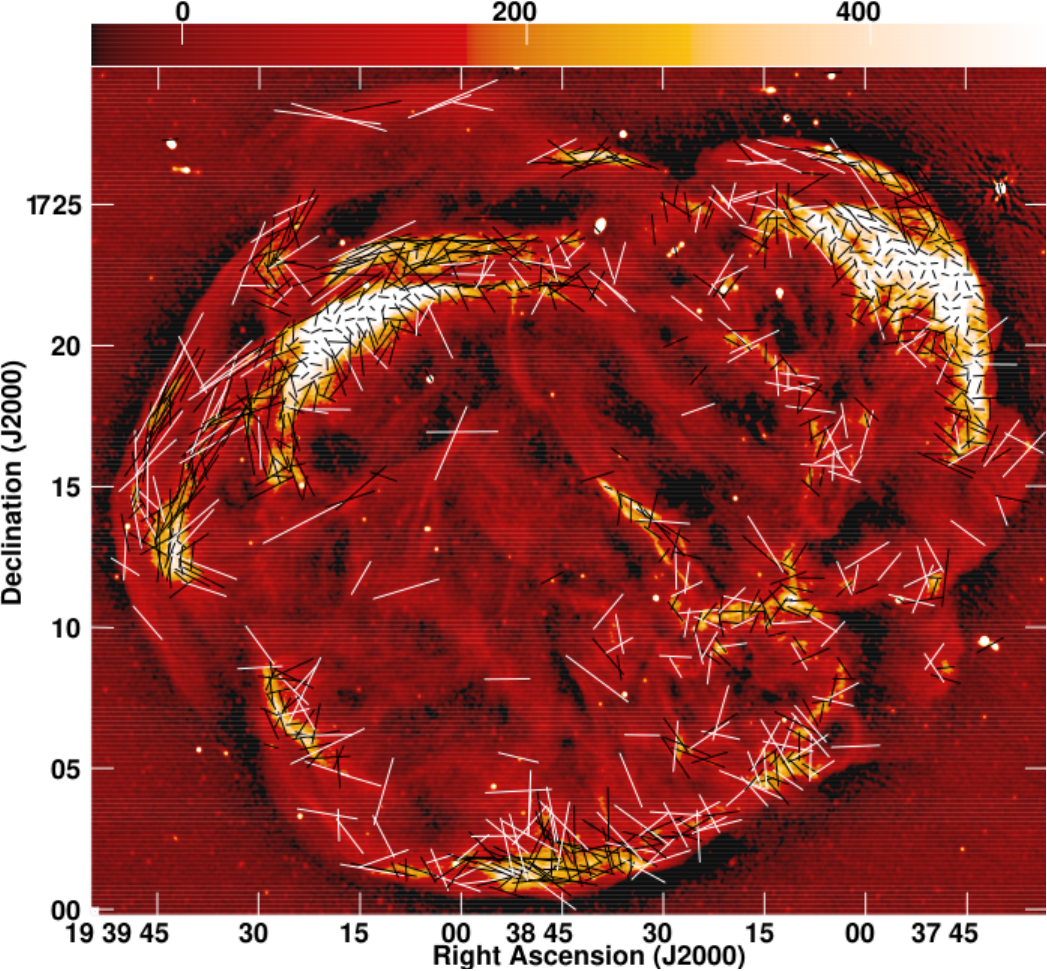}
  \includegraphics[width=3.0in]{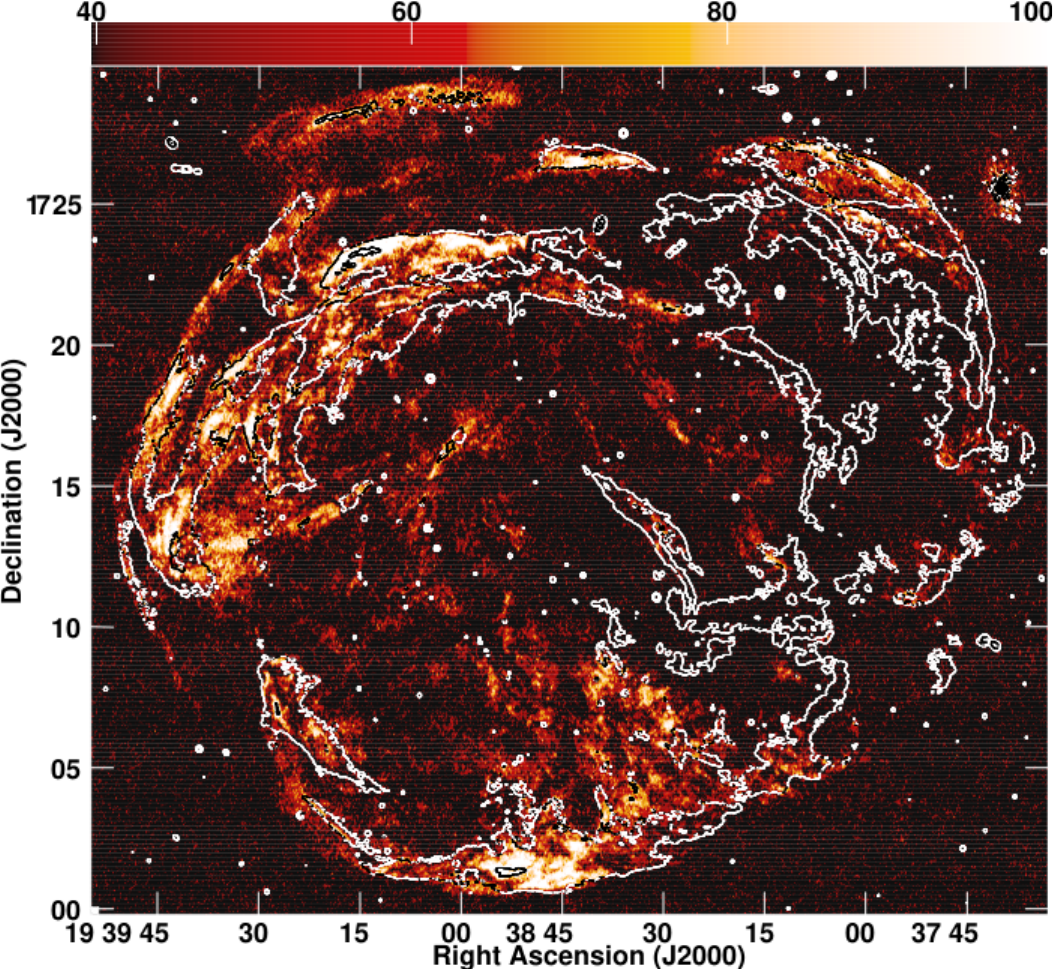}
}
\centerline{
  \includegraphics[width=3.0in]{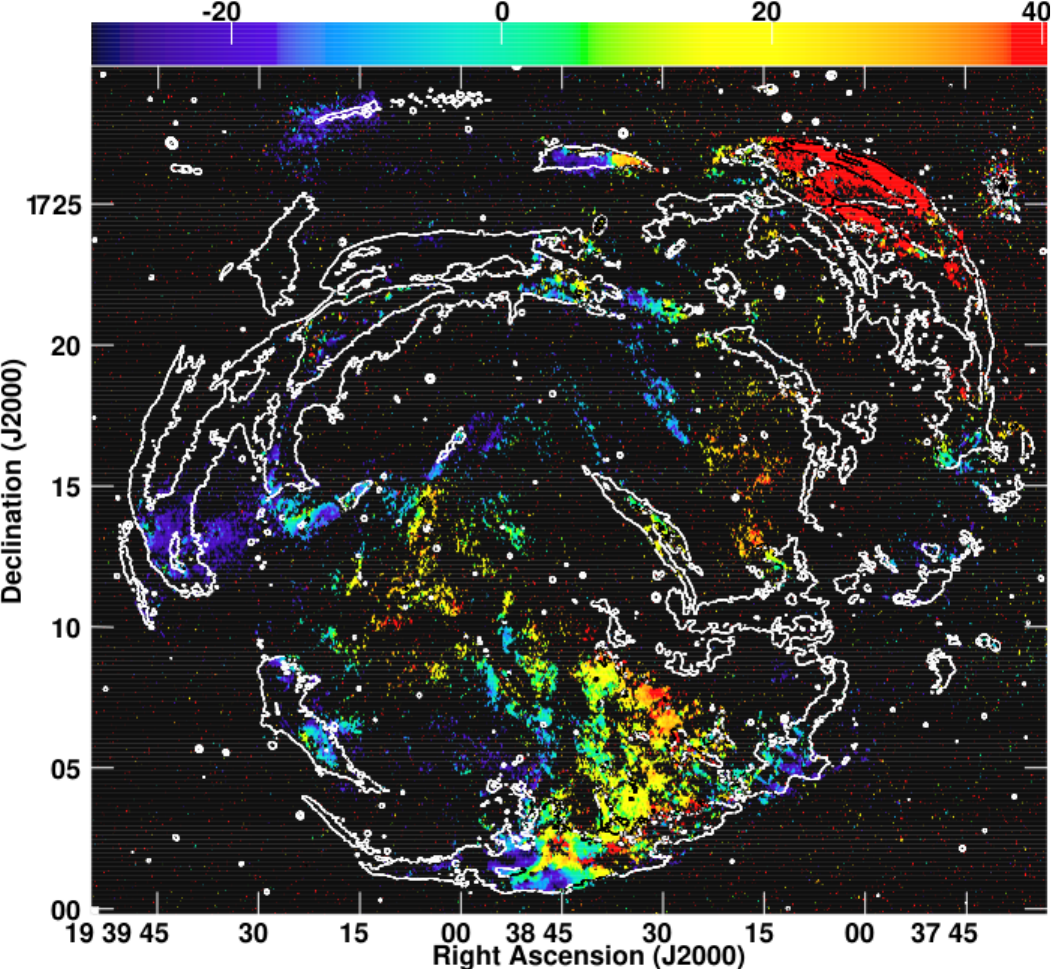}
}
\caption{Like Figure \ref{fig:G315.4Poln} but for G53.6$-$1.8. 
\chg{ The restoring beam is 8.1\asec $\times$ 7.7\asec at position angle = 6.0$^\circ$.}
}
\label{fig:G53.6Poln}
\end{figure*}

The total intensity and polarized emission from G53.6$-$2.2 are dominated by broad filament like structures as shown in Figure \ref{fig:G53.6Poln}.  The magnetic field\chg{s} around the edge of the remnant are largely tangential.

\section{Discussion\label{Discussion}}

\subsection{New radio images of high-latitude SNRs}

For many of the SNRs in our sample there have not been any detailed previous radio images published. In fact, for some of them no previous radio images can be found in the literature. Therefore, it is not surprising that we can give more precise cent\chg{er} coordinates and sizes for many of these sources (Table~\ref{tab:snrchar}). We should also consider giving some of the SNRs new names in Galactic coordinates, in particular those where we found additional significant emission components outside the published dimensions. Significant changes were found for the following SNRs. Archived dimensions were taken from \citet{Green2019}. G17.4$-$2.3 is listed with a diameter of 24\amin\ and we found an extent of more than 1\degr\ and clear indications that there is more emission outside our field of view.  G284.3$-$1.8 is listed as a relatively compact source with a diameter of 24\amin\ and we found numerous related low surface brightness filaments outside the known SNR expanding its diameter to 45\amin. For G332.5$-$5.6 we found a new faint shell outside the known SNR which increased the dimensions from 35\amin\ to 43\amin\ $\times$ 38\amin. G351.0$-$5.4 is listed with a diameter of 30\amin\ and again we found a few faint outer shells that increase the size to 48\amin\ $\times$ 50\amin. 

Most of the supernova remnants in our sample are rather large in angular dimensions, probably because they are at relatively high latitudes and therefore on average they should be closer to the Earth. Sources with large angular dimension have of course missing large scale emission in our MeerKAT observations. Simulating mature regular homogenous SNRs by \citet{Kothes2009} indicates that the missing large-scale emission is coming from  shells that are moving towards and away from us and that typically at least 45--50\% of the emission should be in the small-scale emission from the narrow shells and filaments that we do see. Only if the ambient magnetic field of the SNR is close to the line of sight \chg{do we} not see smooth emission coming from the central area, as the swept-up magnetic field here is almost parallel to the line of sight and therefore does not radiate much synchrotron emission in our direction. Therefore, there are a few simple conclusions we can draw for the nature of the SNRs from the amount of missing large-scale emission.

We were able to integrate 34 flux densities for 32 of our SNRs. There are 2 SNRs, G292.0+1.8 and G326.3$-$1.8, for which we were able to integrate the flux for the PWN and the whole SNR. We believe that we were able to recover 16 full flux densities. That is mainly the small diameter sources and the PWNe. There is of course no clear prescription to determine whether a flux density determined with an interferometer recovers all flux of an extended nebula. This is even worse when there are no good flux densities available in the literature. We deemed a flux density to include all the flux if it agrees within uncertainty with literature values. 

10 of our SNRs have flux densities between 40\% and 60\% of the expected value, which would naively fit into the above mentioned model for SNRs where we miss the emission from the shells that are expanding towards us and away from us. Three SNRs are still above or equal to 70\%, which could indicate a small angle between the line of sight and the ambient magnetic field. This leaves 4 SNRs with less than 40\% recovered flux. Among those are the three largest SNRs in our sample with measured flux densities, G296.5+10.0, G350.0$-$2.0, and G315.4$-$2.3 with 10\%, 20\%, and 30\% recovered flux, respectively. Those are just too large to explain the missing flux with just the shells moving towards us and away from us. G261.9+5.5 is with 35\% recovered flux an odd duck, which shows a lot of substructure and may not fit into our simple model. There is one SNR, G299.2$-$2.9, for which there is no published flux density available, therefore we do not know how much of the flux is recovered. 

\subsection{Bilateral or barrel-shaped SNRs and the Galactic Magnetic Field}

\begin{table}
    \centering
    \caption{\label{tab:bilat} List of bilateral SNRs and their angle $\Psi$ between the axis of symmetry and the Galactic plane. Uncertainties in the determination of $\Psi$ are about 5\degr.}
    \begin{tabular}{|lc|}
    \hline \hline
    SNR & $\Psi$ \\ \hline
    G4.2$-$3.5 & 50\degr \\
    G4.8+6.2 & 36\degr\\
    G5.2$-$2.6 & 25\degr \\
    G6.4+4.0 & 34\degr \\
    G8.7$-$5.0 & 40\degr \\
    G16.2$-$2.7 & 22\degr\\
    G17.8$-$2.6 & 40\degr \\
    G21.8$-$3.0 & 3\degr \\
    G36.6+2.6 & 29\degr \\
    G55.7+3.4 & 84\degr \\
    G57.2+0.8 & 85\degr \\
    G296.5+10.0 & 80\degr \\
    G312.5$-$3.0 & 75\degr \\
    G327.6+14.6 & 82\degr \\
    G332.5$-$5.6 & 90\degr \\
    G350.0$-$2.0 & 5\degr \\
    G351.0$-$5.4 & 50\degr \\
    G353.9$-$2.0 & 3\degr \\
    G356.2+4.5 & 38\degr \\
    G358.0+3.8 & 73\degr \\
    \hline
    \end{tabular}
\end{table}

Mature SNRs typically show a bilateral or barrel-shaped structure, indicative of the expansion inside an approximately uniform ambient medium with a relatively uniform magnetic field \citep{vanderLaan1962,Whiteoak1968}. Those bilateral SNRs can be used to probe the large-scale Galactic magnetic field \citep{Kothes2009} and the orientation of the symmetry axis can give an approximate distance \citep{West2016}. \citet{West2016} showed that for most Galactic SNRs, this bilateral structure can indeed be tied to the large-scale Galactic magnetic field. They used the Galactic magnetic field model of \citet{Jansson2012}, which includes a vertical halo component and a magnetic field parallel to the Galactic plane in the plane of our Galaxy. 

Seven out of the 36 SNRs we observed are in their list of bilateral SNRs (G8.7$-$5.0, G16.2$-$2.7, G36.6+2.6, G296.5+10.0, G350.0$-$2.0, G353.9$-$2.0, and G356.2$+$4.5), and we can add 13 more (see Table~\ref{tab:bilat}). For those 7 bilateral SNRs the angle between the axis of bilateral symmetry and the Galactic model magnetic field agree very well with each other, which confirms a connection between the Galactic field and the bilateral symmetry axis of the SNRs. In our sample of 20 bilateral SNRs we find 11 with $\Psi \le 40\degr$, which we naively call Galactic plane SNRs, 7 SNRs with $\Psi \ge 70\degr$, which we locate in the Galactic halo, and only 2 in between, which might be close to the interface of those two regions. 

\subsection{``Ears'', blowouts, and other protrusions}

In our sample of high-latitude supernova remnants we found that at least half of them show blowouts or protrusions. And for the complex SNRs we cannot really say if they do show blowouts, because of their complexity. This discovery was only possible due to the unprecedented sensitivity and high fidelity of the MeerKAT images to extended emission, as most of these blowouts show an extremely low radio surface brightness. The best-known blowouts in our sample are the ears of Kepler's SNR, which are actually very bright and prominent. There are several scenarios proposed for the formation of ears. Several of these scenarios attribute it to launch of two jets during or post supernova explosion causing protrusion in the forward shock \citep[e.g.,][]{Gaensler1998,Bear2017}. Another model has been proposed by \citet{Chiotellis2021}, in which the two ears are formed through the interaction of the SNR with a bipolar circum-stellar environment. Similarly bright ears can be seen in the SNR G5.2$-$2.6 in the east and maybe in G4.2$-$3.5 in the north-east. Other SNRs that show fainter blowouts along their perimeter are G4.8+6.2, G7.7$-$3.7, G8.7$-$5.0, G17.8$-$2.6, G36.6+2.6, G57.2+0.8, G296.5+10.0, G299.2$-$2.9, G312.5$-$3.0, G315.4$-$2.3, G326.3$-$1.8, G327.6+14.6, and G356.2+4.5. G261.9+5.5 looks like its outer perimeter consists mainly of outflows and G353.9$-$2.0 seems to have an additional faint shell in the north-east outside the main shell. But this could also be the result of a well-structured environment. 

Most of the blowouts seem to indicate that something is breaking through the outer edge of the shell of the SNR. It could be the result of a weakening magnetic field at that place or something from the inside is breaking through like the jet of a pulsar. But the latter would produce at most two breakouts in opposite directions and this is certainly not the case for most of our examples. Another reason, of course, could be a well structured environment. This blowout phenomenon is clearly not well understood, but with the help of the highly sensitive MeerKAT observations and their comparison with other wavelengths we may be able to shed more light on this.

\subsection{Unassociated(?) Faraday Screens}
The two remnants in our sample closest to the Galactic center, G356.2+4.5 and G358.0+3.8, show polarized emission with large and variable Faraday depth near the remnant but which is not clearly related to the remnant.  These appear to be cases of polarized Galactic disk emission viewed through a Faraday screen with sufficiently fine scale structure not to be filtered out by the interferometer array as the corresponding Stokes I emission is.  It is unclear if the Faraday screen involved is associated with, or even physically close to, the remnants.

\subsection{Foreground Faraday Screens}
A number of remnants (G7.7$-$3.7, G326.3$-$1.8, G355.9$-$2.5) have strong gradients and/or local variation in the foreground Faraday screen of hundreds of radians m$^{-2}$.  It is unclear if these screens are related to the remnants.

\subsection{Individual Remnants}
\subsubsection{G327.6+14.6}
This remnant of SN 1006 is only a thousand years old and still in the free expansion phase.  This results in the mostly radial polarization vectors seen in Figure \ref{fig:G327.6Poln}.  Furthermore, this remnant is partly in front of an extended background AGN whose polarized emission can be used to probe Faraday effects inside the remnant.

A uniform Faraday screen inside the remnant would impose a Faraday rotation that would increase with path length through the remnant.  The ring of emission around the edge of the remnant provides polarized emission to probe the foreground Faraday screen.  On many sightlines both the AGN and remnant components are visible in Faraday spectra (Figure \ref{fig:G327.6+14.6_RMSpectra}). Variations in the difference between the peak Faraday depths of the AGN jet and the remnant will map out the difference between the total Faraday depth through the Galaxy and that in front of the remnant.  This should be dominated by any differences inside the remnant. Except for hints of an effect near the edge of the remnant, Figure \ref{fig:G327.6+14.6_AGN_Rem_RM} shows the AGN--remnant difference to be relatively constant.

The other extreme would be if the remnant were filled with a chaotic magnetized plasma with fine scale variations in total Faraday depth, on a smaller scale than the resolution of the images and large enough to cause beam depolarization.  The portion of the AGN jets seen through the remnant would have systematically lower fractional polarization than regions outside the remnant.  Figure \ref{fig:G327.6+14.6_FPol} show this not to be the case.  There is no evidence in our data for a magnetized thermal plasma capable of significant Faraday rotation inside the remnant.
\subsubsection{G326.3$-$1.8}
Both the total intensity and polarized emission from this remnant is dominated by its pulsar wind nebula. A prominent ridge in total intensity on the south--east corner of the PWN is totally depolarized (see Figure \ref{fig:G326.3Poln}).  This must be due to a particularly dense and variable Faraday screen.

\section{Summary\label{Summary}}
We present MeerKAT full-Stokes observations of 36 supernova remnants from the catalog of \citet{Green2019}, all but one at relatively high Galactic latitudes.  The bulk of these are well imaged and many represent major improvement over previously available radio images.  We found that G30.7$-$2.0 is not a SNR but three relatively bright sources appearing to form an arc, while G15.1$-$1.6 appears more likely to be an HII region.

Imaging in Stokes Q and U allows probing the magnetic fields inside the radio emitting region of the remnants as well as magnetized thermal plasma in front of polarized emission. A prominent feature of the PWN in G326.3$-$1.8 is totally depolarized, apparently due to a particularly dense Faraday screen. In the special case of the thousand year old remnant G327.6+14.6, a very extended background AGN with polarized jets allows testing for Faraday rotation interior to the remnant.  No evidence for such an effect is found.

The magnetic field inside G327.6+14.6 has a largely radial magnetic field (Figure \ref{fig:G327.6Poln}) whereas G4.8+6.2 has a magnetic field which is mostly tangential except in the blowout regions where it is radial (Figure \ref{fig:G4.8Poln}).  The magnetic field structure in other remnants is less well defined but where there is strong filamentary structure, the magnetic field runs along the filaments.

Several of our sources have revealed a bilateral or barrel-shaped structure. Such structures have been ubiquitous in mature SNRs and have been modeled as the SNR expanding into an ambient Galactic magnetic field including a vertical halo component supporting the presence of an off-plane vertical component to the Galactic magnetic field \citep{West2016}.
This indicates the presence of magnetic fields in the evolution of SNRs and can potentially be used to determine approximate SNR distances.

\section*{Acknowledgments}
\chg{We would like to thank the anonymous reviewer for comments resulting in an improved paper.}
The MeerKAT telescope is operated by the South African Radio Astronomy Observatory, which is a facility of the National Research Foundation, an agency of the Department of Science and Innovation.
The National Radio Astronomy Observatory is a facility of the National Science Foundation, operated under a cooperative agreement by Associated Universities, Inc.
This research has made use of the NASA/IPAC Extragalactic Database (NED), which is operated by the Jet Propulsion Laboratory, California Institute of Technology, under contract with the National Aeronautics and Space Administration.

\vspace{5mm}
\facilities{MeerKAT}

\software{Obit \cite{OBIT}}
\bibliography{MK_SNR}{}
\bibliographystyle{aasjournal}


\end{document}